\documentclass[aps,pre,twocolumn,amsfonts,amssymb,amsmath,showpacs,floatfix, superscriptaddress, nofootinbib]{revtex4}

\ifx\pdftexversion\undefined
\usepackage[dvips]{graphicx}
\else
\usepackage[pdftex]{graphicx}
\fi
\usepackage{url}        

\usepackage{amsfonts,amsthm}
\usepackage{amsmath}

\usepackage{bbm}
\graphicspath{{pics/}{}}

\newcommand{\hs}{\hspace{6pt}}
\newcommand{\eps}{\epsilon}
\newcommand{\beps}{\boldsymbol{\epsilon}}
\newcommand{\bdel}{\boldsymbol{\delta}}
\newcommand{\dd}{\partial}

\newcommand{\uu}{\mathbf{u}}

\newcommand{\VV}{\mathbf{V}}
\newcommand{\WW}{\mathbf{W}}
\newcommand{\MM}{\mathbf{M}} 

\newcommand{\SSS}{\mathbf{S}}
\newcommand{\QQ}{\mathbf{Q}}
\newcommand{\XX}{\mathbf{X}} 

\newcommand{\VO}{ \mathbf{V}_{(0)} }
\newcommand{\WO}{ \mathbf{W}^{(0)} }

\newcommand{\tuu}{\tilde{\mathbf{u}}}
\newcommand{\OO} {\mathcal {O}}

\newcommand{\s}{\sigma}
\newcommand{\la}{\lambda}
\newcommand{\LA}{\Lambda}

\newcommand{\HP}{\hat{\mathbf{P}}}

\newcommand{\HL}{\hat{\mathbf{L}}}
\newcommand{\HPI}{\hat{\mathbf{\Pi}}}

\newcommand{\bF}{\mathbf{F}}

\newcommand{\RRR}{\mathbf{R}}

\newcommand{\nn}{\nonumber}

\newcommand{\bP}{\mathbf{P}}
\newcommand{\tP}{\tilde{\mathbf{P}}}
\newcommand{\bpsi}{\boldsymbol{\psi}}
\newcommand{\bsig}{\boldsymbol{\sigma}} 
\newcommand{\bone}{\boldsymbol{1}} 
\newcommand{\bY}{\mathbf{Y}}
\newcommand{\BLA}{\boldsymbol{\Lambda}}

\newcommand{\braket}[2]{\langle #1 \mid #2 \rangle  }
\newcommand{\bra}[1]{\langle #1 \mid  }
\newcommand{\ket}[1]{\mid #1 \rangle  }

\newcommand{\bsub}{\begin{subequations}}
\newcommand{\esub}{\end{subequations}}
\newcommand{\ba} {\begin{align} }
\newcommand{\ea}{ \end{align} }

\newcommand{\zeroket}{| \boldsymbol{0} \rangle}
\newcommand{\bzero}{ \boldsymbol{0} }

\newcommand{\etal}{\textit{et al.}}

\newcommand{\hmu}{{\hat{\mu}}}
\newcommand{\hnu}{{\hat{\nu}}}

\newcommand{\BTH}{\boldsymbol{\Theta}}
\newcommand{\XI}{\boldsymbol{\Xi}}
\newcommand{\bom}{ \boldsymbol{\omega}}
\newcommand{\bka}{ \boldsymbol{\kappa}}
\newcommand{\bxi}{ \boldsymbol{\xi}}
\newcommand{\KK}{\mathbf{K}}
\newcommand{\BB}{\mathbf{B}}
\newcommand{\TT}{\mathbf{T}}

\newcommand{\dxspiral}{\textsc{dxspiral}}

\newcommand{\komega}{\bar{\omega}}

\newcommand{\wfr}{w_\mathrm{fr}}

\newcommand{\rmd}{\mathrm{d}}
\newcommand{\rmi}{i}
\newcommand{\rme}{e}

\begin{document}

\title[Effective dynamics of scroll waves]{Effective dynamics of twisted and curved scroll waves using virtual filaments}
\author{Hans Dierckx} \author{Henri Verschelde}
\affiliation{Department of Mathematical Physics and Astronomy, Ghent
   University, Krijgslaan 281 S9 WE05, 9000 Ghent, Belgium}

\begin{abstract}
Scroll waves are three-dimensional excitation patterns that rotate around a central filament curve; they occur in many physical, biological and chemical systems.  We explicitly derive the equations of motion for scroll wave filaments in reaction-diffusion systems with isotropic diffusion up to third order in the filament's twist and curvature. The net drift components define at every instance of time a virtual filament which lies close to the instantaneous filament. Importantly, virtual filaments obey simpler, time-independent laws of motion which we analytically derive here and illustrate with numerical examples. Stability analysis of scroll waves is performed using virtual filaments, showing that filament curvature and twist add as quadratic terms to the nominal filament tension. Applications to oscillating chemical reactions and cardiac tissue are discussed.
\end{abstract}


\maketitle


\section{Introduction}

\begin{figure}[b] \centering
   \includegraphics[width=0.5\textwidth]{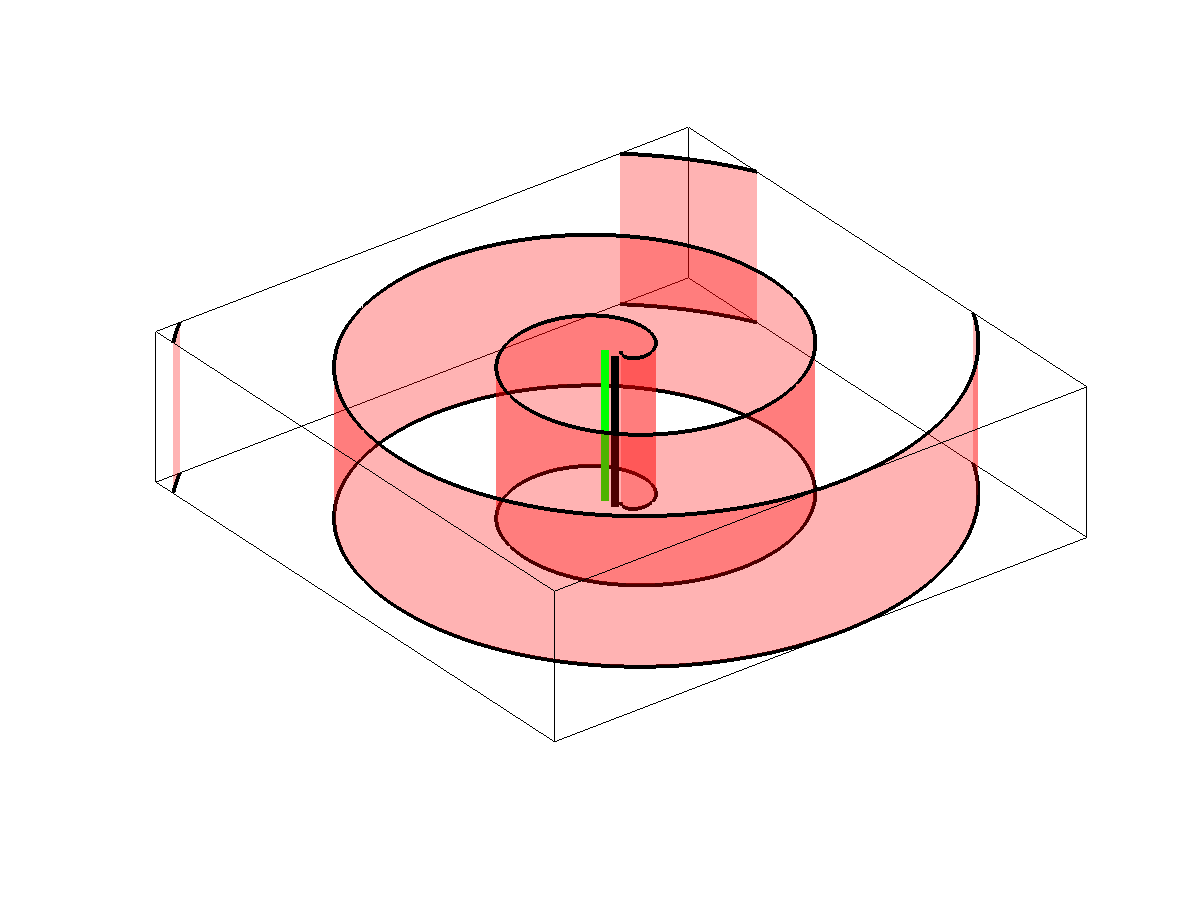}\ \
   \caption[scrollwave example]{(Color online) Example of straight scroll wave for a system with equal diffusion of variables. The red surface is the wave's activation front, the tip line is shown in black and filament curve in green (grey). For the unperturbed scroll wave that is shown here, the gauge and virtual filaments defined below coincide.
}\label{fig:scrollexample}
\end{figure}

Many natural systems exhibit spiral-shaped waves of self-activation. Important examples include the propagation of signaling waves in biological tissues \cite{Winfree:1980, Lechleiter:1991}, propagation of concentration waves in chemical systems \cite{Winfree:1973, Kapral:1995} and cAMP waves during the aggregation of a social amoeba \cite{Siegert:1992}. 
The occurrence of non-linear traveling waves in these media leads to complex wave dynamics, which has been intensively studied by theoretical, experimental and numerical means \cite{Zykov:1987, Keener:1988, Biktashev:1994, Mikhailov:1995, Gray:1995, Fenton:2002}. Already in two spatial dimensions, wave dynamics becomes very interesting, since broken wave fronts will develop into rotating spiral-shaped patterns, called spiral waves. Similar activation patterns have been   observed in experimental recordings of cardiac tissue during atrial tachycardia \cite{Allessie:1973, Davidenko:1992}, which motivated numerous investigations on spiral waves as the underlying mechanism for atrial arrhythmias \cite{Gray:1998,Witkowsky:1998}. In fact, the oscillatory BZ reaction has been used for a decade as a experimental model for spiral waves in cardiac tissue. 
In three spatial dimensions, a broken wave front will develop into a so-called scroll wave, which is the three-dimensional generalization of a spiral wave. Scroll waves rotate around a center line or filament \cite{Clayton:2005}, an example of which is shown in figure~\ref{fig:scrollexample}. 
It moreover appeared that the evolution of a scroll wave is mainly governed by the dynamics of its filament, which is why filaments are sometimes called `organizing centers' for scroll wave activity \cite{Henze:1990, Biktashev:1994}. The concept of a scroll wave filament can therefore simplify the mathematical description of excitation processes, which is a much desired quality in the description of complex systems. In terms of practical applications, scroll wave filaments are frequently being used to describe time-evolution of vortex patterns in the chemical BZ reaction \cite{Jahnke:1988}, and the degree of complexity during ventricular fibrillation \cite{Clayton:2002, Sambelashvili:2003, Clayton:2005, Yamazaki:2012}. Unfortunately, analytical derivations have so far only covered the leading order dynamics of scroll wave filaments \cite{Keener:1988, Biktashev:1994, Verschelde:2007, Dierckx:2009}, not capturing the twist-induced destabilization of filaments. Previous theoretical extensions of filament dynamics in this direction have been limited to phenomenological models \cite{Echebarria:2006, Dierckx:2012}. In this paper, we build further on the seminal works by Keener \cite{Keener:1988} and Biktashev and co-workers \cite{Biktashev:1994}, to rigorously derive laws of motion for scroll wave filaments that also include twist-curvature coupling.

The first asymptotic description of scroll wave dynamics was furnished by Keener~\cite{Keener:1988}. In his work, Frenet-Serret coordinates were constructed around the filament curve to map the standard spiral wave solutions to the three-dimensional medium. Thereafter, he applied the Fredholm alternative theorem to obtain necessary conditions on the filament motion, which are the desired equations of motion. The introduction of critical adjoint eigenmodes of the system, which are also known as `response functions' (RFs) \cite{Biktasheva:1998} formed the basis of many subsequent analytical results on pattern evolution in excitable systems \cite{Biktashev:1994, Biktashev:1994b, Barkley:1992, Barkley:1994, Henry:2002, Biktasheva:2003, Biktasheva:2006, Verschelde:2007, Biktashev:2010, Dierckx:2010, Dierckx:2012, Dierckx:2013}. Strikingly, the RFs were observed to be strongly localized near the spiral wave's rotation centre, which can be rightfully called `particle-wave dualism of spiral waves' \cite{Biktasheva:2003}. The localized sensitivity of scroll waves to external perturbations justifies the effective description of scroll waves in terms of their filaments. This work will make use of RFs to derive the effective laws of motion for scroll wave filaments.

Keener's law of filament motion, which already involved temporal averaging over one rotation period, was later simplified by Biktashev and co-workers, who noticed that some of the coefficients vanish when averaged in time \cite{Biktashev:1994}. The resulting equation of motion (EOM) for a filament curve $\vec{X}(s,t)$ with phase $\phi$, arc length $s$ and time $t$ reads:
 \begin{eqnarray}
\dd_t \phi &=& \komega_0 + a_0 w^2 + d_0 \dd_s w, \nn \\
 \dd_t \vec{X} &=&  \gamma_1 \dd_s^2 \vec{X} + \gamma_2 \dd_s \vec{X} \times \dd_s^2 \vec{X}. \label{eom_bik}
\end{eqnarray}
The latter equation can be equivalently written $\dd_t \vec{X} =   \gamma_1 k \vec{N} + \gamma_2 k \vec{B} $, with $k$ the filament's curvature and $ \vec{N}, \vec{B}$ the normal and binormal vector to the filament curve. The system-dependent coefficient $\gamma_1$ was identified as the filament tension \cite{Biktashev:1994}; it is the foremost important characteristic of a scroll wave. For, in systems where $\gamma_1$ is positive, curved filaments tend to minimize their length and therefore straighten up. If the tension is negative, however, the filament may become unstable and break-up. This mechanism has been named as a possible pathway to ventricular fibrillation \cite{Fenton:2002, Alonso:2007}. We will show below that the filament's curvature and twist will alter the effective tension of a filament, and may therefore also affect its stability. 

One may note in Eq. \eqref{eom_bik} that there is no coupling between the translational and rotational degrees of freedom. Nevertheless, it was known from numerical experiments that  sufficiently twisted scroll waves yield helical filaments \cite{Margerit:2002, Hakim:2002}. In \cite{Echebarria:2006}, the so-called ribbon model was proposed to describe this `sproing' phenomenon as a supercricital pitchfork bifurcation. In the present paper, we improve on the ribbon model in several ways: with our ab initio calculation we show that some terms were missing and provide explicit expressions for the occurring coefficients in terms of RFs. Our method allows to predict quantitatively the critical twist beyond which filaments take a helical shape.

Not all analytical approaches to asymptotic filament dynamics involve RFs. A first complementary description is known as the kinematic approach \cite{Zykov:1987, Mikhailov:1994, Zykov:1996}, in which the motion of spiral and scroll waves is fully determined by the curvature of the wave front near the wave break. In the large-core limit, response functions for broken wave fronts may be used to establish the basic assumptions of the kinematic approach \cite{Mikhailov:1994}. A second approach consists of mapping known spiral solutions to the local neighborhood of the filament curve to find the leading order dynamics. Particular successes were achieved with this method to describe filament drift in excitable media with anisotropic diffusion, which has an important application in cardiac tissue modeling \cite{Berenfeld:1999, Berenfeld:2001, Setayeshgar:2002, Wellner:2002, Wellner:2010}. After Wellner's discovery that such anisotropic systems can be efficiently treated by a curved-space formalism \cite{Wellner:2000, Wellner:2002, tenTusscher:2004, Verschelde:2007, Young:2010}, more results on wave dynamics under generic local anisotropy followed \cite{Verschelde:2007, Dierckx:2009, Dierckx:2010, Dierckx:2012, Dierckx:2013}. For simplicity, we choose not to consider filament dynamics in anisotropic media here; such generalization can be found in \cite{Dierckx:2010}. From the curved-space viewpoint on anisotropy, wave dynamics may be locally approximated by an isotropic medium as long as spatial variations in anisotropy occur on a scale larger than the spiral wave's core size. Our present working hypothesis is thus highly similar to choosing a local inertial frame in general relativity theory. The results are therefore also expected to well approximate scroll wave dynamics in anisotropic cardiac tissue. 

In this paper, we derive the instantaneous equations of motion for scroll wave filaments starting from the reaction-diffusion equations. However, since the coefficients in the EOM depend on the phase of rotation, they form a non-linear quasi-periodic set of partial differential equations (PDEs) which is hard to analyze. Here, one may be inclined to average the coefficients in the PDEs over a full rotation cycle as was performed before \cite{Keener:1988, Biktashev:1994, Verschelde:2007, Dierckx:2010, Dierckx:2012}. However, the exact solution to an averaged non-linear differential equation can not be expected to be exactly equal to the time-average of the exact solution (see, e.g. \cite{Coakley:1969}). In the works \cite{Keener:1988, Biktashev:1994, Verschelde:2007, Dierckx:2012}, this difference was not a concern since only the lowest order dynamics was pursued. Here, we analytically solve the EOM for small time intervals around an arbitrary moment of time and demonstrate that the gauge filament performs an epicycle motion around a curve which we shall call the `virtual' filament. This curve satisfies simpler laws of motion, with coefficients that do not depend on the absolute phase of rotation. 
Our final result can therefore be stated as follows:  
starting from a reaction-diffusion equation (RDE)
\begin{equation}
   \dd_t \uu(\vec{x}, t) = D_0 \Delta \bP \uu(\vec{x}, t) +
   \bF(\uu(\vec{x}, t) ) \label{RDE1}
\end{equation}
that supports rigidly rotating spiral wave solutions in two spatial dimensions, we prove that the scroll wave filament lies at all times close to a so-called virtual filament curve with twist $w$ and curvature $k$ that are of order $\la$, which obeys 
\bsub\label{eomvir_intro} \begin{eqnarray}
\dd_t \phi &=& \komega_0 + a_0 w^2 + b_0 k^2 + d_0 \dd_s w + \OO(\la^4),\\
 \dd_t \vec{X} &=&  \left( \gamma_1 + a_1 w^2 + b_1 k^2 + d_1 \dd_s w \right) \dd_s^2 \vec{X} \\ && + \left( \gamma_2 + a_2 w^2 + b_2 k^2 + d_2 \dd_s w \right) \dd_s \vec{X} \times \dd_s^2 \vec{X} \nn \\
&& +\, c_1 w \dd_s^3 \vec{X}+ c_2 w \dd_s \vec{X} \times \dd_s^3 \vec{X} \nn \\&& -\, e_1 \dd_s^4 \vec{X} - e_2 \dd_s \vec{X} \times \dd_s^4 \vec{X} + \OO(\la^5), \nn
\end{eqnarray}\esub
where $\phi$ and $s$ are the phase and arc length along the virtual filament. These laws of motion extend the previous analytical results \cite{Keener:1988, Biktashev:1994, Henry:2002, Verschelde:2007} and the phenomenological models of \cite{Echebarria:2006, Dierckx:2012}. Importantly, the coefficients that govern filament dynamics emerge in our theory as overlap integrals of response functions. Therefore, we can predict time-evolution of three-dimensional scroll waves based on the properties of the unperturbed spiral wave solution in two spatial dimensions. Our results hold for systems with equal diffusion of variables $(\bP=\bone)$ such as the BZ reaction, as well as for unequal diffusion systems. The last class contains the monodomain models of cardiac tissue, where the first variable $u^1$ denotes the transmembrane potential of cardiac cells, and $\bP = \mathrm{diag}(1,0,\ldots,0)$.

This paper is structured as follows. In the next section, we discuss the mathematical tools needed in the subsequent calculations, including reference frames attached to moving curves and response function theory. In relation to scroll waves, a distinction is made between tip lines and gauge filaments. In the third section, we derive the instantaneous EOM for scroll wave filaments starting from the RDE \eqref{RDE1}. In section~\ref{sec:virtual}, the instantaneous EOM is solved for small times, which shows that a gauge filament performs an epicycle motion around the virtual filament curve that obeys Eq. \eqref{eomvir_intro}. In the fifth section, the novel laws of motion up to third order in twist and curvature are discussed. Simplifications are given for the case of chemical reactions, where all species have equal diffusion coefficients. Finally, linear stability analysis of scroll rings is performed, leading to an expression for the effective tension of scroll wave filaments. In the last section, we verify some of the coefficients for the equal diffusion case with numerical simulations. 

\section{Methods and terminology}

\subsection{Tip line and scroll wave core}

The spiral wave tip is commonly defined as the intersection of two isolines of state variables \cite{Krinsky:1992, Biktashev:1998, Fenton:1998, Clayton:2005}. Sometimes, the pair $(u^1 =u_c, \dd_t u^1 =0)$ is chosen, such that the spiral tip lies in the region where the front and tail of the spiral wave meet \cite{Zykov:1987, Clayton:2002, Fenton:1998}. In three dimensions, the intersection of two isosurfaces delivers what we will call the \textbf{tip line} of the scroll wave. In numerical simulations of reaction-diffusion systems, the tip line can be easily tracked \cite{Fenton:1998}; therefore, most renderings of the scroll wave rotation center are strictly speaking tip lines. Note, however, that the definition of the tip line clearly depends on the chosen isosurfaces. As different observers may not agree on the thresholded variables and values, each observer will see his or her own tip line. If the scroll wave is stationary, the tip line trajectory will be periodic or quasi-periodic, marking a tubular region known as the scroll wave's \textbf{core}. 

When the effects of incomplete recovery (i.e. refractoriness) do not come into play, the motion of a spiral's tip is identical for all time frames. This simplest case results in a circular movement of the spiral tip. Our present derivation is strictly valid only in this circular core regime. In a different parameter regime, the spiral tip motion may not be circular but quasi-periodic, leading to flower-like or star-like trajectories in the absence of external perturbations \cite{Winfree:1991, Krinsky:1992, Braune:1993, Qu:2000}. This regime is called meandering and has been related to the symmetries of the Euclidean plane in \cite{Barkley:1994, Biktashev:1996, Sandstede:1997, Sandstede:1999}. In some detailed models of cardiac excitation, the diffusive properties around the spiral tip are even more pronounced, such that eventually only the refractoriness of previously excited tissue governs the tip trajectory; this regime delivers so-called linear cores \cite{Krinsky:1992}.

\subsection{Coordinate frames adapted to filament shape \label{sec:fermi}}

\begin{figure*}[t] \centering
   \raisebox{5.5cm}{a)\ }\includegraphics[width=0.23\textwidth]{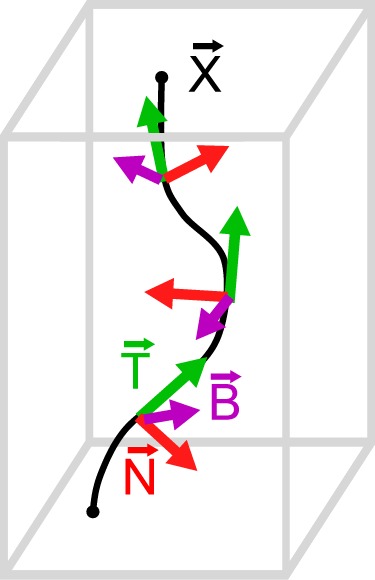}
   \raisebox{5.5cm}{b)\ }\includegraphics[width=0.23\textwidth]{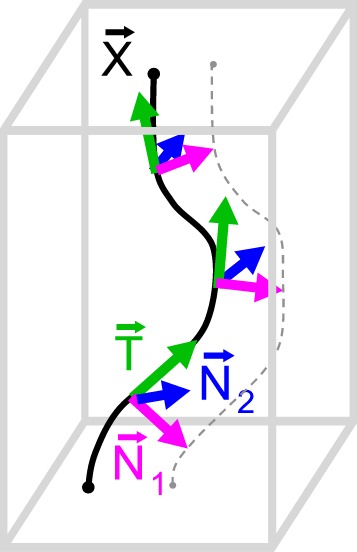}
   \raisebox{5.5cm}{c)\ }\includegraphics[width=0.23\textwidth]{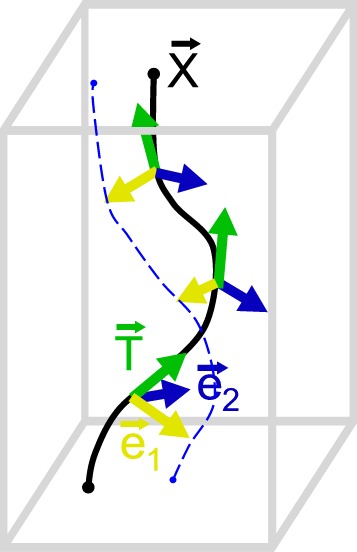}
   \caption[Orthonormal triads adapted to the filament]{(Color online) Orthonormal frames adapted to the shape of the filament curve $X^i(s)$. a)~Frenet-Serret frame, in which the frame twist follows the geometric torsion $\tau_g$ of the filament curve. b)~Relatively parallel adapted frame (Fermi frame), where the transverse vectors $\vec{N_a}$ are parallel transported along the filament. The dashed curve connecting the end points of $\vec{N}_1$ is therefore relatively parallel to the filament. c)~Phase-adapted frame, where the transverse basis vectors $\vec{e}_1$ point towards the scroll wave's tip line (dashed curve). Here, the filament and tip line span the ribbon from the model presented in \cite{Echebarria:2006}.
}\label{fig:frames} 
\end{figure*}

To start, we will consider filaments as being smooth curves in three-dimensional Euclidean space, endowed with Cartesian coordinates $x^i$  $(i \in \{1,2,3 \})$. Such curves exhibit geometric curvature $k$ and torsion $\tau_g$, as is well known from their description in the Frenet-Serret frame \cite{Keener:1988, Biktashev:1994} displayed in Fig. \ref{fig:frames}a. Denoting the filament position at a given instance of time as $x^i = X^i(s)$ ($i \in \{1,2,3 \}$) and assuming sufficient smoothness of the curve, one may subsequently differentiate with respect to arc length $s$ to obtain a right-handed orthonormal triad given by the tangent vector $\vec{T} = \dd_s \vec{X}$, normal vector $\vec{N}$ and binormal vector $\vec{B}$:
\begin{eqnarray}\label{Frenetframe}
 \dd_s \left(
    \begin{array}{c}
      \vec{T} \\
      \vec{N} \\
      \vec{B} \\
    \end{array}
  \right) =
  \left(
    \begin{array}{ccc}
      0 & k & 0 \\
      -k & 0 & \tau_g \\
      0 & -\tau_g & 0 \\
    \end{array}
  \right)
    \left(
    \begin{array}{c}
      \vec{T} \\
      \vec{N} \\
      \vec{B} \\
    \end{array}
  \right).
\end{eqnarray}
Noteworthily, the Frenet-Serret frame is degenerate in points where the filament curvature $k$ vanishes.

In this work, however, we choose to work with orthonormal frames $(\vec{T}, \vec{N_1}, \vec{N_2})$ that are adapted to the scroll wave's phase rather than the geometric torsion of the filament. In general, the coefficient matrix that can be used to frame a curve will be skew-symmetric \cite{Bishop:1975}:
\begin{eqnarray}\label{parallelframe}
 \dd_s \left(
    \begin{array}{c}
      \vec{T} \\
      \vec{N_1} \\
      \vec{N_2} \\
    \end{array}
  \right) =
  \left(
    \begin{array}{ccc}
      0 & \LA^1 & \LA^2 \\
      -\LA^1 & 0 & \wfr \\
      -\LA^2 & -\wfr & 0 \\
    \end{array}
  \right)
    \left(
    \begin{array}{c}
      \vec{T} \\
   \vec{N_1} \\
      \vec{N_2} \\
    \end{array}
  \right).
\end{eqnarray}
The components $\LA^a$ $(a\in\{1,2\})$ are simply the projections of $k \vec{N}$ onto the reference vectors $\vec{N_a}$, i.e.
\begin{eqnarray}
 \LA^a = \vec{N}^a \cdot \dd_s^2 \vec{X}, \qquad(a\in\{1,2\}). 
\end{eqnarray}
The frame twist $\wfr$ describes the rotation rate of the basis vectors $\vec{N_a}$ as one moves along the filament curve: 
\begin{eqnarray}
\wfr = \vec{N_2} \cdot \dd_s \vec{N_1} =  \left( \vec{N_1} \times \dd_s \vec{N_1} \right) \cdot \vec{T}. 
\end{eqnarray}
Various choices for $\wfr$ are possible. First, the choice of a Frenet-Serret frame as in \cite{Keener:1986, Biktashev:1994} corresponds to $\wfr = \tau_g$; see Fig. \ref{fig:frames}.  

Secondly, if one sets $\wfr=0$, the basis vectors $\vec{N_a}$ are parallel transported along the filament curve, such that they form a relatively parallel adapted frame \cite{Bishop:1975} which is also known as a Fermi frame and shown in Fig. \ref{fig:frames}b. Here, the dashed curve connects the end points of the $\vec{N}_1(s)$ vectors; this curve is relatively parallel to the filament. 

Thirdly, one may direct the $\vec{N_1}$ such that it points to a reference point of the scroll wave, such as its tip line. This choice is displayed in Fig. \ref{fig:frames}c, where the dashed line represents the scroll wave's instantaneous tip line. In this case, $\vec{N_1}$ can be identified with the `fiducial vector' of \cite{Echebarria:2006}, such that $\wfr$ equals the scroll wave twist $w$. 

The reference frame which has so far only been defined on the filament curve can be extended to the immediate vicinity of the filament using Fermi coordinates $(\rho^1, \rho^2, \s)$ \cite{Fermi:1922, Verschelde:2007, Dierckx:2009}, at a fixed moment of time:
\begin{eqnarray}
 x^i(\rho^a, \s) = X^i(\s) + \rho^a N_a^i(\s). \label{FCa}
\end{eqnarray}

\subsection{Response functions \label{sec:RF}}

In this section, we briefly review response function theory and introduce sign conventions for going from the Cartesian basis of \cite{Keener:1988, Verschelde:2007, Dierckx:2012} to the complex-valued representation of \cite{Biktashev:1994, Biktasheva:2009}.

When a small stimulus is applied to a spiral or scroll wave, the wave usually only reacts by spatial or rotational drift. If one knows all elementary responses of a spiral wave solution in two dimensions to small external perturbations $-\mathbf{h}(x,y)$ to the right-hand side of Eq. \eqref{RDE1}, one may represent the total drift as an overlap integral of the perturbation $\mathbf{h}$ with the so-called response functions \cite{Biktashev:1994, Biktasheva:1998, Biktasheva:2003, Verschelde:2007} $\bY^x, \bY^y, \bY^\theta$:
\begin{eqnarray}
  \dot{X}  = \braket{\bY^x}{\mathbf{h}}, \nn\\
  \dot{Y}  = \braket{\bY^y}{\mathbf{h}},  \label{RFh} \\
  \dot{\phi}-\omega_0 = \braket{\bY^\theta}{\mathbf{h}},  \nn 
\end{eqnarray}
where the inner product is given by
\begin{eqnarray}
\braket{\mathbf{f}(x,y) }{\mathbf{g}(x,y)} = \iint_{\mathbb{R}^2} \rmd x \rmd y\ \mathbf{f}(x,y)^H
\mathbf{g}(x,y).
\end{eqnarray}
Importantly, it was found that response functions (RFs) are closely related to the symmetries of the RDE \eqref{RDE1} in two spatial dimensions \cite{Keener:1988, Biktashev:1994b}, which allows to compute them numerically for given diffusion matrix $\bP$ and reaction kinetics $\bF(\uu)$ \cite{Hakim:2002, Biktasheva:2009}. 

Before studying the RDE in a rotating frame of reference, it is necessary to reconcile sign conventions for the angular frequency, since we will be an extension of the \dxspiral\ programme of Biktasheva \etal. \cite{dxspiral, Biktasheva:2009} to compute RFs. In \dxspiral, the angular frequency of spirals $\omega_0$ is always taken positive; the program possesses a sign flag $K$ to denote whether rotation is clockwise ($K=1$) or counterclockwise ($K=-1$). Therefore, the polar angle $\theta$ of a point fixed to the rotating spiral has a time derivative $\komega_0$ which is positive  only when the spiral rotates counterclockwise:
\begin{eqnarray}
   \dd_t \theta = \komega_0 = - K \omega_0. \label{defsignflag}
\end{eqnarray}
In the works \cite{Verschelde:2007, Dierckx:2009, Dierckx:2012}, the quantity $\komega_0$ was used, whereas in \cite{Biktasheva:2009} a clockwise rotating spiral was assumed ($K=1$). To preserve full compatibility with those earlier works, we will explicitly write the sign flag $K \in\{-1,+1\}$ throughout this paper.

In a reference system that co-rotates with an unperturbed spiral wave solution $\uu_0(x',y')$ that has angular frequency $\komega_0$, one thus finds
\begin{multline}
   \dd_{t'} \uu(\vec{x}', t') = \\\HP \Delta' \uu(\vec{x}', t') +
   \komega_0 \dd_{\theta'} \uu(\vec{x}', t') + \bF(\uu(\vec{x}', t') ). \quad\label{RDEcomoving}
\end{multline}
Since $\uu_0(x',y')$ is a time-independent solution to this equation, one may differentiate with respect to either $x',y',\theta'$ to find
\begin{eqnarray} \label{GMcart}
\HL \ket{\dd_{x'} \uu_0} &=& - \komega_0 \ket{\dd_{y'} \uu_0}, \nn \\
\HL \ket{\dd_{y'} \uu_0} &=& \komega_0 \ket{\dd_{x'} \uu_0}, \\
\HL \ket{\dd_{\theta'} \uu_0} &=& \ket{\bzero} \nn 
\end{eqnarray}
with the linear operator ($\Delta'_2 = \dd_{x'}^2+ \dd_{y'}^2$):
\begin{eqnarray}
 \HL = \HP \Delta'_2  + \komega_0 \dd_{\theta'}  + \bF'(\uu_0).
\end{eqnarray}
In the works \cite{Biktashev:1994, Biktasheva:2003, Biktasheva:2009}, including \dxspiral, a complex-valued basis of eigenmodes to $\HL$ was used, given by
  \begin{eqnarray}\label{GMcompl}
  \VV_\pm &=& - \frac{1}{2} \left(\dd_{x'} \uu_0 \mp \rmi K \dd_{y'} \uu_0 \right), \nn \\
\VV_{(0)} &=& - \dd_\theta \uu_0.
\end{eqnarray}
Hence, the sign of eigenvalues is independent of $K$:
\begin{eqnarray}
 \HL \VV_{(n)} = \rmi n \omega_0 \VV_{(n)}. \qquad (n \in \{-1,0,1\}).
\end{eqnarray}
Since the modes $\ket{\VV_{(n)}}$ possess eigenvalues with real part equal to zero, they are also known as critical eigenfunctions of $\HL$, or Goldstone modes (GMs). They originate from the translational and rotational symmetry of the RDE \eqref{RDE1}.

Using integration by parts, the adjoint operator to $\HL$ can be found, which satisfies $\braket{\HL^\dagger \mathbf{f}}{\mathbf{g}} = \braket{\mathbf{f}}{ \HL\mathbf{g}} $ for a suitable space of test functions $\mathbf{f}, \mathbf{g}$:
\begin{eqnarray}
 \HL^\dagger = \HP^H \Delta'_2 - \komega_0 \dd_{\theta'}  + \bF'^H(\uu_0).  
\end{eqnarray}
If all entries to $\HP$ and $\bF$ in the original RDE \eqref{RDE1} are real-valued, the spectrum of $\HL^\dagger$ is complex conjugate to the spectrum of $\HL$, such that there exist critical eigenfunctions \cite{Biktasheva:2009}
\begin{eqnarray}
 \HL^\dagger \WW^{(n)} = - \rmi n \omega_0 \WW^{(n)}. \qquad (n \in \{-1,0,1\})
\end{eqnarray}
that can be made biorthonormal to the GMs:
\begin{eqnarray}
\braket{\WW^{(m)}}{\VV_{(n)}} =\delta^{(m)}_{(n)}. \qquad (m,n \in \{-1,0,1\})
\end{eqnarray}
One furthermore shows (see, e.g. \cite{Biktashev:1995b}) that the $\WW^{(n)}$ are precisely a complex representation of the response functions in Eq. \eqref{RFh}: 
\begin{eqnarray} \label{RFdef}
   \WW^{\pm} &=& - \left( \bY^x \mp \rmi K \bY^y \right) ,\nn \\     \WW^{(0)}   &=& - \bY^\theta.
\end{eqnarray}
In the Cartesian basis, the RFs and GMs also satisfy 
$ \braket{\bY^\hmu}{\dd_\hnu \uu_0} =\delta^\hmu_\hnu$ for $ (\hmu, \hnu \in \{x',y',\theta'\})$. When writing left-brackets to the RFs, we implicitly assume complex conjugation, i.e. $\bra{a \WW} = a^* \bra{\WW} $. Therefore, we may write Eq. \eqref{RFdef} as
\begin{eqnarray} \label{RFcompl}
   \bra{\WW^{\pm}} &=& - \left(\bra{\bY^x} \pm \rmi K \bra{\bY^y} \right), \nn \\     
\bra{\WW^{(0)}}   &=& - \bra{ \bY^\theta}.
\end{eqnarray}

Since the complex basis \eqref{GMcompl}-\eqref{RFcompl} is practical in computations, we will write other tensorial quantities as well in this representation, see Appendix \ref{app:complex}. 

\subsection{The gauge filament \label{sec:gauge}}

Scroll wave filaments are often referred to as the instantaneous axis of scroll wave rotation. When the filament is straight and stationary, such definition indeed yields the unique rotation axis of the system. For non-stationary scroll waves, however, it is less clear how to define the filament and various options are possible. In our derivation for the EOM, we will work with a so-called \textbf{gauge filament}, introduced as follows. Given an approximate filament curve (e.g. the tip line) and Fermi coordinates $(\rho^1,\rho^2,\s)$ around it, one may in each plane transverse to the filament curve demand that the difference $\tuu(\rho^1, \rho^2,\s, t) = \uu(\rho^1, \rho^2, \s, t) - \uu_0(\rho^1, \rho^2)$ should be minimal in some sense for the filament curve. From the response function theory in the previous section, we recall that infinitesimal shifts of the solution $\uu_0$ in the plane corresponds to additions of the Goldstone modes $\dd_\hmu \uu_0$ ($\hmu \in\{ \rho^1, \rho^2, \theta\}$) to the unperturbed solution $\uu_0$. Therefore, we may define the filament's position and rotation phase to satisfy the gauge condition
\begin{eqnarray}
\braket{\bY^\hmu}{\tuu}=0   &\qquad&(\hmu \in\{ \rho^1, \rho^2, \theta\}), \nn \\ 
\Leftrightarrow \braket{\WW^{(n)}}{\tuu}=0   &\qquad& ( n \in\{-1,0,+1\}) \label{gaugefil}
\end{eqnarray}
 for all $\s,t$. Eq. \eqref{gaugefil} is the formal definition of the gauge filament, which is commonly used in response function approaches to scroll wave dynamics \cite{Keener:1988, Biktashev:1994, Verschelde:2007, Dierckx:2012}. 

Another curve which lies in the vicinity of the scroll wave's core will be given by the `virtual filament', which we formally introduce in section \ref{sec:virtualdef}.

\section{Derivation of the instantaneous laws of filament motion \label{sec:eomproof}}

\subsection{Moving and rotating Fermi frames}

In our derivation, we consider a moving gauge filament $x^i = X^i (\s,t)$ and will impose that the scroll wave which surrounds it should satisfy the RDE \eqref{RDE1}. Readers who are interested in the resulting law of motion for the gauge filament rather than the technical derivation are referred to section \ref{sec:eomproof_result}.

In order to construct moving Fermi coordinates, we take the following steps. At a given time $t=0$, we erect a Fermi frame $(\vec{T}(\s,0),\vec{e_a}(\s,0))$ along the gauge filament curve $X^i(\s, 0)$, such that automatically $\wfr(\s,0)=0$. For small times $t \neq 0$, we impose that points move orthogonally to the initial filament, which fixes the parameterization $\s$. The tangent vector $\vec{T}(\s,t)$ to the filament is now unambiguously defined, and the vectors $\vec{e_a}(\s,t)$ are tilted such that the remain orthogonal to $\vec{T}(\s,t)$. The only degree of freedom left is rotation around the filament; we shall impose here that $\dd_t \vec{e_1}(\s,t) \cdot \vec{e_2}(\s,t) = 0$ for all $t$. (More details can be found in Appendix \ref{app:frame}.)

Next, we will accommodate for scroll wave twisting using a rotating reference that follows the local phase $\phi(\s,t)$ of the scroll wave. Using the rotation matrix
\begin{eqnarray}
R_A^{\hs a} = \mathbf{R} = \left(  \begin{array}{cc} 
\cos \phi & \sin \phi \\ -\sin \phi & \cos \phi
\end{array}\right) = \exp (\beps \phi). \label{RAa}
\end{eqnarray}
we define for all $\s, t$:
\begin{eqnarray}
  \vec{e_A}(\s,t) = R_A^{\hs a}(\s,t) \vec{e_a}(\s,t). 
\end{eqnarray}
The uppercase letters $A,B,...$ indices will be used throughout this paper to denote a rotating frame of reference, in contrast to the indices $a,b,...$ which refer to frames with minimal rotation. Furthermore, the Einstein summation convention is assumed for all index sets in this paper. In the complex basis for GMs and RFs, implicit summation will only run over $ \{ -1,1\}$, excluding the rotational modes.

We may now introduce Fermi coordinates $(\rho^A, \s, \tau)$ analogous to  Eq. \eqref{FCa} which are fully adapted to the scroll wave's evolution; the coordinate transformation is given by
\bsub \begin{eqnarray}
 x^i(\rho^A, \s, \tau) &=& X^i(\s,\tau) + \rho^A N_A^i(\s,\tau), \label{xepx} \\
t(\tau) &=& \tau. 
\end{eqnarray} \label{FCA} \esub
The coordinates $(\rho^A, \s)$ are the simpler Euclidean version of Fermi coordinates 
in curved space that were also used to describe scroll wave filaments in anisotropic media \cite{Verschelde:2007, Dierckx:2009}. For a straight, untwisted scroll wave, they coincide with Cartesian coordinates that rotate around a fixed axis. 

\subsection{Scroll wave as a stack of spiral waves}

From translational invariance in the Z direction, one knows that the spiral wave solution $\uu_0(x', y')$ exactly solves the RDE \eqref{RDEcomoving} in three spatial dimensions. We now propose a perturbative series of successive approximations to the scroll wave solution to the RDE \eqref{RDEcomoving} for the case of slightly bent and twisted filaments. For, in the moving Fermi coordinates of  Eq. \eqref{FCA}, the solution is expected to be close to the unperturbed spiral wave solution $\uu_0$:
\begin{eqnarray}  \label{ansatzu}
  \uu(\rho^A, \s, \tau) =  \uu_0(\rho^A) + \tuu(\rho^A, \s,\tau),                                    
\end{eqnarray}
where $\tuu$ is of order $\la$. The formal expansion parameter $\la$ is bounded by the magnitude of scroll wave twist and filament curvature, relative to the typical radius $d$ of the scroll wave's core:
\begin{eqnarray}
 \la = \max( kd, |w|d).
\end{eqnarray}
We furthermore assume that that no external stimuli were given in recent history, such that the correction field $\tuu$ only depends on $\s$ and $\tau$ through twist $w$ and curvature components $\LA_A$, i.e. $\tuu (\rho^A, \s, \tau) = \mathbf{\tilde{\uu}} (\rho^A, w(\s, \tau), \LA^A(\s, \tau))$. Our scheme serves to track successive orders of approximation for the scroll wave modification $\tuu$, the angular frequency $\komega$ and the drift velocity $v^A = \dd_t \vec{X} \cdot \vec{e^A}$:
\bsub \label{ansatzomv} \begin{align}
 \tuu =& \sum_{n=1}^{+\infty} \uu_n, & \uu_n = \OO(\la^n),  \\
 \komega =& \komega_0 + \sum_{n=1}^{+\infty} \komega_n, \qquad &  \komega_n = \OO(\la^n),\\
 v^A =& \sum_{n=1}^{+\infty} (v_n)^A, & (v_n)^A = \OO(\la^n).
\end{align}\esub
The definition of the gauge filament \eqref{gaugefil} here implies that the corrections $\uu_n$ of all orders should be orthogonal to the three RFs:
 \begin{eqnarray}  \label{gaugetuu}
   \braket{\WW^{(n)}}{\uu_j} = 0, \qquad ( n \in \{ -1, 0, 1\}\  \text{and} \ j \geq 1). 
\end{eqnarray} 

\subsection{RDE in moving Fermi coordinates}

To express the time-derivative term of  Eq. \eqref{RDE1} in the moving and rotating Fermi coordinates \eqref{FCA}, we use the chain rule to find
\begin{eqnarray}
 \dd_\tau = \dd_t + \komega \dd_\theta + v^A \dd_A
\end{eqnarray}
where $(r,\theta)$ are the polar coordinates corresponding to the $\rho^A$ and $v^A = \vec{e^A} \cdot \dd_t \vec{X}$. The expansion \eqref{ansatzu} then delivers
\begin{eqnarray}
 \dd_t \uu
                &=& \sum_{j=1}^{3}  \dd_\tau \uu_j  - \sum_{j=1}^{3} \sum_{m=0}^j \komega_m \dd_\theta \uu_{j-m}  
 \nn \\&& 
- \sum_{j=1}^{3}  \sum_{m=1}^j (v_m)^A \dd_A \uu_{j-m} + \OO(\la^4).  \label{ttermformal} 
\end{eqnarray}


The reaction term can be expanded in a Taylor series. With indices $\alpha, \beta, ...$ referring to state variables and the summation convention assumed, one finds
\begin{eqnarray}
 F^\alpha (\uu) &=& F^\alpha (\uu_0) + \frac{\dd F^\alpha}{\dd u^\beta}(\uu_0) \tilde{u}^\beta 
\nn  \\&& 
+ \frac{1}{2} \frac{\dd^2 F^\alpha}{\dd u^\beta \dd u^\gamma}(\uu_0) \tilde{u}^\beta \tilde{u}^\gamma + \OO(\la^3). \end{eqnarray}
The explicit calculation up to third order in twist and curvature brings:
\begin{eqnarray} \label{reactermexpl}
  \bF(\uu) = \bF(\uu_0) +  \bF'(\uu_0) \uu_1  
+ \bF'(\uu_0) \uu_2 + \frac{1}{2}\bF''(\uu_0) \uu_1^2  \nn \\
    +\bF'(\uu_0) \uu_3 +\bF''(\uu_0) \uu_1 \uu_2 + \frac{1}{6}\bF'''(\uu_0) \uu_1^3 + \OO(\la^4).\ 
\end{eqnarray}
 
The Fermi coordinates \eqref{FCA} are not orthogonal off the filament due to twist corrections, even at $t=\tau=0$. Therefore,  the Laplacian from the diffusion term in RDE \eqref{RDE1} now involves the metric tensor associated to the coordinate transformation. Using $\rho^3 = \s$ and with $\mu \in \{1,2,3\}$, one has that
\begin{eqnarray} \label{defmetric}
 g_{\mu  \nu} = \dd_\mu \vec{x} \cdot \dd_\nu \vec{x}.
\end{eqnarray}
For the isotropic reaction-diffusion media considered, the Laplacian thus  becomes
\begin{eqnarray}
 \Delta \uu &=& g^{-1/2} \dd_\mu \left( g^{1/2} g^{\mu \nu} \dd_\nu \uu\right) \nn \\
 &=& g^{-1/2} \dd_\mu \left( g^{1/2} g^{\mu \nu}\right)  \dd_\nu \uu + g^{\mu \nu} \dd^2_{\mu \nu}\uu. \label{Lapl2}
\end{eqnarray}
Further elaboration of this term requires evaluation of the metric tensor. Its components easily follow from Eqs. \eqref{parallelframe}, \eqref{FCa} and \eqref{defmetric} at $t=0$. With $\dd_A \vec{x} = \vec{N_A}$ and $\dd_\s \vec{N}_A = (1- \rho^B\LA_B) \vec{T} + \rho^B w_B^{\hs C} \vec{N_C}$, one finds without approximation that\begin{eqnarray}\label{gcov_iso}
 \left( g_{\mu \nu} \right) &=&  \left(
                  \begin{array}{cc}
                    g_{AB} &  g_{A\s} \\
                    g_{\s B}  & g_{\s \s} \\
                  \end{array}
                \right) \\  \nn 
          &=& \left(
                  \begin{array}{cc}
                    \delta_{AB} &  \rho^C w_{CA} \\
                    \rho^C w_{CB}  & (1- \rho^C \LA_C) ^2  + \rho^C \rho^D w^2 \delta_{CD} \\
                  \end{array}
                \right).\ 
\end{eqnarray}
The size of an infinitesimal volume element is related to the determinant of this matrix, denoted $g$. Its square root is found to be
\begin{eqnarray}
   \sqrt{g} = 1 - \rho^A \LA_A.
\end{eqnarray}
That is, only filament curvature, not twist, affects the size of elementary volume elements away from the filament. The covariant metric tensor can be exactly obtained by taking the matrix inverse of  Eq. \eqref{gcov_iso}, e.g. by the cofactor method:
\begin{eqnarray} \label{gcon_iso}
 \left( g^{\mu \nu} \right) &=&  \left(
                  \begin{array}{cc}
                    g^{AB} &  g^{A\s} \\
                    g^{\s B}  & g^{\s \s} \\
                  \end{array}
                \right)\\ 
\nn &=& \left(
                  \begin{array}{cc}
                    \delta^{AB} + \frac{w^A_{\hs C} w^B_{\hs D} \rho^C \rho^D}{\left(1 - \rho^C \LA_C\right)^2} &  \frac{w^A_{\hs C}\rho^C}{\left(1 - \rho^C \LA_C\right)^2} \\
                    \frac{w^B_{\hs C}\rho^C}{\left(1 - \rho^C \LA_C\right)^2} & \frac{1}{\left(1 - \rho^C \LA_C\right)^2}\\
                  \end{array}
                \right).  
\end{eqnarray}
Expanding in the $\rho^A$, we have found for the diffusion term:
\begin{eqnarray}
  &-& \HP \Delta \uu =- \HP \dd_A\dd^A ( \uu_0 + \uu_1 + \uu_2 +  \uu_3) \nn  + \LA^A \HP \dd_A \uu_0 \\
&& +  \LA^A \HP \dd_A \uu_1 +  \LA^A \LA_B \rho^B \HP \dd_A \uu_0 - w^2 \HP \dd^2_\theta \uu_0 \nn\\ && 
+ \HP \dd_\s w \dd_\theta \uu_0 +  \LA^A \HP \dd_A \uu_2 +  \LA^A \LA_B \rho^B \HP \dd_A \uu_1\nn\\ && 
 - w^2 \HP \dd^2_\theta \uu_1 + \dd_\s w  \HP \dd_\theta \uu_1 + \LA^A \LA_B \LA_C \rho^B \rho^C \HP \dd_A \uu_0 \nn \\ 
 &&- 2 w^2 \LA_B \rho^B \HP \dd^2_\theta \uu_0 + 2  \dd_\s w \LA_B \rho^B \HP \dd_\theta \uu_0 \nn\\ && 
+ \LA_A w^2 \eps^A_{\hs C} \rho^C \HP \dd_\theta \uu_0 - \HP \dd^2_\s \uu_1 \nn\\ && + w \dd_\s \LA_A \rho^A \HP \dd_\theta \uu_0 + 2w \HP \dd^2_{\s \theta} \uu_1 +  \OO(\la^4). \label{difftermexpl}
\end{eqnarray}

\subsection{Perturbation scheme}

Collecting the reaction \eqref{reactermexpl}, diffusion \eqref{difftermexpl} and time-derivative \eqref{ttermformal} parts, the RDE\ \eqref{RDE1} can be written as
\begin{eqnarray}
 (\komega-\komega_0)\ket{\dd_\theta \uu_0} + v^A \ket{\dd_A \uu_0} + ( \HL - \dd_\tau ) \ket{ \tuu} = \ket{\SSS}, \ \label{mastereq}
\end{eqnarray}
where the expansion terms are only present in the source term $\ket{\SSS}$. Equation \eqref{mastereq} is the cornerstone of our perturbative approach:
projection onto different subspaces of eigenfunctions of $\HL$ will allow to calculate the higher order corrections to the scroll wave's rotation frequency, drift velocity and wave profile.

We now list the source terms per order in $\la$:
\begin{eqnarray}
 \ket{\SSS} = \ket{\SSS_1} + \ket{\SSS_2} + \ket{\SSS_3} + \OO(\la^4),
\end{eqnarray}
where
\bsub \label{inventory} \begin{eqnarray}
 &\ket{\SSS_1} &= \LA^A \HP \ket{\dd_A \uu_0}, \\
& \ket{\SSS_2} &=  - \komega_1 \ket{\dd_\theta \uu_1} + \left(\LA^A \HP - (v_1)^A \right) \ket{\dd_A \uu_1} \nn \\
&& -\frac{1}{2} \ket{\bF'' \uu_1^2}  + \LA^A \LA_B \rho^B \HP \ket{\dd_A \uu_0} \nn \\
&& - w^2 \HP \ket{\dd^2_\theta \uu_0} + \dd_\s w \HP\ket{\dd_\theta \uu_0} \\
 &\ket{\SSS_3} &= - \komega_1 \ket{\dd_\theta \uu_2} + \left(\dd_\s w \HP -\komega_2 \right) \ket{\dd_\theta \uu_1} \nn \\ &&
 + \left(\LA^A \HP - (v_1)^A \right) \ket{\dd_A \uu_2} - \HP \ket{\dd^2_\s \uu_1} 
 \nn \\
&&  -\frac{1}{6} \ket{\bF^{(3)} \uu_1^3}   - \ket{\bF''  \uu_1 \uu_2} \nn\\
 && + \left(\LA^A \LA_B \rho^B \HP - (v_2)^A \right) \ket{\dd_A \uu_1} - w^2 \HP \ket{\dd^2_\theta \uu_1} \nn\\
 && + \LA^A \LA_B \LA_C \rho^B \rho^C \HP \ket{\dd_A \uu_0}  + 2 \dd_\s w \LA_B \rho^B \HP \ket{\dd_\theta \uu_0}   
 \nn\\
 && + w^2 \left( \LA_A \eps^A_{\hs C}  \rho^C \HP \ket{\dd_\theta \uu_0}  - 2 \LA_B \rho^B \HP \ket{\dd^2_\theta \uu_0} \right) \nn \\
&& + w \left(\dd_\s \LA_A \rho^A \HP \ket{\dd_\theta \uu_0} + 2 \HP \ket{\dd^2_{\theta \s} \uu_1} 
\right). \label{inventoryS3}
\end{eqnarray} \esub
The general idea of the task ahead is simple: given a source term of given order $j$, we can evaluate $(v_j)^A$ and $\omega_j$ by projecting on the translational and rotational RFs, as in \cite{Verschelde:2007}. The remaining components of the master equation \eqref{mastereq} yield a linear system, which can be solved formally or numerically to reveal the field correction $\uu_j$. Once the corrections of given order are known, one can proceed to the next order. Note that we used a similar approach before to obtain the quadratic curvature relation for wave fronts in RD systems \cite{Dierckx:2011}; for scroll waves we now present results to third order in curvature and twist. 

\subsection{First order solution}

The equation \eqref{mastereq} contains no zeroth order terms in $\la$. When only first order terms in $\la$ are kept, it becomes:
\begin{multline} \label{ma1}
    \komega_1 \ket{\dd_\theta \uu_0}  + (v_1)^A \ket{\dd_A \uu_0} +  (\HL - \dd_\tau) \ket{\uu_1} \\ =  \LA^A \HP \ket{\dd_A \uu_0}.
\end{multline}
First, taking the overlap integral with the rotational RF yields
 \begin{eqnarray}
 \komega_1 =  \braket{\bY^\theta}{\SSS_1} = \LA^A \bra{\bY^\theta} \HP \ket{\dd_A \uu_0}. 
\end{eqnarray}
The rotational correction can be renamed to denote it is the first order curvature ($k$) correction: $\omega_1 := \LA^A \omega^k_A$. This term was also found in Keener's work \cite{Keener:1986} where it was named $c_1$; in \cite{Biktashev:1994} its time-average over one rotation period was shown to vanish. We shall postpone the time-averaging process to section \ref{sec:virtual}, as in higher order the products of oscillatory terms may nevertheless contribute to the net filament motion.

Secondly, projection onto the translational RF delivers
\begin{eqnarray}
 (v_1)^B =  \braket{\bY^B}{\SSS_1} 
= \LA^A \bra{\bY^B} \HP \ket{\dd_A \uu_0} .
\end{eqnarray}
Due to frequent use, we will introduce the notations 
\begin{eqnarray} \label{defPBA}
 \bra{\bY^B} \HP \ket{\dd_A \uu_0} =  (v_k)^B_A = P^B_{\hs A}.
 \end{eqnarray}
The isotropic parts of this tensor are \cite{Verschelde:2007}
\begin{align}
 \gamma_1 =& \frac{1}{2} \bra{\bY^A} \HP \ket{\dd_A \uu_0}, & 
 \gamma_2 =& \frac{1}{2} \eps_A^{\hs B} \bra{\bY^A} \HP \ket{\dd_B \uu_0}. 
\end{align}
The pair $(\gamma_1, \gamma_2)$ corresponds to the pair $(b_2, c_3)$ in \cite{Keener:1988}, which were interpreted as filament tension in \cite{Biktashev:1994}; both works implied integration over a full period of rotation to find the net drift of scroll waves. We will perform the averaging over one period in section \ref{sec:virtual}; for now we verify using Eqs. \eqref{GMcompl}, \eqref{RFcompl} that
\begin{eqnarray}
  b_2 + \rmi c_3 := \bra{\WW^+} \HP \ket{\VV_+} = \gamma_1 + \rmi K \gamma_2
\end{eqnarray}
with $K$ the sign flag in \dxspiral. 

Thirdly, we will project  Eq. \eqref{ma1} onto the space orthogonal to the the GMs to find the wave profile correction $\uu_1$. Hereto, we introduce the projection operator $\HPI$, which suppresses all GM or RF components of a given state vector:
\begin{eqnarray} \label{defPI}
 \HPI &=& \mathbbm{1} - \ket{\dd_\theta \uu_0} \bra{\bY^\theta} - \ket{\dd_A \uu_0} \bra{\bY^A}\  \nn \\
 &=& \mathbbm{1} - \sum_{n \in \{-1,0,+1 \} } \ket{\VV_{(n)}} \bra{\WW^{(n)}}. \nn  
\end{eqnarray}
Application to Eq. \eqref{ma1} generates a condition for the wave profile modification $\uu_1$:
 \begin{eqnarray} \label{equu1}
 (\HL-\dd_\tau) \ket{\uu_1} = \HPI \ket{\SSS_1} = \LA^A \HPI  \HP \ket{\dd_A \uu_0}. 
\end{eqnarray} 
Due to the unceasing rotation of the scroll wave around the filament, the above equation contains oscillatory terms regardless of the chosen frame: in the co-rotating frame, $\LA^A$ is a rotating vector, whereas the field correction $\uu_1$ changes periodically when viewed in the lab frame. 


The field correction $\uu_1$ can be obtained from Eq. \eqref{equu1} by Fourier transformation with respect to time, although finding the correct signs can be a bit cumbersome. Recall that from the generalization of the transformation rule \eqref{RFcompl}, we may define
\begin{eqnarray}
  \rho^+ &:= -(x+\rmi K y), \qquad & \LA^+ := -(\LA^x + \rmi K \LA^y).
\end{eqnarray} 
In the rotating frame, the filament normal vector $(\LA^x(\tau), \LA^y(\tau), 0)$ is counter-rotating with frequency $\komega(\s, \tau)$, such that it can be written $\LA^+(\tau) = - \LA^0 \rme^{\pm i \omega \tau}$, with the sign still to be established and possibly depending on the sign flag $K$. 

Let us consider the two cases, starting with $K=+1$ and positive $\omega$. The spiral then rotates clockwise, delivering counter-clockwise rotation for the vector $(\LA^x(\tau), \LA^y(\tau))$. Therefore, $\LA^+(\tau) = - (\LA^x(\tau) + \rmi\LA^y(\tau)) = - \LA^0 \rme^{+ \rmi \omega \tau}$. In the second case where $K=-1$ and $\omega>0$, 
the spiral rotates counter-clockwise such that the normal vector turns clockwise in the co-rotating frame. That's why $\LA^+(\tau) = - (\LA^x(\tau) - \rmi \LA^y(\tau)) = - \LA^0(\tau) \rm\rme^{+ \rmi \omega \tau}$. In both cases, we have found that 
\begin{eqnarray} \label{LA+sign}
 \LA^+(\tau) = - \LA^0 \rme^{+ \rmi \omega \tau} 
\end{eqnarray}
regardless of the sign flag $K$. 

Writing the phase angle as $\phi(\tau)$, with $\omega = \dd_\tau \phi$, the time-derivative of this quantity becomes
 $\dd_\tau \LA^+ = \rmi \omega \LA^+ + \rme^{+ \rmi \omega \tau} \dd_\tau \LA^0$. 
The second term represents the net change of curvature due to filament dynamics, which is shown in Eq. \eqref{laexp3app} to be of order $\la^3$, whence $\dd_\tau \LA^{(m)} = \rmi m \omega_0 \LA^{(m)} + \OO(\la^2)$. 
In solving Eq. \eqref{equu1}, we rewrite $\uu_1$ as
\begin{eqnarray}
 \uu_1(\rho^A, \s, \tau)  &=& \LA^A(\s, \tau) \uu^k_A(\rho^A, \s, \tau) \nn \\ &=&  \LA^{(m)} (\s, \tau)\uu^k_{(m)} (\rho^A, \s, \tau),
\end{eqnarray}
where the implicit summations run over translational indices: $A\in \{x,y\}$ and $m \in \{-1,1\}$. We now need to solve
\begin{eqnarray} \label{Ldtcalc}
(\HL- \rmi m\omega) \LA^{(m)}  \ket{ \uu^k_{(m)} } = \LA^{(m)} \HPI \HP \ket{\VV_{(m)}} +\OO(\la^3)
\end{eqnarray}
for any $\LA^{(m)}$, whence, without summation over $m$, 
\begin{multline}
 (\HL- \rmi m \omega_0)  \ket{ \uu^k_{(m)} } =\\  \HPI \HP \ket{\VV_{(m)}}  \label{recursion_uk} + \rmi m\omega_1  \ket{ \uu^k_{(m)} } + \OO(\la^2).     
\end{multline}
Without loss of generality, we can now put $m=1$, since the case $m=-1$ follows from complex conjugation. The operator $\HL - \rmi \omega_0$ on the left hand side of  Eq. \eqref{recursion_uk} has a  right-hand zero mode $\VV_+$ and is therefore singular. Nevertheless, the solution $\uu^k_{(m)}$ can be found, because the right-hand side of  Eq. \eqref{recursion_uk} is orthogonal to the null-space of $\HL - \rmi \omega_0$. It was precisely the vanishing of the null-space component in the right-hand side that delivered the lowest order equation of motion. This necessary condition is equivalent to the Fredholm alternative theorem, which was also used for scroll waves in \cite{Keener:1988, Biktashev:1994, Verschelde:2007}. Here we may formally invert the last equation to find 
\begin{eqnarray}\ket{ \uu^k_{(m)} } =  (\HL- \rmi\, m\, \omega_0)^{-1} \HPI \HP \ket{\VV_{(m)}} +\OO(\la).
\label{u1}
\end{eqnarray} 
An easy way to remember the sign in $\HL\pm \rmi \omega_0$ in  Eq. \eqref{u1} is that the operator should operate on its zero mode, preceded by $\HPI \HP$. As a corollary to  Eq. \eqref{recursion_uk}, the field correction linear in filament curvature $\uu_1$ completely vanishes for equal diffusion systems: when $\HP = D_0 \mathbf{I}$ one has that $\HPI \ket{\VV_{(m)}}= \zeroket$, whence $\ket{\uu_1} = \zeroket$ altogether.  

It is worth noting in Eq. \eqref{u1} that the field correction $\uu^k$ of the given order is independent of $\s$ and $\tau$. This confirms our former hypothesis that $\uu_1$ only depends on $\s, \tau$ through curvature $k$ and/or twist $w$. It was therefore justified to suppose that $ \dd_\s \tuu_1 = \tuu_1 . \OO(\la) $. 

Finally, we remark that substituting the solution \eqref{u1} into Eq. \eqref{recursion_uk} to find a $\OO(\la)$ correction to $\uu_k$ via the $\omega_1$ term is not a good idea, since it was assumed that $\uu_k$ only depends on time though $\LA^{(m)}$. Hence, the time-derivative of the correction term would wrongfully have been neglected if we try to improve on $\uu_1$ right here. Instead, we will include the additional term in  Eq. \eqref{recursion_uk} in our second order calculation, which we now start.

\subsection{Second order solution}
The second order terms in  Eq. \eqref{mastereq} read
\begin{multline}
  \komega_2 \dd_\theta \uu_0 + (v_2)^A \dd_A \uu_0 
+ (\HL - \dd_\tau) \uu_2  \\ =  \rmi m \omega_1 \LA^{(m)} \uu^k_{(m)} + \SSS_2. \label{ma2}
\end{multline}
The $\omega_1$ term originates from  Eq. \eqref{recursion_uk}; since it is orthogonal to the GMs, it does not contribute to the second order motion. To proceed, we write the source term as 
\begin{eqnarray}
 \ket{\SSS_2} =  w^2 \ket{\SSS_{ww}}  + \LA^A \LA^B \ket{\SSS^{kk}_{AB}}  + \dd_\s w \ket{\SSS_{dw}}
\end{eqnarray}
such that from  Eq. \eqref{inventory} follows
\begin{eqnarray}
 \ket{\SSS_{ww} }&=& -  \HP \ket{ \dd^2_\theta \uu_0}, \nn \\
 \ket{\SSS^{kk}_{AB} } &=& - \komega^k_A \ket{\dd_\theta \uu^k_B}  \label{res_om2} 
 + (\HP\delta^C_A - P^C_{\hs A}) \ket{\dd_C \uu^k _B} \nn \\ &&
+ \rho_B \HP \ket{\dd_A \uu_0} - \frac{1}{2} \ket{\bF''  \uu^k_A \uu^k_B},  \\
 \ket{\SSS_{dw}} &=&  \HP \ket{\dd_\theta \uu_0}.\nn
\end{eqnarray} 
Likewise, the second order motion corrections can be written
\begin{eqnarray}
\komega_2 &=&  w^2 \komega_{ww} + \LA^A \LA^B \komega^{kk}_{AB} + \dd_\s w\, \komega_{dw}, \nn \\
(v_2)^C &=& w^2 v^C_{ww} + \LA^A \LA^B (v_{kk})^C_{AB} + \dd_\s w\, v^C_{dw}.
\end{eqnarray}
Projecting  Eq. \eqref{ma2} onto the rotational and translational RFs then delivers
\begin{align}
     \komega_{ww} &= \braket{\bY^\theta}{\SSS_{ww}},  & v^C_{ww} &= \braket{\bY^C}{\SSS_{ww}}, \nn\\
    \komega^{kk}_{AB} &= \braket{\bY^\theta}{\SSS^{kk}_{AB}}, & (v_{kk})^C_{AB} &= \braket{\bY^C}{\SSS^{kk}_{AB}},\label{vom2}\\
    \komega_{dw} &= \braket{\bY^\theta}{\SSS_{dw}}, & v^C_{dw} &= \braket{\bY^C}{\SSS_{dw}}.\nn
\end{align} 
The terms $v^C_{ww}$ and $v^C_{dw}$ correspond to the couples $(-a_2,-a_3)$ and $(-c_2,-c_4)$ in Keener's work \cite{Keener:1988, Keener:1992}. Even though these terms vanish when averaged over one rotation period \cite{Biktashev:1994}, they will contribute to the higher order filament dynamics, as we show below.

In the case of equal diffusion $\HP = D_0 \mathbf{I}$, one has that $\HP\delta^C_A = D_0 P^C_{\hs A}$ and $\ket{\uu^k_A} = \zeroket$, such that the instantaneous motion corrections reduce to

 \begin{align}
\komega_{ww} &= - D_0\braket{\bY^\theta}{ \dd^2_\theta \uu_0}, & \komega_{dw} &= D_0,
 \nn \\
v^C_{ww}& = - D_0 \braket{\bY^C}{ \dd^2_\theta \uu_0}, & v^C_{dw} &= 0, \nn\\   
 \komega^{kk}_{AB} &=  D_0 \bra{\bY^\theta} \rho_B \ket{\dd_A \uu_0}, &&\nn \\
(v_{kk})^C_{AB} &=  D_0 \bra{\bY^C} \rho_B \ket{\dd_A \uu_0}. && 
\end{align} 


Next, let us consider the second order field corrections $\uu_2$, which can be decomposed as
 \begin{eqnarray} \label{u2dec}
 \ket{\uu_2} &=& w^2  \ket{\uu_{ww}} + \LA^A \LA^B\ket{\uu^{kk}_{AB}} + \dd_\s w\,  \ket{\uu_{dw}}  \\
  &=& w^2  \ket{\uu_{ww}} +  \LA^{(m)} \LA^{(n)} \ket{\uu^{kk}_{(mn)}} + \dd_\s w\,  \ket{\uu_{dw}}. \nn  
\end{eqnarray} 
The time-derivative in Eq. \eqref{ma2} will also operate on the geometric factors $w$ and $\LA^A$. However, from Eq. \eqref{laexp3} we deduce that this will increase the order of perturbation by at least one order in $\la$.
The twist-induced field corrections in Eq. \eqref{u2dec} thus follow from projecting Eq. \eqref{ma2} onto the subspace orthogonal to the RFs using $\HPI$ defined in Eq. \eqref{defPI}:
\bsub \begin{eqnarray} \label{u2eqs}
(\HL- \dd_\tau) \uu_{dw} + \OO(\la) =& \HPI \ket{\SSS_{dw}} = \HPI \HP \ket{\dd_\theta \uu_0}, \\
(\HL- \dd_\tau) \uu_{ww} + \OO(\la) =& \HPI \ket{\SSS_{ww}} = - \HPI \HP \ket{\dd^2_\theta \uu_0}. \qquad 
\end{eqnarray} \esub
As the right-hand sides do not depend on time, the time-derivative drops out, whence
\bsub \label{u2wsol}
\begin{eqnarray} 
 \ket{\uu_{dw} }=& \HL^{-1} \HPI \HP \ket{\dd_\theta \uu_0}, \\
 \ket{\uu_{ww} }=& - \HL^{-1} \HPI \HP \ket{\dd^2_\theta \uu_0}.
\end{eqnarray} 
\esub
For the quadratic curvature correction, we recall from the analysis preceding  Eq. \eqref{LA+sign} that 
\begin{eqnarray}
 \LA^{(m)} = - \LA^0 \rme^{m \rmi \omega \tau}.
\end{eqnarray} 
Taking into account the additional $\omega_1$ term which arose in  Eq. \eqref{recursion_uk} and recalling that $\HPI \uu^k_{(m)} = \uu^k_{(m)}$, we find that $\uu^{kk}$ follows from
\begin{multline} \label{u2kkcalc}
 (\HL - \rmi (m+n) \omega_0 ) \ket{\uu^{kk}_{(mn)} } = \HPI \ket{\SSS^{kk}_{(mn)}} \\ + m \rmi \omega^k_{(n)} \ket{\uu^k_{(m)}} + \rmi (m+n) \omega_1 \ket{\uu^{kk}_{(mn)} } + \OO(\la). \end{multline} 
This equation can be formally solved to
\begin{multline} \label{u2kksol}
\ket{\uu^{kk}_{(mn)} }= \\ (\HL - \rmi (m+n) \omega_0 )^{-1} \HPI \left( \ket{\SSS^{kk}_{(mn)}} + m \rmi \omega^k_{(n)} \ket{\uu^k_{(m)}} \right).  
\end{multline} 

For the simpler case of equal diffusion systems, $\uu^k$ and $\uu_{dw}$ fully vanish, leaving the lowest order contributions
\bsub \begin{eqnarray}
 \ket{\uu_{ww} }&=& - D_0 \HL^{-1} \HPI \ket{\dd^2_\theta \uu_0 }, \\
\ket{\uu^{kk}_{(mn)} }&=& D_0 (\HL - i (m+n) \omega_0 )^{-1} \HPI \ket{ \rho_{(m)}  \VV_{(n)} }.\qquad 
\end{eqnarray}  \esub
From  Eq. \eqref{defddp}, one may additionally notice that $\rho_+ \VV_- = (1/4)(r \dd_r \uu_0 + \rmi K \dd_\theta \uu_0)$, whence, in the case of equal diffusion, $\ket{\uu^{kk}_{+-} }=\ket{\uu^{kk}_{-+} }= (D_0/4) \HL^{-1} \HPI \ket{r \dd_r \uu_0}$.

\subsection{Third order rotation and drift} 

Having arrived at the highest order in curvature and twist which we intend to treat here, we will not consider the field corrections $\uu_3$; only $\omega_3$ and $v^C_3$ will be sought for. 
To start, we note that the approximation of $\HL- \rmi \omega$ by $\HL- \rmi\omega_0$ in the estimation of $\uu_1, \uu_2$ gives additional source terms $\rmi m \omega_2 \uu^k_{(m)}$ and $ \rmi (m+n) \omega_1 \uu^{kk}_{(mn)}$ that are of third order. Since they are orthogonal to the translational and rotational RFs, filament motion will only be affected from fourth order in $\la$ on. The third order source terms are thus given by Eq. \eqref{inventoryS3}, which we here write as 
\begin{eqnarray} \label{decS3}
 &&\ket{\SSS_3} = \LA^A w^2 \ket{\SSS_A^{kww}} +  \LA^A \LA^B \LA^C \ket{\SSS^{kkk}_{ABC}} \qquad \quad \\
&& \quad + w \dd_\s \LA^A \ket{\SSS_A^{dkw}} + \LA^A \dd_\s w \ket{\SSS_A^{kdw}} + \dd^2_\s \LA^A \ket{\SSS_A^{ddk}}. \nn
\end{eqnarray}
By projection of Eq. \eqref{mastereq} on the translational RF, the following drift contributions are found:
\begin{eqnarray} \label{decv3}
 && v^F_3 =\LA^A w^2 (v^{kww})^F_A
+  \LA^A \LA^B \LA^C  (v^{kkk})^F_{ABC} \\
&& \quad + w \dd_\s \LA^A (v^{wdk})^F_A+ \LA^A \dd_\s w (v^{kdw})^F_A  + \dd^2_\s \LA^A (v^{ddk})^F_A. \nn 
\end{eqnarray}
More explicitly, we obtain
\begin{widetext}
\bsub \label{V3real} \begin{eqnarray}
 (v^{kww})^F_{A} &=&- \bra{\bY^\theta} \HP \ket{\dd_A \uu_0 } \braket{\bY^F}{\dd_\theta \uu^{ww}}  + \bra{\bY^\theta} \HP \ket{\dd^2_\theta \uu_0} \braket{\bY^F}{\dd_\theta \uu^k_A} + \bra{\bY^F}(\HP\delta^B_A - P^B_{\hs A} ) \ket{\dd_B \uu^{ww}} \nn  \\
&&  + \bra{\bY^B} \HP \ket{\dd^2_\theta \uu_0} \braket{\bY^F}{\dd_B \uu^k_A} - \bra{\bY^F} \HP \ket{\dd^2_\theta \uu^k_A} + \bra{\bY^F} \HP \rho_A  \ket{ r \dd_r \uu_0}  - \bra{\bY^F} \HP r^2 \ket{\dd_A \uu_0}  \nn \\
&& -2 \bra{\bY^F} \HP \rho_A \ket{\dd^2_\theta \uu_0}  - \braket{\bY^F}{\bF'' \uu^k_A \uu^{ww}},  \\
 (v^{kkk})^F_{ABC} &=& - \bra{\bY^\theta} \HP \ket{\dd_A \uu_0 } \braket{\bY^F}{\dd_\theta \uu^{kk}_{BC}}  -\komega^{kk}_{AB} \braket{\bY^F}{\dd_\theta \uu^{k}_{C}} - (v_{kk})^D_{AB}  \braket{\bY^F}{\dd_D\uu^{k}_{C}}\nn \\
&& + \bra{\bY^F}(\HP\delta^D_A - P^D_{\hs A} ) \ket{\dd_D \uu^{kk}_{BC}} +  \bra{\bY^F} \HP \rho_A \ket{\dd_B \uu^k_C}  \\ 
&& +  \bra{\bY^F} \HP \rho_A \rho_B \ket{\dd_C \uu_0}  - \braket{\bY^F}{\bF'' \uu^k_A \uu^{kk}_{BC}}  - \frac{1}{6} \braket{\bY^F}{\bF''' \uu^k_A \uu^k_B \uu^k_C}, \nn \\
(v^{dkw})^F_{A} &=& \bra{\bY^F} \rho_A \HP \ket{\dd_\theta \uu_0} + 2 \bra{\bY^F} \HP \ket{\dd_\theta \uu^k_A},\\
(v^{kdw})^F_{A} &=& - \bra{\bY^\theta} \HP \ket{\dd_A \uu_0 } \braket{\bY^F}{\dd_\theta \uu^{dw}} 
 - \bra{\bY^B} \HP \ket{\dd_\theta \uu_0} \braket{\bY^F}{\dd_B \uu^k_A} 
 + \bra{\bY^F}  (\HP -  \komega^{dw} ) \ket{\dd_\theta \uu^k_A} 
 \nn \\ 
&& + \bra{\bY^F}(\HP\delta^B_A -  P^B_{\hs A} ) \ket{\dd_B \uu^{dw}}  - \braket{\bY^F}{\bF'' \uu^k_A \uu^{dw}} + 2 \bra{\bY^F} \HP \rho_A \ket{\dd_\theta \uu_0},  \\
(v^{ddk})^F_{A} &=& - \bra{\bY^F} \HP \ket{\uu^k_A}.
\end{eqnarray} \esub
\end{widetext}

Remark that in $v^{kww}$, we have simplified a term using the identity $\eps_{AB} \eps_{CD} = \delta_{AC} \delta_{BD} - \delta_{AD} \delta_{BC},$ from which follows that
$\epsilon_A^{\hs C} \bra{\bY^F} \HP \rho_C \ket{\psi_\theta} =\bra{\bY^F} \HP \rho_A  \ket{ r \dd_r \uu_0}  - \bra{\bY^F} \HP r^2 \ket{\dd_A \uu_0} 
$. 

Once more, the expressions become more manageable when restricted to equal diffusion systems:
\bsub \begin{eqnarray}
 (v^{kww})^F_{A} &=& D_0  \bra{\bY^F} \rho_A  \ket{ r \dd_r \uu_0}  - D_0 \bra{\bY^F} r^2 \ket{\dd_A \uu_0}\nn \\
&&  - 2 D_0 \bra{\bY^F} \rho_A \ket{\dd^2_\theta \uu_0},  \\
 (v^{kkk})^F_{ABC} &=&  D_0 \bra{\bY^F} \rho_A \rho_B \ket{\bpsi_C},  
\end{eqnarray}
\begin{eqnarray} 
 (v^{dkw})^F_{A} &=& D_0\bra{\bY^F} \rho_A \ket{\dd_\theta \uu_0},  \\
  (v^{kdw})^F_{A} &=& 2 D_0\bra{\bY^F} \rho_A \ket{\dd_\theta \uu_0}, \\
 (v^{ddk})^F_{A} &=& 0. 
\end{eqnarray} \label{v3eqdiff} \esub
For the case of equal diffusion of state variables, it appears that feedback loops through modifications $\tuu$ of the wave profile do not come into play in the given order. Only the extent of the spiral's core matters here, as seen by the factors $r$ and $\rho^A$ in the third order coefficients. Note, moreover, that $v^{ddk}$ in this case  vanishes and $v^{kdw} = 2 v^{dkw}$.  

\subsection{Instantaneous motion of the gauge filament \label{sec:eomproof_result}}

The final result of the previous section is the following. If one defines the scroll wave filament in the sense that it is the center line of the solution which has no Goldstone-mode component, this gauge filament curve is proven to obey the dynamical laws:\\
\bsub \label{eominstantrot}
\begin{eqnarray}
 \dd_\tau \phi &=& \komega_0 +\komega^k_A \LA^A +  \komega^{ww} w^2 + \komega^{kk}_{AB} \LA^A \LA^B + \komega^{dw} \dd_\s w \nn \\
 && +  \komega^{kww}_A \LA^A w^2  +  \komega^{kkk}_{ABC} \LA^A \LA^B \LA^C  + \komega^{kdw}_A \LA^A \dd_\s w \nn  \\
 && +  \komega^{dkw}_A \dd_\s \LA^A w + \komega^{ddk}_A \dd_\s^2 \LA^A + \OO(\la^4), 
\\ 
 \dd_\tau X^F &=& (v^k)^F_A \LA^A + (v^{ww})^F w^2 + (v^{kk})^F_{AB} \LA^A \LA^B  \\
 && + (v^{dw})^F \dd_\s w  + (v^{kww})^F_A \LA^A w^2 \nn \\
&& +  (v^{kkk})^F_{ABC} \LA^A \LA^B \LA^C  + (v^{kdw})^F_A \LA^A \dd_\s w 
\nn  \\ &&  + (v^{dkw})^F_A \dd_\s \LA^A w + (v^{ddk})^F_A \dd_\s^2 \LA^A + \OO(\la^4).  \nn
\end{eqnarray} \esub
The equations do not simplify when going to the laboratory frame of reference. For, at time $t=0$, the quantities above do not depend on $\rho^A$, whence $\dd_t = \dd_\tau$. Secondly, the rotating reference vectors $\vec{N_A}$ may be transformed to the non-rotating triad vectors $\vec{N_a}$, using a simple rotation matrix. Therefore, the transformation to the laboratory frame of reference merely involves a change of indices: 

\bsub \label{eominstantlab}
\begin{eqnarray}
 \dd_t \phi &=& \komega_0 +\komega^k_a \LA^a +  \komega^{ww} w^2 + \komega^{kk}_{ab} \LA^a \LA^b + \komega^{dw} \dd_\s w  \nn \\
 &&
+  \komega^{kww}_a \LA^a w^2 
 +  \komega^{kkk}_{abc} \LA^a \LA^b \LA^c  + \komega^{kdw}_a \LA^a \dd_\s w  \nn \\
 && +  \komega^{dkw}_a \dd_\s \LA^a w + \komega^{ddk}_a \dd_\s^2 \LA^a + \OO(\la^4),  \label{eominstantlaba} \\
 \dd_t X^f &=& (v^k)^f_a \LA^a + (v^{ww})^f w^2 + (v^{kk})^f_{ab} \LA^a \LA^b  \label{eominstantlabb} \\
 && + (v^{dw})^f \dd_\s w  + (v^{kww})^f_a \LA^a w^2  \nn \\
&& + (v^{kkk})^f_{abc} \LA^a \LA^b \LA^c  + (v^{kdw})^f_a \LA^a \dd_\s w   \nn \\
 && + (v^{dkw})^f_a \dd_\s \LA^a w + (v^{ddk})^f_a \dd_\s^2 \LA^a + \OO(\la^4).  \nn
\end{eqnarray}  \esub

For a laboratory observer, the $\LA^a$ are slowly varying vectors, but all coefficients $\omega^{\ldots}_{\ldots}$, $(v^{\ldots})^f_{\ldots}$ that have spatial indices are oscillating in time due to the scroll wave's rotation. Not surprisingly, the instantaneous motion of gauge filaments is therefore strongly phase-dependent. Unfortunately, this implicit oscillatory behavior makes further analysis rather difficult.  

To perform rigorous stability analysis, we were forced to introduce some notion of a mean filament position. First, we tried to perform temporal averaging over one period, but the concept of a period is hard to define when the phase itself varies in time on both long and short time scales. Therefore, we have constructed a mean filament by filtering out the phase-dependent oscillations instead. We call these filaments `virtual filaments', in contrast to the gauge filament which we used so far. The definition of virtual filaments and their equation of motion are presented in the next section. 

 \section{Simple dynamics using the virtual filament \label{sec:virtual}}

\subsection{Gauge filament trajectory in lowest order}

We now aim to find the leading order expression for $X^a(\s,t)$, i.e. the solution to the quasi-periodic, non-linear differential equations \eqref{eominstantlaba}-\eqref{eominstantlabb}. 
First, we have attempted to compute the net motion of the curve during one rotation period. Alas, the period of rotation itself is hard to define if a filament is curved and twisted. Now, we find it more useful to isolate the `fast' components of the filament motion, i.e. those contributions with cyclic dependency on $n \omega_0 t$ for integer $n$. In the limit of small filament twist and curvature, the oscillations at a frequency $n \omega_0 t$ occur on a much faster timescale than the filament drift induced by curvature and twist. These will provide only a slow amplitude modulation to the oscillating terms. 

Let us first illustrate the proposed approach by performing the analysis in linear order in the filament's curvature. In the process, it is convenient to single out the phase evolution of an unperturbed scroll wave, which is why we define the unperturbed phase $\zeta_0$ and phase correction $\zeta$ as
 \begin{eqnarray}
\zeta_0(\s, t) &=& \phi_0(\s) + \komega_0 t, \nn\\
 \phi(\s, t) &=&  \zeta_0(\s, t) + \zeta(\s, t), \label{defzeta} 
\end{eqnarray} 
where $\zeta$ is expected to be small in the case of low curvature and twist. In first order in $\la$, the laws of motion for the gauge filament simplify to
\bsub \label{linsys1} \begin{eqnarray}
\dd_t \zeta &=&  \komega^k_a \LA^a + \OO(\la^2),\label{veomphase1} \\
\dd_t X^b &=& (v^k)^b_a \LA^a + \OO(\la^2). \label{veomtrans1}
\end{eqnarray} \esub
For a given initial filament position with phase $\phi_0(\s)$ and curvature components $\LA_0(\s)$, we may now explicitly integrate the linear system \eqref{linsys1}. If the phase evolution had been independent of the filament shape, the coefficients in Eq \eqref{veomtrans1} had been periodic in time, such that one would have obtained a Floquet problem. In the present case, we will therefore look for a Fourier series solution with slow amplitude modulation:
\bsub \label{timesep} \begin{eqnarray}
   \label{Zexp1} &&\zeta(\s, t) = Z_0(\s, t) \\
&& \quad +  Z_c(\s, t) \cos \zeta_0(\s, t) +    Z_s(\s, t) \sin \zeta_0(\s, t) \nn \\ 
&& \quad +   Z_{cc}(\s, t) \cos 2\zeta_0(\s, t) +  Z_{ss}(\s, t) \sin 2 \zeta_0(\s, t) + \ldots \nn \\ 
  \label{XXexp}   &&X^a(\s, t) = X^a_0(\s, t)  \\
&& \quad + X^a_c(\s, t) \cos \zeta_0(\s, t) + X^a_s(\s, t) \sin \zeta_0(\s, t) \nn \\ 
&&\quad + X^a_{cc}(\s, t) \cos 2\zeta_0(\s, t) + X^a_{ss}(\s, t) \sin 2 \zeta_0(\s, t) + \ldots \nn 
\end{eqnarray}
\esub
In linear order in $\la$, $\LA^a = \dd_\s^2 X^a = \LA_0^a + \OO(\la^2)$. Furthermore, the rotation matrices that relate the co-rotating and laboratory frame are
\bsub \label{Rrule1} \begin{eqnarray}
 x^a &= &R^a_{\hs A} x^A, \\
  &&R^a_{\hs A} = (\RRR)^a_{\hs A} = \delta^a_A \cos \phi + \eps^a_{\hs A} \sin \phi, \nn \\
 x^A &= &R^A_{\hs a} x^a,  \\
 &&R^A_{\hs a}= (\RRR^T)^A_{\hs a} = \delta^A_a \cos \phi + \eps_a^{\hs A} \sin \phi. \nn 
\end{eqnarray} \esub
Thus, putting the form \eqref{Zexp1} into Eq. \eqref{veomphase1} brings
\begin{multline}
  \dot{Z_0} + (\dot{Z_c} + \komega_0 Z_s) \cos \zeta_0 + (\dot{Z_s} - \komega_0 Z_c) \sin\zeta_0 \\ = \komega^k_A \LA^a_0 \left(\delta^A_a  \cos \zeta_0 + \eps_a^{\ A} \sin \zeta_0 \right) + \OO(\la^2). \label{Zeq1}  
\end{multline}
We will also write, in obvious matrix notation, 
\begin{eqnarray}
\komega^k_A \LA^a_0 \delta_a^A &= \bom \BLA_0,  \qquad \komega^k_A \LA^a_0 \eps_a^{\hs A} = - \bom \beps \BLA_0.  
\end{eqnarray}
Next, identifying the oscillating components in Eq. \eqref{Zeq1}, we see that $\dot{Z_0}= 0$, together with
\begin{eqnarray} \label{linsysrot1}
\dot{Z_c} + \komega_0  Z_s &= \bom \BLA_0, \qquad
\dot{Z_s} -  \komega_0  Z_c = - \bom \beps \BLA_0. 
\end{eqnarray}
Elimination of $Z_s$ delivers $\ddot{Z_c} + \komega_0^2 Z_c = \komega_0 \bom \beps \BLA_0$. In the general solution $Z_c(t) = \alpha_1 \cos \omega_0 t  + \alpha_2 \sin \omega_0 t  + \frac{\bom \beps \BLA_0 }{\komega_0}$, the coefficients $\alpha_1,\alpha_2$ must vanish, since we defined $Z_c$ to be a slow amplitude modulation; the same argument holds for $Z_s$. Therefore, we find
\begin{eqnarray}
 Z_s &=& \frac{\bom \BLA_0}{\komega_0} + \OO(\la^2),
\nn\\
 Z_c &=& \frac{\bom \beps \BLA_0 }{\komega_0} + \OO(\la^2),\label{Zsol1}\\
 Z_0 &=&  -Z_c \cos \phi_0 -Z_s \sin \phi_0 + \OO(\la^2). \nn
\end{eqnarray}
The constant value of $Z_0$ follows from the requirement that the phase correction $\zeta$ should vanish for $t=0$. We conclude from Eqs. \eqref{Zexp1}, \eqref{Zsol1} that in lowest order in $\la$ and for small times $t$, the scroll wave's phase oscillates at a frequency $\omega_0$ with constant amplitude around its unperturbed value, with amplitude proportional to the filament curvature $k$:
\begin{eqnarray}
  A_{phase} &=& \frac{\sqrt{(\bom \BLA_0)^2+(\bom \beps \BLA_0)^2} }{|\komega_0|} = \frac{||\bom || k}{\omega_0}
\\ &=&   \frac{ | \omega^k_+ |}{\omega_0} k
 =  \frac{ | \bra{\WW^{(0)}} \HP \ket{\VV_+}|}{\omega_0} k
.
\end{eqnarray}

Next, we consider the filament motion in the plane transverse to the filament at $t=0$, i.e.
\begin{eqnarray}
  X^i(\s,t) = X^i(\s,0) + X^a(\s,t) \vec{N_a}(\s) \label{filframe_gauge}
\end{eqnarray}
with $X^a(\s,t)$ expanded as in  Eq. \eqref{XXexp} and $X^a(\s,0)=0$. Viewed as a vector in the transverse plane to the filament, we will whenever convenient write $X^a$ as $\XX$. 
For the time-dependent coefficient $P^b_{\hs a}$, we find
\begin{eqnarray}
  P^b_{\hs a} =& R^b_{\hs B} P^B_{\hs A} R^A_{\hs a},  \label{Plabind}
\end{eqnarray}
or, in matrix notation, $\bP_{lab} = \RRR \bP \RRR^T$. 
Here, we shall decompose the constant matrix $P^B_{\hs A}$ in a basis of Pauli matrices (see also Appendix \ref{app:pauli}), i.e.
\begin{eqnarray}
  \bP =& P_0 \bdel +  P_1 \bsig^1 +  P_2 \beps + P_3 \bsig^3.
\end{eqnarray}
Furthermore, we split the rotation matrix $\RRR = \RRR_0 \tilde{\RRR}$:
\begin{eqnarray}
\RRR = \exp \beps \phi, \qquad
  \RRR_0 = \exp \beps \zeta_0, \qquad 
\tilde{\RRR} = \exp \beps \zeta. 
\end{eqnarray}
Rotating the tensor $\bP$ according to the unperturbed phase brings for $\tilde{\bP} = \RRR_0 \bP \RRR_0^T$, with $\RRR_0 = \cos \zeta_0 \bdel + \sin \zeta_0 \beps$,
\begin{eqnarray}
 \tP &=& \frac{\bP - \beps \bP \beps}{2} + \frac{\bP + \beps \bP \beps}{2} \cos 2\zeta_0+  \frac{\beps\bP - \bP \beps}{2} \sin 2 \zeta_0 \nn \\
&=&  (P_0 \bdel + P_2 \beps) + \bP_6  \cos 2 \zeta_0 + \bP_5  \sin 2 \zeta_0.
\label{tildePtf}
\end{eqnarray}
Here, we have additionally defined
\begin{eqnarray}
\bP_5 &= (P_1 \bsig^3 - P_3 \bsig^1), \qquad
\bP_6 = (P_1 \bsig^1 + P_3 \bsig^3). \label{defP56}
\end{eqnarray}
Useful properties of these matrices can be found in Appendix \ref{app:pauli}. Finally, we find for the $P^b_{\hs a} = (\bP_{lab})^b_{\hs a} $:
\begin{eqnarray}
   \bP_{lab} &=&  \RRR_0\exp(\beps \zeta) \bP \exp(-\beps \zeta)^T \RRR^T_0   \nn \\
&=&  P_0 \bdel + P_2 \beps + (1-2\zeta^2)\left(  \bP_6   \cos 2 \zeta_0 + \bP_5  \sin 2 \zeta_0 \right) \nn \\
&&  + 2 \zeta  \left(\bP_5  \cos 2\zeta_0 - \bP_6  \sin 2 \zeta_0 \right) + \OO(\la^3). \label{Plab}
\end{eqnarray} 
For translational motion, we find from Eqs. \eqref{tildePtf} and \eqref{Plab} that $\bP_{lab} = \tP +\OO(\la)$, which implies that 
\begin{eqnarray}
 \dot{\XX} = \tP \BLA_0 + \OO(\la^2). 
\end{eqnarray}
Substituting the proposed form \eqref{XXexp} then yields the linear system
\begin{eqnarray}
\dot{\XX}_0 &=& (P_0 \bdel + P_2 \beps) \BLA_0, \nn\\
\dot \XX_c + \komega_0 \XX_s &=& \bzero, \nn \\
\dot \XX_s - \komega_0 \XX_c &=& \bzero, \label{linsystrans1}  \\
\dot \XX_{cc} + 2\komega_0 \XX_{ss} &=& \bP_6 \BLA_0, \nn\\
\dot \XX_{ss} - 2\komega_0 \XX_{cc} &=& \bP_5 \BLA_0. \nn
\end{eqnarray} 
This linear system of equations is uncoupled in the given order and therefore easily solved, after one recalls that the coefficients $\XX_{\ldots}$ cannot oscillate at frequencies $n \omega_0$ with $n$ integer. Moreover, the initial condition $\XX(\s,0) = \mathbf{0}$ fixes the value of $\XX_0$: 
\begin{eqnarray}
 \XX_0 &=& t (P_0 \bdel + P_2 \beps) \BLA_0  \nn \\ 
&& -\XX_{cc} \cos 2\phi_0 - \XX_{ss} \sin 2\phi_0 ++ \OO(\la^2),  \nn\\
 \XX_s &=&  \mathbf{0} + \OO(\la^2),  \nn \\
 \XX_c &=&  \mathbf{0} + \OO(\la^2),  \\
  \XX_{ss} &=& \frac{\bP_6 \BLA_0}{2\komega_0}  + \OO(\la^2), \nn \\
  \XX_{cc} &=& - \frac{ \bP_5 \BLA_0}{2\komega_0}  + \OO(\la^2).  \nn
\end{eqnarray} 
Note that the oscillatory terms $\XX_{ss}, \XX_{cc}$ have not been described elsewhere. In Eq. \eqref{P56ortho} of the appendix, we prove that the vectors $\bP_5 \BLA_0$ and $\bP_6 \BLA_0$ are mutually orthogonal, from which the amplitude of the drift oscillations is given by
\begin{eqnarray}
   \quad A_{tr} &=& \frac{1}{2 |\komega_0|}\sqrt{ \BLA_0^T (\bP_5^2 \cos^2 2\zeta_0+ \bP_6^2 \sin^2 2\zeta_0) \BLA_0} \nn \\
  &=& \frac{\sqrt{(P_1^2 +P_3^2)} \ k}{2 \omega_0} 
= \frac{k}{2 \omega_0}  | \bra{\WW^+} \HP \ket{\VV_-} |. \quad
\end{eqnarray}

In summary, we find in lowest order that the solution to the dynamical laws for filament motion is given by
\bsub \label{filamentmotion1}
\begin{eqnarray}
    \phi(\s, t) &=&  \phi_0 + \komega_0 t \label{filamentmotion1a}
 \\ && + \frac{\bom  \beps\BLA_0}{\komega_0} \left[ \cos(\komega_0 t +\phi_0) - \cos \phi_0\right] \nn \\
&& + \frac{\bom \BLA_0}{\komega_0}  \left[ \sin(\komega_0 t +\phi_0) - \sin \phi_0\right] +\OO(\la^2),\nn 
 \\
  \XX(\s,t) &=& t (P_0 \bdel + P_2 \beps) \BLA_0
 \label{filamentmotion1b} \\ &&  - \frac{ \bP_5 \BLA_0}{2\komega_0}  
\left[ \cos 2( \komega_0 t + \phi_0) - \cos 2 \phi_0\right] \nn \\ 
 && + \frac{ \bP_6 \BLA_0}{2\komega_0}  
\left[ \sin 2( \komega_0 t + \phi_0) - \sin 2 \phi_0\right]
+ \OO(\la^2).  \nn
\end{eqnarray} \esub

\subsection{The virtual filament \label{sec:virtualdef}}
 

Although the solution \eqref{filamentmotion1} may be averaged in time over one rotation period, this procedure becomes increasingly cumbersome when going to higher order. Notably, the rotation period will vary per cycle, and depend on the chosen initial phase $\phi_0$. We take another approach here, based on the observation that the actual filament motion naturally separates in a slow component and a fast oscillatory component. In particular,  Eqs. \eqref{filamentmotion1} reveals that the filament trajectory in every transverse plane takes the shape of a cycloid for small $t$ and $\la$. For this reason, we propose to redefine the filament curve such that all fast cyclic motion is suppressed from it. We shall call this curve the `virtual filament', in contrast to the `gauge filament', which we defined in section \ref{sec:gauge} by demanding that the wave profile around it should possess no component along the Goldstone modes of the problem.

Around every time instance $t$, the gauge filament is known to describe an epicycle trajectory. The \textbf{virtual filament} to a scroll wave is the instantaneous gauge filament, after the epicycle motion component due to the rotation of the scroll wave has been eliminated. 
Direct application of this definition to the lowest order filament solution \eqref{filamentmotion1} immediately brings for the virtual filament that
\bsub  \label{virfilsol1}  \begin{eqnarray}
 \phi_{vir} &=&  \phi_0 + \komega_0 t \label{virfilsol1a} \\
&& - \frac{\bom  \beps\BLA_0}{\komega_0}  \cos \phi_0 -
\frac{\bom \BLA_0}{\komega_0}  \sin \phi_0 +\OO(\la^2),  \nn 
\end{eqnarray}
\begin{eqnarray}
 \XX_{vir} &=& t (P_0 \bdel + P_2 \beps) \BLA_0  \label{virfilsol1b}  \\
&& + \frac{ \bP_5 \BLA_0}{2\komega_0}  
\cos 2 \phi_0
 - \frac{ \bP_6 \BLA_0}{2\komega_0}  
 \sin 2 \phi_0 
+ \OO(\la^2).  \nn
\end{eqnarray}  \esub
The trajectory of the virtual filament still seems to depend on the initial phase $\phi_0$ of the gauge filament. Fortunately, we may recall that our twist-adapted frame is still attached to the gauge filament at the time $t=0$, which causes the residual dependency on $\phi_0$. If one, however, attaches the frame of reference to the virtual filament at $t=0$, one finds instead of Eqs. \eqref{filframe_gauge} that
\bsub \begin{eqnarray}
  \phi(\s,t) = \phi_{vir}(\s,0) + \omega_0 t + \zeta(\s,t). \\
  X^i(\s,t) = X^i_{vir}(\s,0) + X^a(\s,t) (N^{vir})^i_a(\s). \label{filframe_vir}
\end{eqnarray} \esub
In this reference frame, the same linear systems \eqref{linsysrot1},\eqref{linsystrans1} are generated, but the initial conditions $\zeta(\s,0)=0$, $X^a(\s,0)=0$ are replaced by the vanishing of oscillatory components in the solution. In the frame of the virtual filament, we readily find instead of Eqs. \eqref{filamentmotion1} that the gauge filament evolves as 
\bsub \label{filamentmotion1virframe}
\begin{eqnarray}
\phi(\s, t) &=& \phi_0 + \komega_0 t   + \frac{\bom  \beps\BLA_0}{\komega_0}  \cos(\komega_0 t +\phi_0) \nn \\
&&  + \frac{\bom \BLA_0}{\komega_0}  \sin(\komega_0 t +\phi_0) +\OO(\la^2),  \\
 \XX(\s,t) &=& t (P_0 \bdel + P_2 \beps) \BLA_0 
 - \frac{ \bP_5 \BLA_0}{2\komega_0}   \cos 2( \komega_0 t + \phi_0) \nn \\ &&  + \frac{ \bP_6 \BLA_0}{2\komega_0}  \sin 2( \komega_0 t + \phi_0) + \OO(\la^2).  
\end{eqnarray} \esub
By definition, the virtual filament consists of the non-oscillating drift components:
\begin{eqnarray}
  \phi_{vir}(\s, t) =  \phi_0 + \komega_0 t  +\OO(\la^2),  \label{filamentmotion1vir}\\
\XX_{vir}(\s,t) = t (P_0 \bdel + P_2 \beps) \BLA_0 + \OO(\la^2).  \nn
\end{eqnarray} 
Since these relations are satisfied in a small interval of time around an arbitrarily picked instance of time, we may differentiate the solution \eqref{filamentmotion1vir} with respect to time to find the law of motion for the virtual filament:
\begin{eqnarray}  \label{virfil1a} 
  \dd_t \phi_{vir}(\s, t) &=&  \komega_0  +\OO(\la^2),  \\
\dd_t \XX_{vir}(\s,t) &=& (P_0 \bdel + P_2 \beps) \BLA_{vir} + \OO(\la^2). \nn
\end{eqnarray} 
We have now finally established a result that we had been searching for, since Eq. \eqref{virfil1a} can be written with $\gamma_1 \equiv P_0$ and $\gamma_2 \equiv - P_2$ as
 \begin{eqnarray} \label{virfil1}
  \dd_t \phi_{vir} &=&  \komega_0  +\OO(\la^2),  \\
\dd_t \vec{X}_{vir}(\s,t) &=& \gamma_1 \dd_\s^2 \vec{X}_{vir} + \gamma_2 \dd_\s \vec{X}_{vir} \times  \dd_\s^2 \vec{X}_{vir} + \OO(\la^2). \nn
\end{eqnarray}
Our result is formally identical to the rotation-averaged law of motion by Biktashev \textit{et al}.  \cite{Biktashev:1994}. Here, we re-interpret this classical expression as being the instantaneous equation of motion for the virtual filament.


We now continue the derivation of the EOM for the virtual filament up to third order in twist and curvature; the final result will be reached in section \ref{sec:eomvir}, so the reader which is not interested in the technical derivation may continue reading from Eq.\eqref{filamentmotion3}. 
 
\subsection{Higher order rotational dynamics for the virtual filament}

To denote the position of the filament curves, an absolute, stationary frame of reference is now chosen. Since phase dynamics can be treated at the level of the EOM for the gauge filament, there is no need to construct a twisted frame. For simplicity, we thus construct a relatively parallel adapted frame around the virtual filament at time $t=0$, which has position $X^i_0(\s,0)$. That is, the basis vectors $\vec{T}_0(\s), \vec{N^1}(\s)$ and $\vec{N^2}(\s)$ are parallel transported along the virtual filament curve. Relative to this fixed frame, the transverse velocity gauge is imposed on time evolution of both the gauge filament $X^i$ and the virtual filament $X_0^i$:
\begin{eqnarray}
  X^i(\s,t) &= X^i_0(\s,0) + X^a(\s,t) \vec{N_a}(\s,0), \nn \\
  X^i_0(\s,t) &= X^i_0(\s,0) + X^a_0(\s,t) \vec{N_a}(\s,0).
\end{eqnarray}
The lower cap indices $a,b,c,...$ are used here since the frame is non-rotating. As before, we introduce a boldface matrix notation in the transverse plane, and write $\XX$ and $\XX_0$ instead of $X^a$ and $X^a_0$.

A frame of reference directions on the gauge filament, annotated $(\vec{e^\s}(\s,t)$, $\vec{e^1}(\s,t)$, $\vec{e^2}(\s,t))$ is obtained from the set $\vec{N^a}(\s)$, by performing the following procedure at all times $t$. 
The fixed orthonormal vectors $\vec{N^1}(\s)$, $\vec{N^2}(\s)$ are parallel transported along a straight line to the intersection point of the gauge filament and the given transverse plane. Afterwards, they are minimally rotated to become orthogonal to the gauge filament. These conditions completely fix the time-dependent frame on the gauge filament, such that we may compute (see Eq. \eqref{laexp3app} in Appendix \ref{app:frame}):
\bsub \label{laexp3} \begin{eqnarray}
 \LA^a(\s,t) &=&  \vec{e^a}(\s,t) \cdot \dd_s^2 \vec{X}(\s,t) \nn \\&=& \LA^a_0 + \LA^a_0 X_b \LA^b_0 + X''^a +\OO(\la^4), \\
\wfr (\s,t) &=& \frac{\eps^{ab}}{2}\dd_s \vec{e_a}(\s,t) \cdot \vec{e_b}(\s,t) \nn \\ &=& \eps_a^{\hs b} X'_b \LA^a_0 + \OO(\la^4). 
\end{eqnarray} \esub

In second order, the instantaneous, quasi-periodic non-linear differential equation for the gauge filament phase is given by Eq.\eqref{eominstantlaba}. 
For the twist and curvature terms it can be calculated that (with $f' = \dd_\s f $)
\bsub \begin{eqnarray}
 \dd_s \phi &=& \phi_0' + Z_0'+ (Z_c' + Z_s \phi_0') \cos \zeta_0 \\ &&\nn + (Z_s'-Z_c \phi_0') \sin \zeta_0 +  \OO(\la^3),\quad\ \ \\
(\dd_s \phi)^2 &=& \phi_0'^2 + \OO(\la^3), \\
 \dd^2_s \phi &=& \phi_0'' + \OO(\la^3),\\
\LA^a &=& \LA^a_0 + \OO(\la^3). 
\end{eqnarray} \esub 
We also decompose the matrix element $\komega^{kk}_{AB} = R_A^{\hs a} \komega_{ab}^{kk} R^b_{\hs B}$ in the Pauli matrix basis:
\begin{eqnarray}
  (\komega^{kk}_{AB}) &=& \Theta_0 \bdel + \Theta_1 \bsig^1 +  \Theta_2 \beps + \Theta_3 \bsig^3, \\
\BTH_5 &=& \Theta_1 \bsig^3 - \Theta_3 \bsig^1,  \quad 
\BTH_6 \ =\ \Theta_1 \bsig^1 + \Theta_3 \bsig^3,  \nn
\end{eqnarray} 
such that in analogy to the transformation rule for $(P^b_{\hs a})$, i.e. Eq. \eqref{tildePtf} follows
\begin{multline}
  \komega_{ab}^{kk} \LA^a \LA^b = \Theta_0 k_0^2 + \BLA_0^T\BTH_6 \BLA_0 \cos 2 \zeta_0 \\ + \BLA_0^T\BTH_5 \BLA_0 \sin 2 \zeta_0 + \OO(\la^3).
 \label{omab2}
\end{multline}
The second order contribution within $\omega_a \LA^a$ is found as
\begin{eqnarray}
   \komega_a^{k} \LA^a  &=& \bom \RRR^T \BLA_0 + \OO(\la^3) \nn \\
&=& \bom \RRR_0^T(\cos \zeta \bdel - \sin \zeta \beps ) \BLA_0 + \OO(\la^3)\nn\\
  &=& - \frac{(\bom \beps \BLA_0)^2 + (\bom \BLA_0)^2}{2 \komega_0} 
+ \cos \zeta_0 \bom \BLA_0 \nn \\&&
 -  \sin \zeta_0  \bom \beps \BLA_0\nn \\
  && + \cos 2 \zeta_0 \frac{(\bom \BLA_0)^2- (\bom \beps \BLA_0)^2 }{2 \komega_0} \nn \\&&
- \sin 2 \zeta_0 \frac{(\bom \BLA_0)( \bom \beps \BLA_0)}{\komega_0} + \OO(\la^3). \label{omalama}
\end{eqnarray} 
Here, we introduce the notation  
\begin{eqnarray}
 \frac{\bom \otimes \bom}{\komega_0} &=& \bar{\BTH} = \bar{\Theta}_0 \bdel + \bar{\Theta}_1 \bsig^1 + \bar{\Theta}_3 \bsig^3,  \\
\bar{\BTH}_5 &=& \bar{\Theta}_1 \bsig^3 - \bar{\Theta}_3 \bsig^1, \quad 
\bar{\BTH}_6 = \bar{\Theta}_1 \bsig^1 + \bar{\Theta}_3 \bsig^3. \nn
\end{eqnarray} 
such that
\begin{multline}
   \komega_a^{k} \LA^a  = - \bar{\Theta}_0 k^2_0 + \cos \zeta_0  \bom \BLA_0 -  \sin \zeta_0 \bom \beps \BLA_0 \\  
+ \cos 2 \zeta_0\ \BLA_0^T \BTH_6 \BLA_0 + \sin 2 \zeta_0 \ \BLA_0^T \BTH_5 \BLA_0
 + \OO(\la^3). 
\end{multline} 
At this point, we can define the coefficients that will appear in the final phase equation:
\begin{eqnarray}
a_0 &=& \komega_{ww}, \nn \\
b_0 &=& \Theta_0- \bar{\Theta}_0 = \frac{\delta^{AB}}{2} \left( \komega_{AB}^{kk}  - \frac{\komega_A^{k} \komega_B^{k}}{\komega_0} \right), \label{rot2uneq}  \\
d_0 &=& \komega_{dw}. \nn
\end{eqnarray} 
Then, the system for the phase oscillations up to second order becomes, with $w_0  = \phi_0'$,
\begin{eqnarray}
\dot{Z}_{0} &=& a_0 w_0^2 + b_0 k^2_0 + d_0 w_0', \nn \\
\dot{Z_c} + \omega_0  Z_s &=& \bom \BLA_0, \nn \\
 \dot{Z_{cc}} + 2\omega_0  Z_{ss} &=& \BLA_0^T (\BTH_6 + \bar{\BTH}_6 ) \BLA_0, \label{Zvar2}\\
\dot{Z_s} -  \omega_0  Z_c &=& - \bom \beps \BLA_0, \nn \\
\dot{Z_{ss}} -  2\omega_0  Z_{cc} &=& \BLA_0^T (\BTH_5 + \bar{\BTH}_5) \BLA_0. \nn
\end{eqnarray} 
Hence follows
\begin{eqnarray}  \label{ZcZs2}
 Z_{0} &=& ( a_0 w_0^2 + b_0 k^2_0  + d_0  w_0' ) t, \nn \\
  Z_{s} &=& \frac{\bom \BLA_0 }{\omega_0}, \qquad\ 
  Z_{ss} = \frac{\BLA_0^T (\BTH_6 + \bar{\BTH}_6 )  \BLA_0}{2\komega_0}, \\
Z_{c} &=& \frac{ \bom \beps \BLA_0}{\omega_0},
\qquad
Z_{cc} = -  \frac{ \BLA_0^T (\BTH_5 + \bar{\BTH}_5)  \BLA_0}{2\komega_0}.\nn 
\end{eqnarray} 
For the evolution of virtual phase, it could be even concluded from the first equation of \eqref{Zvar2} alone that, with $\dot{\phi}_{vir} - \komega_0 = \dot{Z}_0$, 
\begin{equation}
\dot{\phi}_{vir} = \komega_0 + a_0 w_{vir}^2 + b_0 k_{vir}^2 + d_0 w' _{vir} +  \OO(\la^3).\label{eomrot2}
\end{equation}
We thus recover the time-averaged phase equation of Biktashev \etal. \cite{Biktashev:1994}, supplemented with the term $b_0 k^2$, accounting for the differential rotation rate of scroll rings of unequal radii. This correction is responsible for the shift in the resonant window that was observed in numerical experiments to push scroll waves to the domain boundary, where they become rings of small radius and therefore of high curvature \cite{Morgan:2009}.
Since the rotational correction will be needed in what follows, we introduce 
\begin{eqnarray}
\komega_2 = a_0 w_0^2 + b_0 k^2_0 + d_0 w_0' . 
\end{eqnarray}

In third order, no contributions to the virtual phase are found. Therefore,  Eq. \eqref{eomrot2} is our most advanced result on phase evolution. This PDE is an inhomogeneous Burger's equation, with local filament curvature acting as a source term. Such equation is known to support shock waves and rarefaction waves and deserves to be studied further in the present context.

\subsection{Higher order translational dynamics for the virtual filament}

Similarly to the phase equation, the translational dynamics for the virtual filament can be analyzed in second order in curvature and twist. However, our calculations show that there is no net contribution to the virtual filament motion, eventually leading to
\begin{eqnarray}
 \dot{\XX}_0 = (P_0 \bdel + P_2 \beps) \BLA 
 + \OO(\la^3). \label{eomtrans1}
\end{eqnarray}
Nonetheless, nonzero Fourier amplitudes  $\XX_c, \XX_s, \XX_{ccc}, \XX_{sss}$ are obtained, which demonstrates that the gauge filament exhibits non-trivial second order dynamics with temporal frequency content $\omega_0$ and $3\omega_0$. 


For the third order translational dynamics, we recall the EOM \eqref{eominstantlabb} that needs to be solved for the gauge filament:
\label{eominstantlab}
\begin{eqnarray}
 \dd_t X^f &=& (v^k)^f_a \LA^a + (v^{ww})^f w^2 + (v^{kk})^f_{ab} \LA^a \LA^b  \label{eom3tr} \\
 && + (v^{dw})^f \dd_\s w  + (v^{kww})^f_a \LA^a w^2  \nn \\
&& + (v^{kkk})^f_{abc} \LA^a \LA^b \LA^c  + (v^{kdw})^f_a \LA^a \dd_\s w   \nn \\
 && + (v^{dkw})^f_a \dd_\s \LA^a w + (v^{ddk})^f_a \dd_\s^2 \LA^a + \OO(\la^4).  \nn
\end{eqnarray}

The task at hand is to determine all contributions to the virtual filament motion $\dot{\XX_0}$ that are generated in third order in curvature and twist. We now start treating all nine terms in  Eq. \eqref{eom3tr}, one at a time. The outcome will be gathered in the system \eqref{sys3tr}. 

As the first term, we need to evaluate $(v^k)^f_{a} \LA^a$, with $(v^k)^f_{a}$ previously defined as $P^f_{\hs a}$. To find $\LA^a$, we take the lowest order gauge filament solution \eqref{filamentmotion1}.
Substitution in Eq. \eqref{laexp3} 
generates 
\begin{eqnarray}
   \BLA = \BLA_0 + \MM_0 t +  \cos 2 \zeta_0 \MM_{cc} + \sin 2 \zeta_0 \MM_{ss} + \OO(\la^4) \quad
\end{eqnarray}
with 
\bsub \begin{eqnarray} 
\MM_0 &=& P_0 k_0^2 \BLA_0 + (P_0 \bdel + P_2 \beps) \BLA_0'', \\
\MM_{cc} &=&  -4 w_0^2 \XX_{cc} + (\BLA_0^T \XX_{cc})\BLA_0 
\nn \\ && + 4 w_0 \XX_{ss}' + 2 \XX_{ss}w_0' + \XX_{cc}'', \\
\MM_{ss} &=&  -4 w_0^2 \XX_{ss} + (\BLA_0^T \XX_{ss})\BLA_0 
\nn \\ && - 4 w_0 \XX_{cc}'  - 2 \XX_{cc}w_0' + \XX_{ss}''.
\end{eqnarray}  \esub
Multiplication with the tension matrix \eqref{Plab} yields a complicated expression, which is denoted as
\begin{eqnarray}
  \bP_{lab} \BLA  &=& (P_0 \bdel + P_2 \beps) \BLA_0 \label{Qdef} \\ && + \QQ_1 + \QQ_2 + \QQ_3 + \QQ_4 +  ... + \OO(\la^4). \nn
\end{eqnarray}
Henceforth, we write trailing dots ($\ldots$) to hide the terms that are the product of a constant factor in time and $\cos(n \zeta_0)$, $\sin (n \zeta_0)$, since these will only determine oscillation amplitudes and not affect virtual filament motion. The third order terms that could impact on virtual filament dynamics are given by
\begin{eqnarray}
 \QQ_1 &=& (\bP_6 \cos 2 \zeta_0 + \bP_5 \sin 2 \zeta_0 ) \BLA_0 (- 2\zeta^2), \\
 \QQ_2 &=&2 (\bP_5 \cos 2 \zeta_0 - \bP_6 \sin 2 \zeta_0 ) \BLA_0 \zeta, \nn \\
 \QQ_3 &=& (P_0 \bdel + P_2 \beps +  \bP_6   \cos 2 \zeta_0 + \bP_5  \sin 2 \zeta_0 ) \MM_0 t, \nn \\
 \QQ_4 &=&  \left( \bP_6   \cos 2 \zeta_0 + \bP_5  \sin 2 \zeta_0  \right)\left(\MM_{cc} \cos 2 \zeta_0  + \MM_{ss} \sin 2 \zeta_0  \right).  \nn 
\end{eqnarray}
It will turn out below that none of the terms in $\QQ_3$ will contribute to the virtual filament motion, as they contain a factor $t$.  After some arithmetic one finds for the remaining terms, with $\hat{\BTH} = \bar{\BTH} + {\BTH}$, 
\begin{eqnarray}
\QQ_1&=& \frac{k^2_0}{\komega_0} \left(  (\bar{\Theta}_1 P_1 + \bar{\Theta}_3 P_3) \BLA +  (\bar{\Theta}_1 P_3 - \bar{\Theta}_3 P_1) \beps \BLA  \right) + ... \nn \\
 \QQ_2 &=& 2 (\bP_5 \BLA_0 \cos 2 \zeta_0 - \bP_6 \BLA_0 \sin 2 \zeta_0 )  \komega_2 t \nn \\
&& - \frac{k^2_0}{2 \komega_0} \left(  (\hat{\Theta}_1 P_1 + \hat{\Theta}_3 P_3) \BLA +  (\hat{\Theta}_1 P_3 - \hat{\Theta}_3 P_1) \beps \BLA  \right) + ... \nn \\
\QQ_4 &=& \frac{P_1^2+ P_3^2}{2\komega_0}\left( \eps \BLA_0'' - 4 w_0^2 \beps \BLA_0 \right) \nn \\&& + \frac{P_1^2+ P_3^2}{2\komega_0}\left( 4 w_0 \BLA_0' +2 w_0' \BLA_0 \right) + ... \label{sum1}
\end{eqnarray} 

For the second term in Eq. \eqref{eom3tr}, we decompose 
\begin{eqnarray}
(v^{kk})^F_{AB} = K^F_0 \s^0_{AB} + K^F_1 \s^1_{AB} +  K^F_2 \eps_{AB} + K^F_3 \s^3_{AB}, \nn  \\
\KK^F_5 = K^F_1 \bsig^3 - K^F_3 \bsig^1, \quad 
\KK^F_6 = K^F_1 \bsig^1 + K^F_3 \bsig^3. \qquad
\end{eqnarray} 
Thus we find, after some intermediate steps,
\begin{eqnarray}
&&  (v^{kk})^f_{ab} \LA^a \LA^b = R^f_{\hs F} \BLA^T_0 \KK^F_{lab} \BLA_0 + \OO(\la^4) \qquad \label{Kab} \\
&&  = \frac{1}{2\komega_0} \left( \eps^f_{\hs F} \bom \beps \BLA_0 - \delta^f_{\hs F}  \bom \BLA_0 \right) K^F_0 k^2 _0 + ... + \OO(\la^4).  \nn 
\end{eqnarray}
To further simplify the result, consider $K^F_0$ as a vector $\KK_0$ in the transverse plane, attached to the spiral wave solution, i.e. $\KK_0 = K_0^F \mathbf{e}_F$. 
The induced motion is then, from  Eq. \eqref{Kab},
\begin{eqnarray}
  (v^{kk})^f_{ab} \LA^a \LA^b \mathbf{e}_f &= \frac{k_0^2}{2\komega_0} \left( \beps \KK_0  \bom \beps - \KK_0 \bom \right)\BLA_0. 
\end{eqnarray}
The direct product of $\KK_0$ and $\bom$ forms a tensor of rank 2, i.e. $\KK_0 \otimes \bom = \komega_0 \bka$. Subsequently, this quantity too can be decomposed in its Pauli components
 $\bka = \kappa_0 \bdel + \kappa_1 \bsig^1 +  \kappa_2 \beps +\kappa_3 \bsig^3$
to find
\begin{eqnarray}
  K^f_{ab} \LA^a \LA^b \mathbf{e}_f &= - k^2_0 \left( \kappa_0 \bdel + \kappa_2 \beps \right) \BLA_0. \label{sum2}
\end{eqnarray}
Remark that $ \kappa_0 = (1/2) \rm{Tr}(\bka) = \KK_0 \cdot \bom / (2 \komega_0)$ and $\kappa_2 = \TT \cdot (\KK_0 \times \bom) / (2 \komega_0)$, with $\TT$ the tangent vector to the filament curve.  
 
Moving further along the terms of Eq. \eqref{eom3tr}, one encounters the twist terms $(v^{ww})^f (\dd_s \phi)^2$ and $(v^{dw})^f \dd^2_s \phi$. We should take into account that the frame $\vec{e_a}$ has a non-vanishing twist $w_{fr}$ for $t\neq 0$, but since $w_{fr}=\OO(\la^3)$, its products and arc length derivatives will be of higher order and not contribute here. Hence, 
\bsub \label{dsphi3} \begin{eqnarray}
  \dd_s \phi &=& \left(1 + \BLA_0^T \XX \right) w_0 + Z_0' t  + (Z_{c}' + w_0 Z_s) \cos  \zeta_0 \nn  \\
  && + (Z_{s}' - w_0 Z_c) \sin  \zeta_0 +  (Z_{cc}' + 2 w_0 Z_{ss}) \cos 2 \zeta_0 \nn \\
&&  + (Z_{ss}' - 2 w_0 Z_{cc}) \sin 2 \zeta_0  + \OO(\la^4) ,\label{dsphitime}   \\
  \dd^2_s \phi &=& w_0' +\left(- Z_c w_0^2+ w_0' Z_s + 2 w_0 Z_s' + Z_c'' \right) \cos  \zeta_0 \nn \\
 && + ( - Z_s  w_0^2 - w_0' Z_c  - 2 w_0 Z_c'  +Z_{s}''  ) \sin  \zeta_0 \nn \\ && + \OO(\la^4),  \label{dphids2} \\
  (\dd_s \phi)^2 &=& w_0^2 + 2 w_0 (Z_c' +w_0 Z_s )\cos \zeta_0 \nn \\ 
&& + 2 w_0 ( Z_s' - w_0 Z_c) \sin \zeta_0 + \OO(\la^4).  \label{dsphisq}
\end{eqnarray} \esub
Identifying $(v^{ww})^f$ with a vector $\mathbf{a}$ in the transverse plane and $(v^{dw})^f$ with $\mathbf{d}$, one finds
 \begin{eqnarray}
  (\dd_s \phi)^2 a^f \mathbf{e}_f  
&=&  - \frac{ w^2_0}{2 \komega_0}  (\beps \mathbf{a} \bom \beps - \mathbf{a} \bom) \BLA_0 \nn \\ && + \frac{ w_0}{\komega_0} (\mathbf{a} \bom \beps + \beps \mathbf{a} \bom) \BLA'_0 + ... + \OO(\la^4), \\
  \dd_s^2 \phi  d^f \mathbf{e}_f 
&=& \frac{1}{2 \omega_0}\left( \beps \mathbf{d} \bom \beps -  \mathbf{d} \bom \right) 2 w_0 \BLA_0' \nn + ... \\ &&  + \frac{1}{2 \omega_0}\left( \mathbf{d} \bom \beps+ \beps \mathbf{d} \bom \right)\left(  \BLA_0'' - w_0^2 \BLA_0 \right) + \OO(\la^4). \nn
\end{eqnarray} 
We introduce the $2 \times 2$ matrices
\begin{eqnarray}
\boldsymbol{\alpha} &= \frac{\mathbf{a} \otimes \bom}{\komega_0} = \alpha_0 \bdel + \alpha_1 \bsig^1 +  \alpha_2 \beps +\alpha_3 \bsig^3, \nn \\
\mathbf{\Delta} &= \frac{\mathbf{d} \otimes \bom}{\komega_0} = \Delta_0 \bdel + \Delta_1 \bsig^1 +  \Delta_2 \beps +\Delta_3 \bsig^3,
\end{eqnarray} 
 which allows to write
\bsub \begin{eqnarray}
&&   (v^{ww})^f \dd_s^2 \phi \ \mathbf{e}_f 
=  (\alpha_0 \bdel + \alpha_2 \beps) w_0^2 \BLA_0 \\
&&\qquad +2 (- \alpha_2 \bdel + \alpha_0 \beps ) w_0 \BLA'_0  + ... + \OO(\la^4), \nn\\
&&   (v^{dw})^f \dd_s^2 \phi \ \mathbf{e}_f =- w_0 \left(  \Delta_0 \bdel + \Delta_2 \beps\right) \BLA'_0 \\
&&\qquad + \left( -\Delta_2 \bdel + \Delta_0 \beps \right) (\BLA_0''- w_0^2 \BLA_0) + ... + \OO(\la^4).\nn 
\end{eqnarray} \label{sum3} \esub

Returning to Eq. \eqref{eom3tr}, we see that the four terms which tune in only at third order require no expansion of the rotation matrices here, whence easily follows for e.g the twist-curvature coupling term:
\bsub \label{sum4} \begin{eqnarray}
w_0^2 (v^{kww})^f_{\hs a} \LA^a \mathbf{e}_f &=& 
w_0^2 \left(\tilde{A}_0 \bdel + \tilde{A}_2 \beps  \right) \BLA_0 \nn \\ && + ...  + \OO(\la^4),
\end{eqnarray}
where we have split the constant matrix $(v^{kww})^F_{\hs A}$ in its Pauli components $\mathbf{A} = \tilde{A}_0 \bdel+ \tilde{A}_1 \bsig^1+ \tilde{A}_2 \beps + \tilde{A}_3 \bsig^3 $. Similarly, one obtains up to third order in $\la$,
\begin{eqnarray}
\dd_s^2 \phi\, (v^{kdw})^f_{\ a} \LA^a \mathbf{e}_f  &=&  w'_0 \left(\tilde{D}_0 \bdel + \tilde{D}_2 \beps  \right) \BLA_0 + ...  
\ \qquad  \\
 (v^{ddk})^f_{\ a} \dd_s^2 \LA^a \mathbf{e}_f  &=& \left(\tilde{E}_0 \bdel + \tilde{E}_2 \beps  \right) \BLA''_0 + ... 
  \\
\dd_s \phi\, (v^{dkw})^f_{\ a} \dd_s \LA^a \mathbf{e}_f  &=&  w_0 \left(\tilde{C}_0 \bdel + \tilde{C}_2 \beps  \right) \BLA'_0 + ...  
\end{eqnarray} \esub 
 
Ultimately, we consider the term $ (v^{kkk})^f_{abc} \LA^a\LA^b \LA^c $ that appears in Eq. \eqref{eom3tr}. In Appendix \ref{app:vkkkav}, we show that this term can be written as
\begin{multline}
(v^{kkk})^f_{abc} \LA^a\LA^b \LA^c \mathbf{e}_f 
\\ = k^2_0 \left( \tilde{B}_0 \bdel +\tilde{B}_2 \beps \right) \BLA_0 +... +  \OO(\la^4). \label{vkkkav}
\end{multline}

Finally, we have expanded all individual terms of the gauge filament EOM \eqref{eom3tr} in third order in twist and curvature. When gathering all contributions from Eqs. \eqref{sum1},\eqref{sum2},\eqref{sum3},\eqref{sum4},\eqref{vkkkav}, we see that a linear system emerges of the shape: 
\begin{eqnarray}
\dot{\XX}_0 = (P_0 \bdel + P_2 \beps) \BLA_0 &+& \XI_0(\s) + \bxi_0(\s) t +\OO( \la^4), \nn \\
\dot \XX_{c} + \omega_0 \XX_{s} &=& \XI_{c}(\s) + \OO(\la^4),\nn\\
\dot \XX_{s} -\omega_0 \XX_{c} &=&  \XI_{s}(\s) + \OO(\la^4), \nn\\
\dot \XX_{cc} + 2\omega_0 \XX_{ss} &=& \XI_{cc}(\s) + \bxi_{cc}(\s)t + \OO(\la^4),\nn\\
\dot \XX_{ss} - 2\omega_0 \XX_{cc} &=& \XI_{ss}(\s) + \bxi_{ss}(\s)t + \OO(\la^4) ,\nn  \\
\dot \XX_{ccc} + 3\omega_0 \XX_{sss} &=& \XI_{ccc}(\s) + \OO(\la^4),\nn\\
\dot \XX_{sss} -3\omega_0 \XX_{ccc} &=&  \XI_{sss}(\s) + \OO(\la^4), \nn\\
\dot \XX_{cccc} + 4\omega_0 \XX_{ssss} &=& \XI_{cccc}(\s) + \OO(\la^4),\nn\\
\dot \XX_{sssss} -4\omega_0 \XX_{cccc} &=&  \XI_{ssss}(\s) + \OO(\la^4). \label{sys3tr}
\end{eqnarray}
where the secular terms $\bxi_{cc}(\s)t $, $\bxi_{ss}(\s)t$ originate from $\QQ_3$ in \eqref{Qdef}. When solving this linear system, it can be checked that these terms $\bxi_{cc}(\s)t $, $\bxi_{ss}(\s)t$ 
will only modulate the amplitude of oscillations. Since the oscillations are by definition filtered out to find the virtual filament, they do not affect its motion. 

Virtual filament dynamics is encoded in the first equation of \eqref{sys3tr}, which can be readily integrated to
\begin{eqnarray}
 \XX_0 = (P_0 \bdel + P_2 \beps) \BLA_0 t + \XI_0(\s) t + \bxi_0(\s) \frac{t^2}{2} + \OO(\la^4). \quad 
\end{eqnarray}
The integration constant was set to zero here, since at the time $t=0$, the virtual filament coincides with the curve to which the parallel frame was attached. By differentiating with respect to time, it follows for the virtual filament motion at the time $t=0$ that 
\begin{eqnarray}
\dot{\XX}_{vir}(\s,0) = (P_0 \bdel + P_2 \beps) \BLA_0 + \XI_0(\s) + \OO(\la^4).  
\end{eqnarray}
Thus, to find the coefficients in the equation of motion for the virtual filament, we only need $\XI_0(\s)$, i.e. the sum of terms without explicit time dependence that we computed in Eqs.  \eqref{sum1},\eqref{sum2},\eqref{sum3},\eqref{sum4} and \eqref{vkkkav}:
\begin{eqnarray}
&&\XI_0 = \frac{k^2_0}{\komega_0} \left(  (\bar{\Theta}_1 P_1 + \bar{\Theta}_3 P_3) \BLA +  (\bar{\Theta}_1 P_3 - \bar{\Theta}_3 P_1) \beps \BLA  \right) \nn \\
&&- \frac{k^2_0}{2 \komega_0} \left(  (\hat{\Theta}_1 P_1 + \hat{\Theta}_3 P_3) \BLA +  (\hat{\Theta}_1 P_3 - \hat{\Theta}_3 P_1) \beps \BLA  \right) \nn \\
&&+ \frac{P_1^2+ P_3^2}{2\komega_0}\left( \eps \BLA_0'' - 4 w_0^2 \beps \BLA_0 + 4 w_0 \BLA_0' +2 w_0' \BLA_0 \right) \nn \\
&&- k^2_0 \left( \kappa_0 \bdel + \kappa_2 \beps \right) \BLA_0 
+ (\alpha_0 \bdel + \alpha_2 \beps) w_0^2 \BLA_0 \nn \\
&& + 2 (- \alpha_2 \bdel + \alpha_0 \beps ) w_0 \BLA'_0 - w_0 \left(  \Delta_0 \bdel + \Delta_2 \beps\right) \BLA'_0 \nn \\
&&+ \left( -\Delta_2 \bdel + \Delta_0 \beps \right) (\BLA_0''- w_0^2 \BLA_0) +w_0^2 \left(\tilde{A}_0 \bdel + \tilde{A}_2 \beps  \right) \BLA_0 \nn \\ 
&& +  w'_0 \left(\tilde{D}_0 \bdel + \tilde{D}_2 \beps  \right) \BLA_0 +
\left(\tilde{E}_0 \bdel + \tilde{E}_2 \beps  \right) \BLA''_0 \nn \\ &&
+  w_0 \left(\tilde{C}_0 \bdel + \tilde{C}_2 \beps  \right) \BLA'_0+ k^2_0 \left( \tilde{B}_0 \bdel +\tilde{B}_2 \beps \right) \BLA_0. \label{XI}
\end{eqnarray}

\subsection{Law of motion for the virtual filament \label{sec:eomvir}}

From Eq. \eqref{XI} above, we find for the motion of the virtual filament an expression of the form
\begin{eqnarray}
 \dot{\XX}_{vir} &=& (\gamma_1 \bdel - \gamma_2 \beps) \BLA \label{filamentmotion3} 
+ (a_1 \bdel - a_2 \beps)w^2 \BLA \\ &&+  (B_1 \bdel - B_2 \beps) k^2 \BLA  +  (c_1 \bdel - c_2 \beps) w \BLA' \nn \\ &&
+  (d_1 \bdel - d_2 \beps) w' \BLA
-  (e_1 \bdel - e_2 \beps) \BLA'' + \OO(\la^5).   \nn 
\end{eqnarray} 
The minus sign preceding $e_1, e_2$ has been chosen as in \cite{Dierckx:2012}, such that positive $e_1$ yields a positive rigidity of the filament curve. The equality of $e_1, e_2$ in the present context to the expressions given in \cite{Dierckx:2012} will be furnished in Appendix \ref{app:buck}. 
One may check from symmetry that no $\OO(\la^4)$ terms will cause net drift of the virtual filament, such that our result is valid up to terms to the fifth order of twist and curvature. 

An important contribution in this work is that our lengthy calculations explicitly provide the coefficients in terms of response functions, which allows to predict three-dimensional scroll wave dynamics once the perturbative responses of the two-dimensional spiral wave are known. By comparison of Eqs. \eqref{XI} and \eqref{filamentmotion3} we obtain, with 
$\tilde{\Theta}_a =  \frac{1}{2}\hat{\Theta}_a - \bar{\Theta}_a = \frac{1}{2}(\Theta_a - \bar{\Theta}_a)$, $a \in \{1,2\}$:
\bsub \label{coeffs3}  \begin{align}
  \gamma_1 & = P_0,\\ 
\gamma_2 &= - P_2, \nn
\end{align}
\begin{align}
 {a_1} &= \tilde{A}_0+\alpha_0 + \Delta_2, \\
{a_2} &= -\tilde{A}_2-\alpha_2 + \Delta_0 + 2 \frac{P_1^2+ P^2_3}{\komega_0}, \nn \\
 {B_1} &=  \tilde B_0 - \frac{\tilde{\Theta}_1 P_1 + \tilde{\Theta}_3 P_3}{2 \komega_0} - \kappa_0, \\
{B_2} &=  -\tilde B_2 + \frac{ \tilde{\Theta}_1 P_3 - \tilde{\Theta}_3 P_3}{2 \komega_0} + \kappa_2, \nn\\
 {c_1} &=  \tilde{C}_0 - 2 \alpha_2 -\Delta_0 + 2 \frac{P_1^2+P_3^2}{\komega_0}, \\
{c_2} &=  -\tilde{C}_2 -2\alpha_0 + \Delta_2, \nn\\
 {d_1} &= \tilde{D}_0 +  \frac{P_1^2+P_3^2}{\komega_0}, \\
{d_2} &= -\tilde{D}_2, \nn \\
 {e_1} &= -\tilde{E}_0 + \Delta_2, \label{e12new} \\
{e_2} &= \tilde{E}_2 + \Delta_0  + \frac{P_1^2 + P_3^2}{2 \komega_0}.  \nn
\end{align} \esub

An alternative expression of the EOM \eqref{filamentmotion3} is found using subsequent arc length differentiation of the filament position, following the style of \cite{Biktashev:1994, Echebarria:2006}. 
From $\BLA=\LA_x \mathbf{e}_x +\LA_y \mathbf{e}_y $ follows that $\beps \BLA=\LA_y \mathbf{e}_x - \LA_x \mathbf{e}_y$, whereas $\vec{T} \times \dd_\s^2 \vec{X} = - \LA_y \vec{e_x} + \LA_x \vec{e_y}$. Therefore, $-\beps \BLA$ corresponds to $ +\vec{T} \times \dd_\s^2 \vec{X}$, which is why the minus sign was included with $\beps$ terms in  Eq. \eqref{filamentmotion3}. We may now use the properties of the relatively parallel frame around the filament, i.e. 
$ \dd_\s \vec{T} = \LA^A \vec{N_A}, 
 \dd_\s \vec{N_A} = - \LA^A \vec{T},\  
\vec{T} \times \vec{N_A} = \eps_A^{\hs B} \vec{N_B},\ $
to find an alternative form of the EOM for virtual scroll wave filaments. Herein, some of the higher order coefficients will recombine to different combinations
\bsub \begin{eqnarray}
B_1 = b_1+e_1,\qquad  B_2 = b_2 + e_2,\\
b_1 =B_1-e_1,  \qquad b _2 = B_2-e_2.  
\end{eqnarray} \esub

At last, we have obtained the law of motion for the virtual filament, which approximates the gauge filament around which the scroll wave rotates:
\begin{eqnarray}
  \dot{\vec{X}} &=& \quad  \left( \gamma_1 + a_1 w^2 + b_1 k^2 + d_1 \dd_s w \right) \dd_s^2 \vec{X}  \nn \\
 &&+ \left( \gamma_2 + a_2 w^2 + b_2 k^2 + d_2 \dd_s w \right) \dd_s \vec{X} \times \dd_s^2 \vec{X} \nn  \\
  &&+ c_1 w \left[\dd_s^3 \vec{X}\right]_\perp + c_2 w \vec{X} \times \dd_s^3 \vec{X} \nn \\
&& - e_1 \left[\dd_s^4 \vec{X} \right]_\perp - e_2 \dd_s \vec{X} \times \dd_s^4 \vec{X} + \OO(\la^5). \label{myribbon} 
\end{eqnarray} 
Together with the evolution equation for the virtual phase, i.e. Eq. \eqref{eomrot2},
\begin{equation}
\dot{\phi} = \komega_0 + a_0 w^2 + b_0 k^2 + d_0 w' +  \OO(\la^4).\label{eomrot2repeated}
\end{equation}
these analytical expressions and their proof form the main result of this manuscript. 


\section{Discussion}

\subsection{Properties of the virtual filament}

\subsubsection{Uniqueness} 
Starting from a given trajectory of the gauge filament, the filtering out of epicycle motion may prove difficult in practice; there is no unique method for doing so. Nevertheless, for a given order in $\la$ of calculation, we may \textit{define} a unique virtual filament by removing the constant terms in Eqs. \eqref{virfilsol1}, which depend on the instantaneous curvature and twist of the gauge filament. Note that the same can be done in higher order, such that every gauge filament by its shape and phase determines a single virtual filament. Conversely, Eqs. \eqref{filamentmotion1vir} allow to reconstruct the gauge filament up to given order in $\la$ for a given virtual filament. With this procedure, we have obtained a mapping between gauge and virtual filaments that is bijective in the regime of small curvature and twist. 

\subsubsection{Proximity of virtual and gauge filament} 
In both cases, the deviation between both filaments is proportional to the local filament curvature, such that their difference is in general bounded; for a straight scroll wave, the virtual and gauge filament are identical in lowest order in curvature and twist. In the trivial case of a straight, untwisted filament, the gauge filament, virtual filament and time-averaged gauge filament all coincide.

\subsubsection{Boundary conditions}
When no-flux conditions are imposed on the RD system, the method of mirror sources guarantees that the gauge filament ends orthogonally to planar medium boundaries; 
twist must also vanish in its end points. For the virtual filament, the same reasoning can be used, such that they also end orthogonally to planar medium boundaries and have vanishing virtual twist in their end points. This remark is essential to the stability analysis of scroll wave filaments which we will perform in section \ref{sec:linstab}. 

\subsubsection{General note on filaments}
Our procedure not only demonstrates the existence and dynamics of the virtual filament, but also broadens the concept of a `filament' in general. To describe the organizing center of a scroll wave, one may use a straight line (see, e.g. \cite{Hakim:2002, Dierckx:2012} ), a tip line that is computed from simulations or experiments (see \cite{Clayton:2005} for various calculations choices), an instantaneous or time-averaged gauge filament \cite{Keener:1986, Biktashev:1994} or the virtual filament presented in this paper. Depending on the problem studied, one may choose a suitable filament definition. Clearly, the advantage of the virtual filament is that its higher order equation of motion becomes independent of the scroll wave's phase. Note, finally, that the definition of the virtual filament (i.e. exhibiting no oscillations depending on rotation phase) is by itself a gauge choice.

\subsubsection{Equal diffusion case}
For equal diffusion systems, the matrix $\bP_{lab}$  is proportional to the identity matrix, such that $P_1=P_3=0$ and $\bP_5 = \bP_6 = \mathbf{0}$. In this case, there is no distinction in lowest order between the virtual and the gauge filament. 

The situation simplifies considerably for systems with equal diffusion, i.e. when $\HP = D_0 \mathbf{I}$. This case arises in the modeling of the BZ oscillating chemical reaction \cite{Zaikin:1970}. From section \ref{sec:eomproof}, we know that in such case, $P^{(m)}_{\hs (n)} := \bra{\WW^{(m)}} \HP \ket{\VV_{(n)} }= D_0 \delta^{(m)}_{(n)}$, whence $d_0=D_0, \gamma_1 = P_0=D_0$. Furthermore, $P_1, P_2, P_3$ will all vanish. Therefore, $\bom,\boldsymbol{\alpha}, \boldsymbol{\Delta}$ and $\bka$ also disappear. Additionally, $
\ket{ \uu^k_A}$ vanishes as the source term is completely removed by application of the projector \eqref{defPI}. Under this circumstance, the coefficients in the phase equation reduce to
\bsub \label{rot2eq} \begin{eqnarray}
a_0 &=& -D_0 \braket{\WO}{ \dd_\theta \VO}, \\
b_0 &=& - D_0 \bra{\WO} \HP \ket{ r \dd_r \uu_0 }, \\
d_0 &=& D_0.
\end{eqnarray} \esub
In the translational law, many higher order contributions drop out for the equal diffusion case and $\tilde{E}_0 = \tilde{E}_2=0$. The simpler set of coefficients in the equal diffusion case is thus
\bsub \begin{align}
\gamma_1  &= D_0, &
\gamma_2  &= 0,  \\
{a_1} &= \tilde{A}_0, &
{a_2} &= -\tilde{A}_2,   \\
b_1 &= {B_1} =  \tilde B_0,  &
b_2 &= {B_2} = -\tilde B_2,   \\
{c_1} &=  \tilde{C}_0, &
{c_2} &= - \tilde{C}_2, \\
{d_1} &=\tilde{D}_0, &
{d_2} &= -\tilde{D}_2, \\
{e_1} &= 0, &
{e_2} &=  0.  
\end{align} \esub

Using the same decomposition as in $\gamma_1 + \rmi K\gamma_2 = \bra{\WW^+} \HP \ket{\VV_+}$, we find, only in the equal diffusion case:
\bsub \label{tr3eq} \begin{eqnarray}
  a_1 + \rmi K  a_2 
&=& D_0  \bra{\WW^+} \rho_+  \ket{ r \dd_r \uu_0}  - D_0 \bra{\WW^+} r^2 \ket{\VV_+} \nn \\ && + 2 D_0 \bra{\bY^+} \rho_+ \ket{\dd_\theta \VV_{(0)}}, \\
  b_1 + \rmi K   b_2 &=&  \frac{D_0}{2} \bra{\WW^+} \rho_+  \ket{ r \dd_r \uu_0}  \nn \\
&& + \frac{D_0}{4} \bra{\WW^+} r^2 \ket{\VV_+},  \label{b12eq}  \\
  c_1 + \rmi K c_2 &=& 
   - D_0\bra{\WW^+} \rho_+ \ket{\VV_{(0)}},  \\
  d_1 + \rmi K d_2 &=& 
 - 2 D_0\bra{\WW^+} \rho_+ \ket{\VV_{(0)}}. 
\end{eqnarray} \esub
We immediately notice that  $(d_1+\rmi Kd_2) = 2 (c_1+\rmi Kc_2)$ here. Moreover, the surviving terms are all related to the extent of the region where the product of GM and RF significantly differs from zero, i.e. related to the effective diameter of the scroll wave's core. 


\subsection{Laws and models of filament motion}

Having rigorously proven the law of filament motion \eqref{myribbon}, we may now compare it to previous models for filament dynamics. A summary is given in table \ref{tab:coeff}.
\begin{table}[b] \centering
 \begin{tabular}{ccccccc}
\hline 
Present & \cite{Keener:1988, Keener:1992, Biktashev:1994} & \cite{Henry:2002} & \cite{Echebarria:2006} &  \cite{Verschelde:2007, Dierckx:2009} &  \cite{Dierckx:2012} \\
\hline
$(\gamma_1, \gamma_2)$* & $(b_2, c_3)$* & $(a_{||}, a_\perp)$* & $(a_1,a_2)$&  $(\gamma_1, - \gamma_2)*$ & $(\gamma_1, \gamma_2)*$\\
$(a_1,a_2)*$ & -&-&-&-&-\\
$(b_1,b_2)*$ &-&- &-&-& $(b_1,b_2)$\\
$(c_1,c_2)*$ &-&- &$(-d_2, d_1)$&-&-\\
$(d_1,d_2)*$ &-&- &-&-&-\\
$(e_1,e_2)*$ &-&- &$(b_1, b_2)$ &-&$(e_1,e_2)*$ \\
\hline
$a_0$* &$a_1$*&$*$&$c$& -&-\\
$b_0$* &-&-&-&-&-\\
$d_0$* & $b_1$* & $*$ & $D$ & -& -\\
\hline
 \end{tabular}
\caption{Reference list of coefficients in the filament equation of motion. The coefficients $a_1,a_2,d_1,d_2,b_0$ have not been described elsewhere. $*$ denotes that an analytical expression in terms of response functions was given in that refence.\label{tab:coeff}}
\end{table}

All terms from the postulated ribbon model \cite{Echebarria:2006} indeed contribute to the motion of filaments in reality. However, it is found that the filament dynamics from Eq. \eqref{myribbon} is considerably richer than assumed in the ribbon model. The explicit twist-curvature coupling coefficients  $a_1, a_2$, $d_1, d_2$ are first presented here. It can be readily seen that both twist ($w^2$) and twist gradients ($\dd_s w$) add to the filament tension $\gamma_1$. Moreover, when filament curvature becomes significant, the coefficients $b_0, b_1, b_2$ describe how respectively the scroll wave's rotation frequency and nominal filament tension are altered. 
The fact that a scroll ring's rotation frequency may change at small radius was already noted in a computational study of low-voltage defibrillation \cite{Morgan:2009}, where resonant external stimuli were used to eradicate scroll wave activity. Therefore, our $b_0$ term may prove useful in the search for low-voltage defibrillation of cardiac tissue. 

\subsection{Linear stability analysis of scroll waves \label{sec:linstab}}

Using the laws of filament motion, one may investigate the stability of a scroll wave given the shape of its filament. In previous studies, e.g., it was found that straight untwisted filaments are only stable if their tension coefficient $\gamma_1$ is positive \cite{Biktashev:1994}. Later, higher order terms were manually added to the equations \cite{Echebarria:2006, Dierckx:2012}, in order to explain the shape taken by a twisted or buckling filaments. 

Having rigorously derived Eq. \eqref{myribbon}, and interpreted it as the motion of a virtual filament, we find that a filament of constant twist $w$ and curvature $k$ is stable with respect to spatial perturbations with wave number $p$ only when
\begin{eqnarray}
 \Gamma_1  = \gamma_1 + a_1 w^2 + 2 b_1 k^2 + c_2 p w + e_1 p^2 > 0. \label{gammaeff}
\end{eqnarray}
This inspires us to call $\Gamma_1$ the effective tension of a scroll wave filament. It is clearly seen here that the filament's curvature and twist affects its stability. Whether twist and curvature render the filament more or less stable will depend on the sign of the coefficients $a_1$, $b_1$ and $c_2$. Even when the combined effect of nominal tension, twist and curvature produces a negative effective filament tension, the rigidity $e_1$ may nevertheless  keep the filament straight in thin domains \cite{Dierckx:2012}.

In the case of low twist, the more general EOM \eqref{myribbon} reduces to 
\begin{eqnarray} \label{myrigid}
  \dot{\vec{X}} &=& \left( \gamma_1 + b_1 k^2  \right) \dd_s^2 \vec{X}   + \left( \gamma_2 + b_2 k^2  \right) \dd_s \vec{X} \times \dd_\s^2 \vec{X}\nn \\&& 
- e_1 \left[\dd_s^4 \vec{X} \right]_\perp  - e_2 \dd_s  \vec{X} \times \dd_s^4 \vec{X}  + \OO(\la^4).  
\end{eqnarray} 
This equation was recently analyzed \cite{Dierckx:2012} to show that scroll wave filaments possess a mechanical rigidity $e_1$, which may prevent them from breaking up in thin media, even when $\gamma_1  < 0$. In that work, we used the stability analysis of a straight scroll wave \cite{Hakim:2002} to show that the rigidity coefficients equal
\begin{eqnarray}
  \tilde{e}_1 + i K \tilde{e}_2 &=& \bra{\WW^+} \bP (\HL-i \omega_0)^{-1} \mathbf{\hat{\pi}}\  \bP \ket{\VV_+} \label{rigidity}
\end{eqnarray}
with $\hat{\pi} = \mathbbm{1} - \ket{\VV_+}\bra{\WW^+}$ different from $\HPI$ defined in Eq. \eqref{defPI}. Alternatively, our present calculations based on virtual filaments state that (see Eq. \eqref{e12new})
\begin{eqnarray}
 e_1 + \rmi K e_2 &=& \bra{\WW^+} \HP (\HL- \rmi \omega_0)^{-1} \mathbf{\hat{\Pi}}\  \HP \ket{\VV_+}  \nn \\ && + (\Delta_2 + \rmi K\Delta_0)   + \rmi K  \frac{P_1^2 + P_3^2}{2 \komega_0}. 
\end{eqnarray}
We show in App. \ref{app:buck} that our expressions \eqref{e12new} are equivalent to Eq. \eqref{rigidity}. Therefore, the low twist limit of our current calculations provides the full technical proof that was promised in \cite{Dierckx:2012}. Moreover, 
the equality of rigidity coefficients using two independent methods of calculation
confirms the validity of the virtual filament concept in this particular case.

\section{Numerical validation}

In this section, we aim to numerically validate the response function expressions predicted by our theoretical framework. Before that, we detail our numerical methods and provide a simple example of a virtual filament. 



\subsection{Numerical methods}

For simplicity, we used the two-variable Barkley kinetics \cite{Barkley:1991}. With $\uu = [u,v]^T$ and $\bF = [f,g]^T$, the reaction functions are given by
\begin{eqnarray}
 f(u,v) &=& \frac{1}{\eps} u(1-u)\left(u- \frac{v+b}{a} \right),
\\
g(u,v) &=& u-v. \nn
\end{eqnarray}
Throughout our simulations, we used the parameter values $a=0.7$, $b=0.01$, $\eps=0.025$, which lie well outside the meander regime. As mentioned above, we took an equal diffusion system with $D_0=1$, i.e. $\HP = D_0 \bP = \mathbf{I}$.


We computed the response functions and overlap integrals for the given parameters of Barkley's model using the publicly available software package \dxspiral\ \cite{dxspiral}. Details on its methods are given in \cite{Biktasheva:2009}. Goldstone modes and response functions in the complex basis were computed on a polar grid of radius $10.0$ with $N_r=300$ elements in the radial direction and $N_t=64$ in the angular direction. We added the formulas \eqref{tr3eq} for the coefficients in the equal diffusion case to the code. In addition to the trivial values $\gamma_1=d_0=1$ and $\gamma_2=e_1=e_2=0$, this leads to (with sign flag $K=1$),
\begin{eqnarray}
(a_1, a_2) &=& (-12.701, -4.056), \nn \\
(b_1, b_2) &=& (-1.210, -2.055),\nn\\
(c_1, c_2) &=& (3.003, 3.029), \label{vals}\\
(d_1,d_2 ) &=& (6.006, 6.057), \nn \\
a_0 &=& -1.505, \qquad
b_0 = 0.283.\nn 
\end{eqnarray}
Forward simulations of scroll waves were performed in a custom-written parallel C++ code using the explicit Euler stepping method for solving the RDE \eqref{RDE1}. 

\subsection{Virtual filament under electroforetic drift}

\begin{figure}[t] \centering
\raisebox{5cm}{a)} \includegraphics[width=0.4\textwidth]{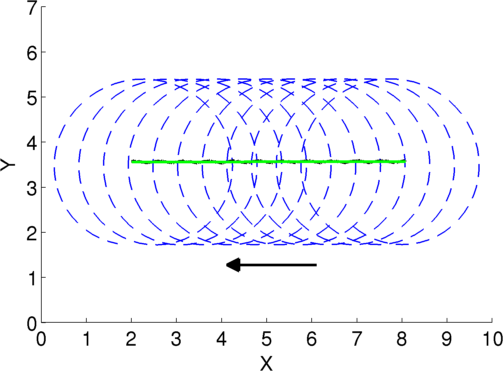}  \\
\raisebox{5cm}{b)} 
\includegraphics[width=0.4\textwidth]{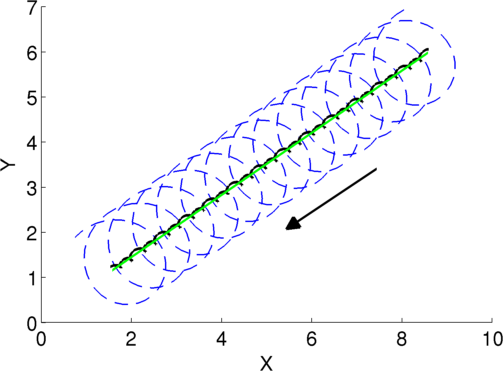} 
   \caption[Drift example]{(Color online) Trajectory of the tip (blue, dashed), gauge filament (black) and virtual filament (green) for a numerical simulation of electroforetic drift with $R=10$. Barkley's model was used ($a=0.07, b=0.01, \eps=0.025, D_u=1$), with equal diffusion $D_v=1$ (a) and single diffusion $D_v=0$ (b).
}\label{fig:efd}
\end{figure}

To make the notion of a virtual filament less abstract, we provide a simple example in Fig. \ref{fig:efd}. Suppose that a small convection term is added to the RDE \eqref{RDE1}, which models the presence of a constant electrical field in chemical reaction-diffusion systems \cite{Agladze:1992, Steinbock:1992}. In cardiac modelling, a convection term with constant amplitude $D_0/R$ applied only to the first variable is typically studied to represents in lowest order the  drift induced by a hypothetical filament curvature $k=1/R$ \cite{Hakim:2000, Henry:2004}:
\begin{equation}
   \dd_t \uu(\vec{x}, t) = D_0 \Delta \bP \uu(\vec{x}, t) +  \bF(\uu(\vec{x}, t) )  + \frac{D_0}{R} \bP \dd_x \uu(\vec{x},t).  \label{RDE_efd}
\end{equation}
Therefore, it is a suitable test bench to easily visualize virtual filament motion. We performed a numerical simulation of Barkley's model \cite{Barkley:1991} with parameters $a=0.07, b=0.01, \eps = 0.025, D_0=1.0$ and $\bP = \mathrm{diag}(1, D_v)$. We measured the tip line by tracing the intersection of the $u=0.5$ and $v=0.5$ isosurfaces using the algorithm described in \cite{Fenton:1998}. For the cases of $D_v =1$, $D_v=0$, tip trajectories are shown in blue in Fig. \ref{fig:efd}. 

From the tip trajectories, the gauge filament was estimated as follows. From a snapshot of a reference simulation of an unperturbed spiral wave, the position of the spiral's exact rotation center was obtained relative to the direction of $\nabla u$, taken at the tip position. This vector was added to the tip positions under electroforetic drift, relative to the current orientation of $\nabla u$. This first approximation to the virtual filament trajectory is given by the black line in Figs. \ref{fig:efd}. When $D_v > 0$, an oscillatory component of motion is  observed with frequency close to $2 \omega_0$, as predicted by Eq. \eqref{filamentmotion1virframe}. 

Since we have picked an example which has constant $\BLA = (- 1/R, 0)$, the secular terms of the motion, which define the virtual filament, can be obtained from linear regression of the $x$ and $y$ components of the gauge filament trajectory. This line is shown in green. In the equal diffusion case, the electroforetic drift term is equivalent to a convection term with velocity $D_0/R$. Hence, the gauge filament will describe a linear trajectory and coincides exactly with the virtual filament. With unequal diffusivities of state variables as in figure \ref{fig:efd}b, a difference between gauge and virtual filaments is observed. 

From this example, we see how two subsequent filtering steps are needed to reconstruct the virtual filament from the more easily accessible tip line. Tracking the virtual filament of non-stationary filaments in numerical simulations or even experiments would be a challenging task. Fortunately, one may in practical cases continue to use tip lines to approximate both the gauge and virtual filaments.  The virtual filament is only meant as a theoretical aid to construct and understand the effective laws of motion for scroll waves.

\subsection{Numerical validation of the drift coefficients}
The expressions for the lowest order coefficients $\gamma_1$, $\gamma_2$, $d_0$ are well known. Recently, the development of numerical methods to compute spiral wave RFs \cite{Biktashev:1996, Hakim:2002, Biktasheva:2009} allowed to evaluate the overlap integrals explicitly. Since then, the basic coefficients $\gamma_1$, $\gamma_2$, $d_0$ have been computed to high accuracy \cite{Hakim:2002, Biktasheva:2009}. The expression for the rotational twist correction $a_0$ was validated in \cite{Henry:2002}; the formula for the rigidity coefficients $e_1, e_2$ has been numerically verified in \cite{Dierckx:2012}. 

To demonstrate and verify our current approach, we have measured some of the coefficients in numerical simulations of scroll wave using the RDE~\eqref{RDE1}. The measured coefficients were compared to the values predicted by our theory. Here, we only report on the equal diffusion case, in which case the overlap integrals \eqref{rot2eq}-\eqref{tr3eq} for the coefficients are not too complicated. 

\subsubsection{Collapse of untwisted scroll rings}

\begin{figure}[b] \centering
\raisebox{5cm}{a)} \includegraphics[width=0.38\textwidth]{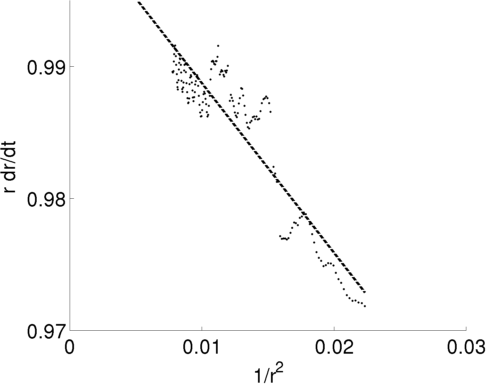} \\ 
\raisebox{5cm}{b)} \includegraphics[width=0.34\textwidth]{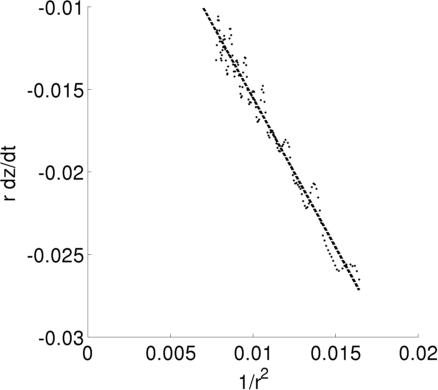}  
   \caption[Measurement of $b_1$, $b_2$]{Measurement of the coefficients $b_1$, $b_2$ in a numerical simulation of  Barkley's model ($a=0.07, b=0.01, \eps=0.025, D_u=D_v = 1$). 
a) Linear regression of the radial velocity of the virtual filament, to yield $b_1 = -1.300$. b) Linear regression of the axial velocity of the virtual filament, delivering $b_2=-1.812$.
}\label{fig:bcoeff}
\end{figure}

\begin{figure*} \centering
\raisebox{4cm}{a)} \includegraphics[width=0.2\textwidth]{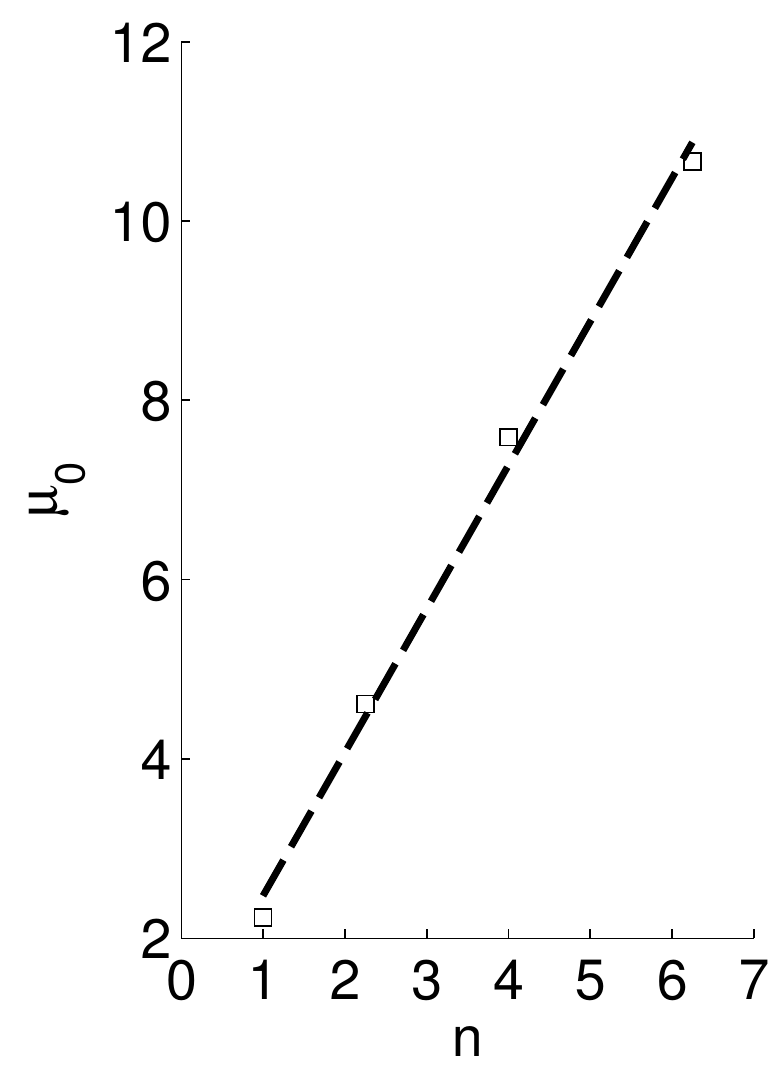} 
\raisebox{4cm}{b)} \includegraphics[width=0.2\textwidth]{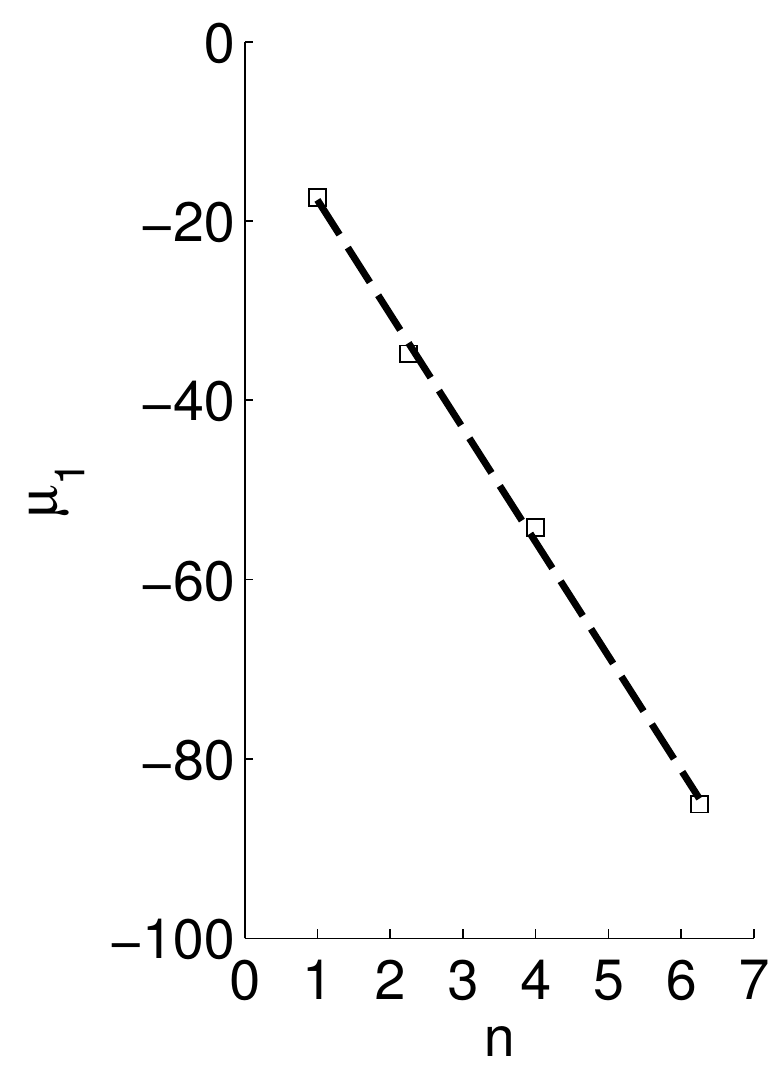}  
\raisebox{4cm}{c)} \includegraphics[width=0.2\textwidth]{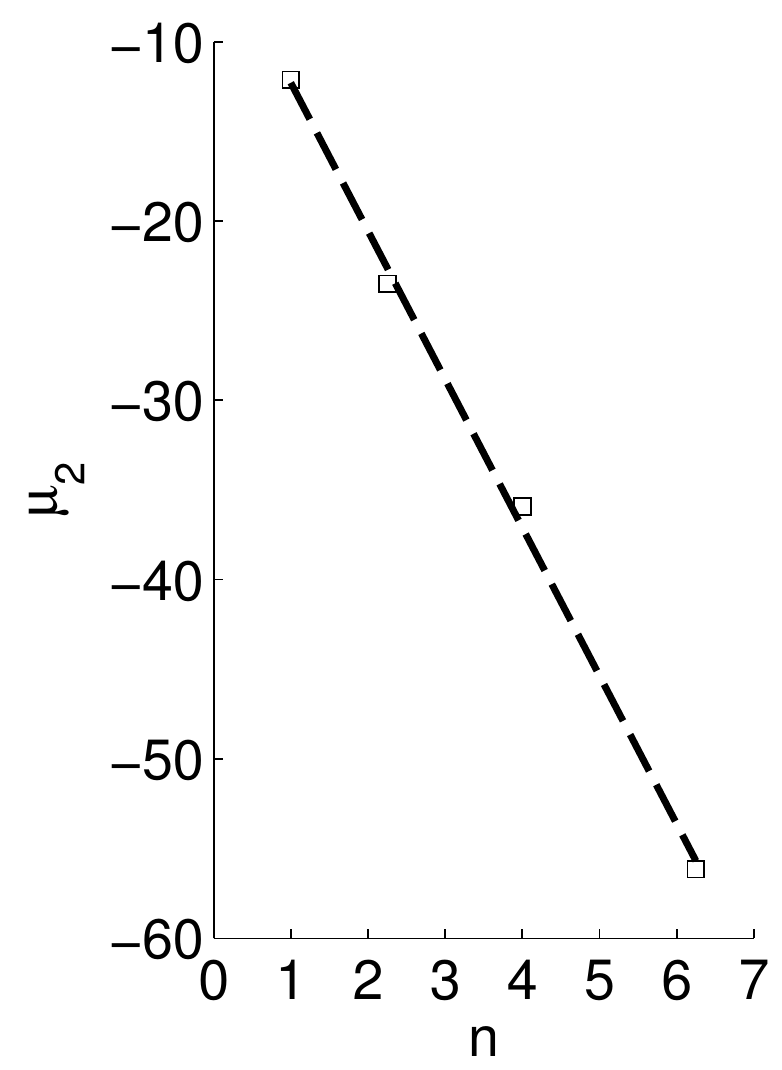}  
   \caption[Measurement of $a_0$, $a_1$, $a_2$]{Drift velocity coefficients $\mu_i$ of twisted scroll rings of radius $R$ was measured as a function of the winding number of the scroll wave $n$ in a numerical simulation of Barkley's model ($a=0.07, b=0.01, \eps=0.025, D_u=D_v = 1$). The slope of the curves $\mu_i(n)$ delivers the values for the coefficients $a_i$, delivering experimental values of 
$a_0 =1.600$ (panel a)
$a_1 = -12.712$ (panel b), $a_2 = -8.264$ (panel c).
}\label{fig:acoeff}
\end{figure*}

Untwisted scroll rings can easily be simulated in two dimensions, since there is no dependence of the angular coordinate along the ring. Hence, we initiated spiral waves in a $(r, z)$-plane 
with $r \in [1, 21]$, $z \in [0, 15]$. Grid resolution was $dx=dz=0.1$ and time step  $0.008$. As initial conditions, we took a numerical spiral wave solution from the Euclidean plane, rotated it over an angle $\phi_0$ around its rotation center and put it on the $(r,z)$ plane. We repeated this procedure for $N_p = 90$ angles $\phi_0$ uniformly distributed over the range $[0, 2\pi]$. The tip position was sampled with intervals $dt_s=0.25$, such that for each time frame set of $N_p$ tip positions was found. From our theory, these points lie approximately on a circle around the gauge filament. Since ellipses may be easily fit by linear regression, we fit an ellipse to the set of $N_p$ tip points, whose central point was then considered as the instantaneous gauge filament position. To find the velocity of the virtual filament, we performed linear regression on overlapping sections of the gauge filament trajectory with a sliding window of width $20\,dt_s$. Finally, linear regression was performed in terms of $1/R^2$ with $R$ the instantaneous scroll ring radius. From our EOM \eqref{myribbon} follows, with $Z$ the axial position of the scroll ring: 
\begin{eqnarray}
- R \dd_t R =  \gamma_1 + b_1 /R^2  + \OO(\la^4),  \label{breg}\\
\ \ R \dd_t Z =  \gamma_2 + b_2 /R^2 + \OO(\la^4). \nn 
\end{eqnarray}
The resulting tip and filament trajectories are presented in Fig. \ref{fig:bcoeff}. Linear regression of Eq. \eqref{breg} yields $(b_1, b_2) = (-1.300, -1.812)$, which lie respectively $7.4\%$ and $11.8\%$ off the predicted values $(-1.210, -2.055)$ which we obtained using response functions and temporal averaging of the filament solution. 

\subsubsection{Drift of twisted scroll rings}

Twisted scroll rings were simulated in a cylindrical coordinates $r, \psi, z$. To avoid small grid elements near the axis $r=0$, the simulation domain had $r \in [ R_{\rm min}, R_{\rm min} + L_r ] $, $\psi\in [ 0, 2\pi/n ]$, $z \in [0, L_z]$. Neumann boundaries were applied in $r$ and $z$ directions, and periodic boundary conditions for $\psi$. 
We initiated twisted scroll rings in a three-dimensional grid by putting an unperturbed spiral wave solution in each $r,z$-plane, and letting it rotate such that it makes one full turn between $\psi=0$ and $\psi= 2 \pi / n$. With periodic boundary conditions at the $\psi$ edges, a twisted scroll ring is simulated that makes $n$ full turns. In the simulations series, we also allowed non-integral values for $n$, such that the imposed twist could be varied continuously and independently from the scroll ring radius $R$. This way, we found
\begin{eqnarray}
 w = n / R, 
\qquad
k = 1/R. \label{wkring}
\end{eqnarray}
The virtual filament of the twisted scroll rings is from Eq. \eqref{myribbon} expected to obey
\begin{eqnarray}
\ \  \dd_t \phi &=& \omega_0 + \mu_0 (n) / R^2 + \OO(\la^4), \nn \\ 
- R \dd_t R &=& \gamma_1 + \mu_1(n) / R^2 + \OO(\la^4),\\ 
\ \  R \dd_t Z &=& \gamma_2 + \mu_2(n) / R^2 + \OO(\la^4), \nn 
\end{eqnarray}
where $\mu_j(n) = a_j n^2 +b_j, (j \in \{1,2,3 \})$.
We chose to work at large radius $R$, such that the circular filament remains stable with respect to the sproing instability \cite{Henze:1990}. While the circular filament had at each time instance $r$ and $z$ constant, the tip line approximately describes a circle in the $r,z$ plane as all values of $\psi$ are traversed. Here too, we drew the best fitting ellipse through this set to find the instantaneous gauge filament position. A linear regression on its position in the $(r,z)$-plane with sliding window in time then yielded $\mu_0, \mu_1$, $\mu_2$ for different values of $n$.

The dependencies $\mu_i(n)$ found in our simulations are depicted in Fig. \ref{fig:acoeff}. Linear regression in terms of $n^2$ then produced $a_0 = 1.600$ close to the predicted value of 1.505. (This coefficient has been validated to higher accuracy using simulations on a disk in \cite{Henry:2002}.) 
For the novel drift coefficients, we find in the numerical experiment that $(a_1, a_2) = (-12.713, -8.264)$. The value of $a_1$ lies very close to the predicted value from Eq. \eqref{vals}, confirming our response function calculations. The measured value for $a_2$, however, is roughly twice as big as the predicted value. A possible explanation is that, due to the absence of the $\gamma_2$ term, the drift velocity in this direction is very small: in one time unit, the scroll wave drifts over a distance of the order $(a_2+b_2)/R^3 \approx 0.006$, which is well below the spatial discretization step of $0.1$. To validate $a_2$ in this setting will thus require future simulations at a much finer resolution. 

\subsubsection{Sproing instability for twisted scroll waves}

The deformation of a twisted scroll wave into a helical shape is well known, and was previously discussed using the ribbon model \cite{Echebarria:2006}. Our present results allow a quantitative analysis of the phenomenon. 

From condition \eqref{gammaeff}, the virtual filament of a straight scroll wave with constant twist will be stable only when $\gamma_1 + a_1 w^2 + c_2 p w + e_1 p^2 > 0$. As usual, we take the filament along the Z-axis of the domain of height $L$ with periodic boundary conditions, whence $w=2\pi/L$ and $p = m p_0 = m 2 \pi/L$, $m \in \mathbbm{Z}$. The ground mode $(m=1)$ will therefore start growing when 
\begin{equation}
 |w| > w_c =  \frac{2\pi}{L_c} = \sqrt{\frac{\gamma_1}{c_2-a_1-e_1}}. \label{wc}
\end{equation}
Assuming that the bifurcation is supercritical, the helical filament will restabilize at a radius $R$ where $\dd_t \vec{X} \cdot \vec{e}_r = 0$. Since the scroll wave's phase difference over the height $L$ is fixed to $2 \pi$, its twist with respect to a parallel frame \cite{Bishop:1975} is found to be $w = p_0/(1+p_0^2R^2)$. Putting this in the stationarity condition $\dd_t \vec{X} \cdot \vec{e}_r = 0$ with $L\approx L_c$, this leads to
\begin{multline}
R^2 = \frac{ \sqrt{\gamma_1 (c_2 -a_1 - e_1)} }{w_c^2 \pi (2 c_2 - 2 a_1 +b_1 - e_1) } (L_c-L) \\ 
 + \OO \left(  (L_c-L)^2 \right).
\end{multline} 
Only when the denominator $2 c_2 - 2 a_1 +b_1 - e_1$ is positive, the bifurcation will be supercritical and lead to a finite helical radius near the instability threshold. 
 \begin{figure}[h] \centering
\raisebox{3.5cm}{a)} \includegraphics[width=0.28\textwidth]{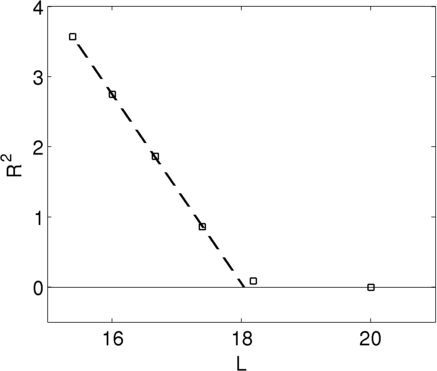} \\
\raisebox{3.5cm}{b)} \includegraphics[width=0.28\textwidth]{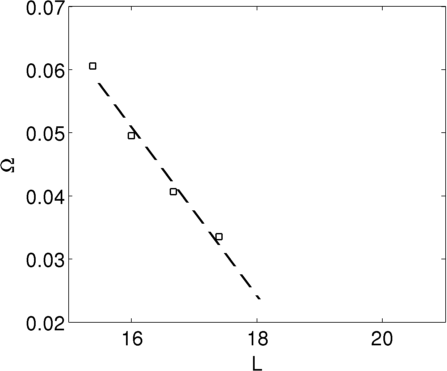}  
   \caption[Sproing simulation]{Sproing instability in a numerical simulation of Barkley's model ($a=0.07, b=0.01, \eps=0.025, D_u=D_v = 1$). a) Bifurcation plot of squared helical radius $R$ versus domain height. b) Precession frequency of the helical state.
}\label{fig:sproing}
\end{figure}

The motion along the binormal will yield a precession frequency around the central axis of the helix. The precession rate follows from projecting the filament velocity onto the circumferential unit vector $\vec{e}_\theta$ from cylindrical coordinates $(r,\theta,z)$:
\begin{equation} \label{omegasproing}
  \Omega = \frac{\dd_t \vec{X} \cdot \vec{e}_{\theta} }{R} = w_c^4(c_1+a_2)
\end{equation}
We performed simulations with Barkley model parameters as detailed above in a rectangular box of size $32\times 32 \times L$ ($dx=dy=0.1$) with $L$ varying between $15.4$ and $20.0$ while keeping $N_z=125$ constant. When $L<L_c$, the projection of the tip line onto the XY-plane remains circular, but its radius will oscillate at the scroll's rotation frequency $\omega$, with amplitude $R$ equal to the helical radius of the restabilized state. A plot of $R^2$ versus $L$ is presented in Fig. \ref{fig:sproing}a, yielding $w_c = 0.346$, $L_c = 18.2$, which deviates only $7.7\%$ from to the predicted value of $w_c = 0.322$ from Eq. \eqref{wc}.  
The precession frequency was measured to be 0.021, which is of the same order as expected value of 0.010 from Eq. \eqref{omegasproing}.

\section{Conclusions}

We have shown how the leading order dynamics of scroll wave filaments can be quantitatively derived from properties of the unperturbed spiral wave. The classical gauge filament is demonstrated to perform a complex motion in which twist-curvature coupling depends on the phase of the scroll wave rotation. Therefore, we advertise a Copernican view on scroll wave dynamics: the gauge filament, which is the instantaneous rotation center for both tip line and scroll wave, revolves around
a previously unobserved companion, the virtual filament.  Using the simplifying concept of a virtual filament, the time-averaged rotation and translation of scroll waves was obtained and analyzed. Virtual filaments lie at all times close to the classical gauge filament, but obey simpler, time-independent laws of motion. 

The dynamics of scroll waves is found to be purely geometrical and largely independent of the reaction kinetics, which only enter the description through the constant coefficients
in the law of motion. This finding offers perspectives to characterize a reaction-diffusion system with isotropic diffusion by its set of 15 coefficients, which capture the effective behavior of scroll wave filaments. Our results put firm mathematical ground under previous models of higher order filament dynamics and can be used to further study and model the rich dynamics of scroll waves. 

 \section*{Acknowledgements}

The authors thank Vadim Biktashev and Irina Biktasheva for helpful discussions, and Alexander Panfilov for advice on writing the manuscript. H.D. is supported by the FWO-Flanders. 

\bibliographystyle{unsrt}

\appendix


\section{Pauli matrices \label{app:pauli}}

In our calculations, we use a real-valued basis of Pauli spin matrices \cite{Bransden:1989} as a basis for the vector space of $2\times 2$ matrices: 
\begin{align} \label{Pauli}
 \bdel &= \left(
              \begin{array}{cc}
                1 & 0 \\
                0 & 1 \\
              \end{array}
            \right), &
 \beps &= \left(
              \begin{array}{cc}
                0 & - 1 \\
                1 & 0 \\
              \end{array}
            \right), \nn \\
 \bsig^1 &= \left(
              \begin{array}{cc}
                0 & 1 \\
                1 & 0 \\
              \end{array}
            \right), 
&  \bsig^3 &= \left(
              \begin{array}{cc}
                1 & 0 \\
                0 & -1 \\
              \end{array}
            \right).
\end{align}
Useful product relations are given by
\begin{align}\label{ess}
    \bsig^1 \beps &= -\bsig_3, & \bsig^3 \beps &= \bsig^1, &\bsig^1 \bsig^3 &= -\beps, \nn  \\
    \beps \bsig^1 &= \bsig^3, &\beps \bsig^3 &= -\bsig^1, & \bsig^3 \bsig^1 &= \beps. 
\end{align}
In our calculations, we repeatedly decompose real-valued $2 \times 2$ matrices as
\begin{eqnarray}
 \mathbf{P} = P_0 \bsig^0 + P_1 \bsig^1 + P_2 \beps + P_3 \bsig^3 \label{paulidec}
\end{eqnarray}
The following combinations often occur, and are therefore assigned their own symbol:
\begin{eqnarray}
\bP_5 = (P_1 \bsig^3 - P_3 \bsig^1), \qquad 
\bP_6 = (P_1 \bsig^1 + P_3 \bsig^3). \label{defP56app}
\end{eqnarray}
Note that they are symmetric 
and satisfy
\begin{eqnarray}
\bP_5^2 &=& \bP_6^2 = (P_1^2+P_3^2) \bdel, \nn \\
\bP_6 \bP_5  &=& -\bP_5 \bP_6 =  -(P_1^2 + P_3^2)\beps. \label{P56ortho}
\end{eqnarray} 
The last property implies that, for a given column matrix $\mathbf{v}$, $\bP_6 \mathbf{v}$ and $\bP_5 \mathbf{v}$ are orthogonal. Furthermore, the matrices $\bP_5, \bP_6$ anticommute with the generator of rotations $\beps$ and obey
\begin{align}
 \beps \bP_5 &= - \bP_6,   &[\beps, \bP_5] &= -2 \bP_6,  &\eps \bP_5 \eps &= \bP_5, \nn \\
 \beps \bP_6 &= \bP_5, &[\beps, \bP_6] &= 2 \bP_5,  &\eps \bP_6 \eps &= \bP_6. \label{propP56}
\end{align}

\section{Complex basis \label{app:complex}}
It is convenient to extend the complex-valued basis from section \ref{sec:RF} to represent quantities that are not Goldstone modes or response functions.
One of them is the fully antisymmetric tensor $\beps$, which has $\eps_{x'y'}=-\eps_{y'x'}=1$:
\bsub 
\label{defddp} \begin{eqnarray}
  \dd_\pm &=& - \frac{1}{2} (\dd_{x'} \mp \rmi K \dd_{y'}), \\
 \rho_\pm &=& - \frac{1}{2} (x' \mp \rmi K y') = \frac{1}{2} \rho^\mp,\\
 \delta_{(m)}^{(n)} &=& \mathbf{1}_{(m)}^{(n)}, \\
 \eps_{(m)}^{\hs\ (n)} &=& - \rmi K (\bsig^3)_{(m)}^{\hs\ (n)}.
\end{eqnarray}
\esub
In this result, the Pauli spin matrices $\bsig^1, \bsig^3$ appear (see e.g. \cite{Bransden:1989} and \ref{app:pauli} ).
While the upper or lower placement of Cartesian indices does not matter because the Euclidean plane has trivial metric $g_{AB} = \delta_{AB}$, indices in the complex basis should be raised and lowered using the metric tensor in the complex basis:
\begin{eqnarray}
\left(  g^{(mn)} \right) &=& 2(\bsig^1), \nn  \\
\left(g_{(mn)}\right)  &=& (1/2)(\bsig^1). \label{complexmetric}
\end{eqnarray}

\section{Reference frame for moving filaments \label{app:frame}}

In section \ref{sec:fermi} of the main text, we have outlined how a minimally twisted Fermi frame $(\vec{T_0}(\s), \vec{N_1}(\s), \vec{N_2}(\s) )$ can be attached to a stationary curve $X_0^i(\s,0)$ with arc length $\s$. We will denote its curvature components here as $\LA^a_0 = \vec{N^a} \cdot \dd_\s^2 \vec{X_0}$; its frame twist is zero by construction. Since filaments evolve in time, one also needs to study how a curve $X^i(\s, t)$ with prescribed movement with respect to the stationary Fermi frame changes its twist and curvature. The frame attached to the moving curve will be denoted $(\vec{T}(\s,t), \vec{e_1}(\s,t), \vec{e_2}(\s,t) )$. To fix the parameterization of the curve $X^i$, we impose that points move in the plane transverse to the reference curve $X_0^i$:
\begin{eqnarray}
  \vec{T_0} \cdot \dd_t \vec{X} = 0. 
\end{eqnarray}
As a consequence, the arc length $s$ on $X^i$ is in general different from $\s$. The relative position of the moving curve is now fully determined by functions $X^a(\s, t)$:     
\begin{eqnarray}
  X^i(\s, t) &= X^i_0(\s) + X^a(\s, t) \vec{N_a}(\s) \label{defXA}
\end{eqnarray}
The lower case indices $a,b,...$ refer to a non-rotating frame.

To preserve generality, we do not impose that both curves coincide at the time $t=0$. Rather, we demand that the initial deviation $X^a(\s, 0)$ and the velocity $\dd_t X^a(\s,t)$ are bounded, say of order $\la$.

In the derivation of the filament EOM, we need to know the curvature components $\LA^a(\s,t)$ and frame twist $\wfr(\s,t)$ of the moving frame $(\vec{T}(\s,t), \vec{e_1}(\s,t), \vec{e_2}(\s,t) )$ for small times $t$. 
In the notation $\dd_\s f = f'$, we find from differentiating Eq. \eqref{defXA} that
\begin{eqnarray}
  \vec{X}'  &=& (1- \LA^a_0 X_a ) \vec{T_0} + X'^a \vec{N_a}. \nn\\ 
  \dd_\s s &=& ||\vec{X}' || = 1 - X^a \LA_a^0 + \OO(\la^4), \nn \\
 \vec{T} &=&  \vec{T_0} + {X'}^a \vec{N_a} + \OO(\la^4), \nn \\
\dd_s \vec{T}(\s, t) &=& \LA^a_0 \vec{N_a} - \vec{T_0}  {X'}^a \LA_a^0 \nn \\ && + \vec{N_b} \left( \LA^b_0 X^c \LA_c^0 + {X''}^b \right). \label{dsTapp} 
\end{eqnarray} 

To investigate explicit time evolution of the transverse basis vectors $\vec{e_a}$, we propose the expansions
\begin{eqnarray}
 \vec{e_a}(\s,t) &= g_a(\s,t) \vec{T_0}(\s) + (\delta_a^b + F_a^{\hs b}(\s,t)) \vec{N_b}(\s), \\
F_{ab}(\s,t) &= \sum_{n=1}^4 (F_n(\s,t) \eps_{ab} + h^{(n)}_{ab}(\s,t)) + \OO(\la^5), \nn
\end{eqnarray} 
with yet unknown coefficients satisfying  $F_n=\OO(\la^n)$, $h^{(n)}_{ab}=\OO(\la^n)$, $h^{(n)}_{ab} = h^{(n)}_{ba}$. We shall now impose orthonormality, i.e. $\vec{e_a} \cdot \vec{T}=0$ and $\vec{e_a} \cdot \vec{e_b} = \delta_{ab}$. The first condition yields 
\begin{eqnarray} 
g_a = -(X'_a + F_a^{\hs b} X'_b)(1+\LA^c_0 X_c) + \OO(\la^5), \end{eqnarray}
while the second requires, with $F = \sum_{n=1}^4  F_n$:
\begin{multline}
 X'_a X'_b + 2 h_{ab} + F^2 \delta_{ab}  + F \left(  h_a^{\hs b} \eps_{bc} + \eps_a^{\hs b} h_{bc} \right) \\ + h_a^{\hs b} h_{bc} = \OO(\la^5), 
\end{multline}
Solving for successive orders of $\la$, we find $h^{(1)}_{ab} =0$, $h^{(2)}_{ab} = \frac{F_1^2}{2}\delta_{ab}$, $h^{(3)}_{ab} =  - F_1F_2 \delta_{ab}$ and $h^{(4)}_{ab} =  - \frac{1}{2} X'_a X'_b - \frac{1}{8} F_1^4 \delta_{ab}$. Apparently, the function $F$ is a degree of freedom that states how the vectors $\vec{e_a}$ are rotated in time. Prescribing minimal rotation of reference vectors, we may thus impose that $F_{ab} = 0 + \OO(\la^4)$, such that for small times $t$ it holds that
\begin{multline}
 \vec{e_a}(\s, t) = - X'_a(1+\LA^c_0 X_c) \vec{T_0} \\ + \left(\delta_a^b - \frac{1}{2} X'_a X'^b\right) \vec{N_b} + \OO(\la^5).
\end{multline}
The curvature components and frame twist of the moving curve are therefore given by
\bsub 
\begin{eqnarray}
 \LA^a(\s,t) &=&  \vec{e^a}(\s,t) \cdot \dd_s^2 \vec{X}(\s,t) \nn \\ &=& \LA^a_0 + \LA^a_0 X_b \LA^b_0 + X''^a +\OO(\la^4), \label{laexp3app}\\
\wfr(\s,t) &=& \frac{\eps^{ab}}{2}\dd_s \vec{e_a}(\s,t) \cdot \vec{e_b}(\s,t) \nn \\ &=& \eps_a^{\hs b} X'_b \LA^a_0 + \OO(\la^4). 
\end{eqnarray} \esub

\section{Net drift components of the cubic curvature contribution \label{app:vkkkav}}

Finding the net filament motion due to the cubic curvature term
$B^f_{abc} \LA^a \LA^b \LA^c \mathbf{e}_f$ requires some intermediate steps which are presented here. 
To facilitate reading, we will write $(v^{kkk})^f_{abc}$ as $ B^f_{abc}$ here. First, let us symmetrize the lower indices: the relation
\begin{multline}
 B^f_{abc} \LA^a \LA^b \LA^c = \frac{1}{6}(B^f_{abc}+B^f_{acb}+B^f_{bca} \\ + B^f_{bac}+B^f_{cab}+B^f_{cba}) \LA^a \LA^b \LA^c 
\end{multline}
leads to 
$ B^f_{abc} \LA^a \LA^b \LA^c =  \tilde{B}^f_{abc} \LA^a \LA^b \LA^c$, with 
\begin{eqnarray}
\tilde{B}^f_{abc} = \frac{1}{6}(B^f_{abc}+B^f_{acb}+B^f_{bca}+B^f_{bac}+B^f_{cab}+B^f_{cba}) \quad 
\end{eqnarray}
fully symmetric in its lower indices. We may now expand
\begin{eqnarray}
&&\tilde{B}^f_{abc} \LA^a\LA^b \LA^c 
= R^f_{\hs F} R_a^{\hs A} R_b^{\hs B}  R_c^{\hs C}  \tilde{B}^F_{ABC}  \LA^a  \LA^b \LA^c \nn \\
  &&= (\delta^f_F \cos \zeta_0 + \eps^f_{\hs F} \sin \zeta_0)  (\delta_a^A \cos \zeta_0 + \eps_a^{\hs A} \sin \zeta_0) \nn \\ 
&& \qquad .(\delta_b^B \cos \zeta_0 + \eps_b^{\hs B} \sin \zeta_0)(\delta_c^C \cos \zeta_0 + \eps_c^{\hs C} \sin \zeta_0)
\nn \\ &&\qquad . \tilde{B}^F_{ABC}  \LA^a_0  \LA^b_0 \LA^c_0 + \OO(\la^4). 
\end{eqnarray}
Next, we rely on the averaging properties
\begin{align}
 \cos^4 \zeta_0  &= \frac{3}{8} + ... , &
 \sin^4 \zeta_0 &= \frac{3}{8} + ...,  \nn\\
 \cos \zeta_0 \sin^3 \zeta_0&= 0+ ..., &
 \sin \zeta_0 \cos^3 \zeta_0&= 0 + ..., \nn \\
 \cos^2 \zeta_0 \sin^2 \zeta_0 &= \frac{1}{8}  + ... \   &&   
\end{align}
to find out that (with $\otimes$ the direct product of tensors)
\begin{eqnarray}
&&  (\cos \zeta_0 \bdel + \sin \zeta_0 \beps) \otimes(\cos \zeta_0 \bdel + \sin \zeta_0 \beps) \\ 
&&\otimes (\cos \zeta_0 \bdel + \sin \zeta_0 \beps) \otimes
(\cos \zeta_0 \bdel + \sin \zeta_0 \beps) \nn \\
&=& \frac{3}{8}(\bdel \otimes \bdel \otimes  \bdel \otimes  \bdel + \beps \otimes  \beps \otimes  \beps \otimes  \beps)  \nn \\
&& +\frac{1}{8} [ \bdel \otimes (\bdel \otimes \beps \otimes  \beps+\beps \otimes  \bdel \otimes  \beps +\beps \otimes  \beps \otimes \bdel) ] \nn \\
&& +\frac{1}{8}[ \beps \otimes (\beps \otimes  \bdel \otimes  \bdel+\bdel \otimes  \beps \otimes  \bdel+\bdel \otimes  \bdel \otimes  \beps) ]  + ... \nn
\end{eqnarray}
This expression will be contracted with the symmetrized tensor $\tilde{B}^F_{ABC}$, after which the terms on the last line can be grouped due to permutation symmetry of the last three indices. Hence one arrives at
\begin{eqnarray}
 &&  B^f_{abc} \LA^a\LA^b \LA^c = \tilde{B}^f_{abc} \LA^a\LA^b \LA^c \nn\\
&=&\quad \frac{3}{8}\left( \delta^f_F \delta_a^A \delta_b^B \delta_c^C +  \eps^f_{\hs F} \eps_a^{\hs A} \eps_b^{\hs B} \eps_c^{\hs C}  \right)\tilde{B}^F_{ABC} \LA^a_0  \LA^b_0 \LA^c_0  \nn \\
&& \quad +\frac{3}{8}\left(\delta^f_F \delta_a^A \eps_b^{\hs B} \eps_c^{\hs C} +  \eps^f_{\hs F} \eps_a^{\hs A} \delta_b^B \delta_c^C \right)\tilde{B}^F_{ABC} \LA^a_0  \LA^b_0 \LA^c_0 \nn \\ && \quad +... +  \OO(\la^4). 
\end{eqnarray}
This relation can be further simplified by the property
$\delta_b^B \delta_c^C + \eps_b^{\hs B} \eps_c^{\hs C} = \delta_{bc} \delta^{BC} + \eps_{bc} \eps^{BC}$,
yielding
\begin{multline}
 B^f_{abc} \LA^a\LA^b \LA^c = \frac{3}{8}\left(\delta^f_F \delta_a^A +   \eps^f_{\hs F} \eps_a^{\hs A} \right)\\ .( \tilde{B}^F_{ABC}  \delta^{BC} ) k^2_0 \LA^a_0 +... +  \OO(\la^4). 
\end{multline}
After defining $\frac{3}{4}\tilde{B}^F_{ABC} \delta^{BC} = (\tilde{B})^F_A$, and decomposing the resulting matrix in Pauli components, one may finally write in matrix notation 
\begin{eqnarray}
B^f_{abc} \LA^a\LA^b \LA^c \ \mathbf{e}_f &= 
 \frac{1}{2}  k^2_0 \left( \tilde{\BB} + \beps \tilde{\BB} \beps^T \right) \BLA_0 +... +  \OO(\la^4) \nn \\
=& k^2_0 \left( \tilde{B}_0 \bdel +\tilde{B}_2 \beps \right) \BLA_0 +... +  \OO(\la^4). \qquad
\end{eqnarray}
This is the result that we quoted in Eq. \eqref{vkkkav}. 

In the case of equal diffusion of state variables ($\bP = D_0 \mathbf{I}$), the expression for $ \tilde B_0,  \tilde B_2$ in terms of RFs is easily found:
\begin{eqnarray}
 \tilde B^F_A &=& \frac{3}{4} \frac{D_0}{3} \delta^{BC} ( \bra{\bY^F} \rho^A\rho^B \ket{\dd_C \uu_0} \\ &&
+ \bra{\bY^F} \rho^B\rho^C \ket{\dd_A \uu_0} + \bra{\bY^F} \rho^C\rho^A \ket{\dd_B \uu_0} ) \nn \\
 &=& \frac{D_0}{4} \left(2  \bra{\bY^F} \rho^A \ket{r \dd_r \uu_0} + \bra{\bY^F} r^2 \ket{\dd_A \uu_0} \right). \nn
\end{eqnarray}
For $b_1+\rmi Kb_2$ in the equal diffusion case, one may subsequently verify that
\begin{eqnarray}
&&  b_1 + \rmi K b_2 = 3 \frac{D_0}{3} \left[  \bra{\WW^+} \rho_+  \ket{(\rho_+ \VV_-) + (\rho_- \VV_+)} \right. \nn \\ 
&&\qquad \qquad \qquad
+\left.  \bra{\WW^+} \rho_+ \rho_- \ket{\VV_+}  \right] \\
&&= \frac{D_0}{4} \left( 2  \bra{\WW^+} \rho_+  \ket{r \dd_r \uu_0} + \bra{\WW^+} r^2 \ket{\VV_+} \right), \nn
\end{eqnarray}
confirming expression \eqref{b12eq}.

\section{Filament rigidity coefficients \label{app:buck}}

In previous work \cite{Dierckx:2012}, we showed that filaments exhibit physical rigidity given by
\begin{eqnarray}
  \tilde{e}_1 + \rmi K \tilde{e}_2 &=& \bra{\WW^+} \HP (\HL-\rmi \omega_0)^{-1} \mathbf{\hat{\pi}}\  \HP \ket{\VV_+}, \label{apprigidity}
\end{eqnarray}
whereas the virtual filament approach in the current manuscript predicts by Eq. \eqref{e12new} that the same coefficients should be given by
\begin{multline}
  e_1 + \rmi K e_2 = -(\tilde{E}_0 - \rmi K \tilde{E}_2) \\ +  (\Delta_2 + \rmi K\Delta_0) + \rmi K  \frac{P_1^2 + P_3^2}{2 \komega_0}
\end{multline}
with $- (\tilde{E}_0 - \rmi K \tilde{E}_2) = - (v^{ddk})^+_+ = \bra{\WW^+} \HP (\HL- \rmi \omega_0)^{-1} \mathbf{\hat{\Pi}}\  \HP \ket{\VV_+}$. Since the full projector $\HPI = \mathbf{\hat{\pi}} - \ket{\VV_-} \bra{\WW^-} - \ket{\VV_{(0)}} \bra{\WW^{(0)}}$ it remains to be shown that the extra terms in $\HPI$ compensate for the additional terms in Eq. \eqref{apprigidity}. Indeed, one finds
 \begin{eqnarray}
 &&\bra{\WW^+} \HP (\HL-\rmi \omega_0)^{-1}\ket{\VV_0} \bra{\WW^0} \HP \ket{\VV_+} \\ 
&=&\frac{\rmi}{\omega_0} \bra{\WW^+} \HP \ket{\VV_0} \bra{\WW^0} \HP \ket{\VV_+} \nn \\
&=& \frac{\rmi}{\omega_0}(\bom \otimes \mathbf{d})^+_+
= \rmi K(\Delta_0 - \rmi K \Delta_2),  \nn \\
&& \bra{\WW^+} \HP (\HL- \rmi \omega_0)^{-1}\ket{\VV_-} \bra{\WW^-} \HP \ket{\VV_+} \\ 
&=&\frac{\rmi}{2\omega_0} \bra{\WW^+} \HP \ket{\VV_-} \bra{\WW^-} \HP \ket{\VV_+} \nn
= \frac{\rmi K}{2 \komega_0} (P_1^2+P_3^2). 
\end{eqnarray} 
which concludes the proof. In summary, three equivalent ways can be used to prove that Eq. \eqref{apprigidity} are the filament's rigidity coefficients: linear perturbation of a straight scroll wave \cite{Dierckx:2012}, the Feynman-Hellman method \cite{Dierckx:2012} and the EOM \eqref{myribbon} for the virtual filament. 

\section{Explicit expressions for the coefficients in the filament EOM}

The promise of this paper was to derive EOM \eqref{myribbon} and present the coefficients in terms of response functions. This way, they can be explicitly computed numerically for given reaction kinetics $\bF(\uu)$ and diffusivity ratios $\HP = D_0 \bP$. For case of equal diffusion, the result considerably simplifies to Eq. \eqref{rot2eq} for the phase evolution and Eq. \eqref{tr3eq} for translation. 

In the general case of unequal diffusion, expressions \eqref{rot2uneq} for the phase evolution lead to
\bsub 
\begin{eqnarray}
a_0 &=& - \bra{\WO} \HP \ket{ \dd_\theta \VO},  
\end{eqnarray}
\begin{eqnarray}
b_0 &=& g^{(np)} \left( 
- \bra{\WO} \HP \ket{\VV_{(n)}}  \braket{\WO}{\dd_\theta \uu^k_{(p)}} 
  \right. \\ 
&&- \bra{\WO} (\HP\delta^{(m)}_{(n)} - P^{(m)}_{\hs\ (n)}) \ket{\dd_{(m)} \uu^k _{(p)} } \nn \\
&& -  \bra{\WO} \rho_{(n)} \HP \ket{\dd_{(p)} \uu_0} + \frac{1}{2} \braket{\WO}{\bF''  \uu^k_{(n)} \uu^k_{(p)}}  \nn \\
&& \left. - \frac{1}{\komega_0}  \bra{\WO} \HP \ket{\VV_{(n)}} \bra{\WO} \HP \ket{\VV_{(p)}} \right), \nn \\
d_0 &=& \bra{\WO} \HP \ket{\VO}. 
\end{eqnarray} 
 \esub
As before, the summation convention for repeated indices $(m),(n),(p)$ runs over $\{-1,1\}$. Further recall from Eq. \eqref{complexmetric} that $g^{++} = g^{--} = 0$ and $g^{+-} = g^{-+}=2$. 

For the translational motion of the virtual filament, the tension and rigidity coefficients are given by
\bsub
 \label{Vall1} \begin{eqnarray}
\gamma_1 + \rmi K \gamma_2 &=& P_0 - \rmi K P_2 = \bra{\WW^+} \HP \ket{\VV_+}, \\
e_1 + \rmi K e_2 &=& \bra{\bY^F} \HP (\HL-i\omega_0)^{-1} \nn \\ &&  \left[ \mathbbm{1} - \ket{\VV_+} \bra{\WW^+}  \right] \HP \ket{\VV_+}.
\end{eqnarray} 
\esub
The other coefficients can be computed from expressions \eqref{coeffs3}. In some of the coefficients below we use underlined indices to denote symmetrizing of lower indices, e.g. $S_{\underline{\,a\,}} T_{\underline{\,b\,}} R_{\underline{\,c\,}} = \frac{1}{6} (S_a T_b R_c + S_a T_c R_b +S_b T_a R_c +S_b T_c R_a +S_c T_a R_b + S_c T_b R_a)$. With implicit summation over $m\in \{-1,1\}$ this yields
\begin{widetext}
\bsub 
 \begin{eqnarray}
\tilde{A}_0 - \rmi K\tilde{A}_2 &=& \bra{\WO} \HP \ket{\VV_+} \braket{\WW^+}{\dd_\theta \uu^{ww}}  + \bra{\WO} \HP \ket{\dd_\theta \VO} \braket{\WW^+}{\dd_\theta \uu^k_+}   \\
&& + \bra{\WW^+}(\HP \delta^{(m)}_{\ +} - P^{(m)}_{\hs\  +} ) \ket{\dd_{(m)} \uu^{ww}}  + \bra{\WW^{(m)}} \HP \ket{\dd^2_\theta \uu_0} \braket{\WW^+}{\dd_{(m)} \uu^k_+} \nn \\
&& - \bra{\WW^+} \HP \ket{\dd^2_\theta \uu^k_+} + \bra{\WW^+} \HP \rho_+  \ket{ r \dd_r \uu_0}  - \bra{\WW^+} \HP r^2 \ket{\VV_+}  
 -2 \bra{\WW^+} \HP \rho_+ \ket{\dd^2_\theta \uu_0}  - \braket{\WW^+}{\bF'' \uu^k_+ \uu^{ww}}, \nn \\
\label{appb12}  
\tilde{B}_0 - \rmi K\tilde{B}_2 &=&3 \left(  + \bra{\WO} \HP \ket{\VV_{\underline{\,+\,}} }\braket{\WW^+}{\dd_\theta \uu^{kk}_{\underline{+-}}} - \braket{\WW^+}{ \SSS^{kk}_{\underline{+-}}  } \braket{\WW^+}{\dd_\theta \uu^{k}_{\underline{\,+\,}} } \right. \\
&& - \braket{\WW^+}{ \SSS^{kk}_{\underline{+-}} }  \braket{\WW^+}{\dd_{(m)}\uu^{k}_{\underline{\,+\,}}  } 
 + \bra{\WW^+}(\HP\delta^{(m)}_{\underline{\,+\,}}   - P^{(m)}_{\hs \underline{\,+\,}} ) \ket{\dd_{(m)} \uu^{kk}_{\underline{+-}} } \nn \\
&& + \bra{\WW^+} \HP \rho_{\underline{\,+\,}}\ket{\dd_{\underline{\,+\,}} \uu^k_{\underline{\,-\,}} } +  \bra{\WW^+} \HP \rho_{\underline{\,+\,}} \rho_{\underline{\,+\,}}  \ket{\VV_{\underline{\,-\,}} } 
\left.  - \braket{\WW^+}{\bF'' \uu^k_{\underline{\,+\,}} \uu^{kk}_{\underline{+-}} }  - \frac{1}{6} \braket{\WW^+}{\bF''' \uu^k_{\underline{\,+\,}}  \uu^k_{\underline{\,+\,}}  \uu^k_{\underline{\,-\,}} } \right) , \nn  \\
\tilde{C}_0 - \rmi K\tilde{C}_2 &=& \bra{\WW+} \rho_+ \HP \ket{\dd_\theta \uu_0} + 2 \bra{\WW^+} \HP \ket{\dd_\theta \uu^k_+},  \\
\tilde{D}_0 - \rmi K\tilde{D}_2 &=& + \bra{\WO} \HP \ket{\VV_+} \braket{\WW^+}{\dd_\theta \uu^{dw}} 
 - \bra{\WW^{(m)} } \HP \ket{\dd_\theta \uu_0} \braket{\WW^+}{\dd_{(m)} \uu^k_+} 
\\ 
&& + \bra{\WW^+}  (\HP -  d_0 ) \ket{\dd_\theta \uu^k_+} 
 + \bra{\WW^+}(\HP\delta^{(m)}_+ -  P^{(m)}_{\hs +} ) \ket{\dd_{(m)} \uu^{dw}}  
- \braket{\WW^+}{\bF'' \uu^k_+ \uu^{dw}} + 2 \bra{\WW^+} \HP \rho_+ \ket{\dd_\theta \uu_0}, \nn \\
\alpha_0 - \rmi K\alpha_2 &=& - \frac{1}{\komega_0} \bra{\WW^+} \HP\ket{\dd_\theta \VO } \bra{\WO}\HP \ket{\VV_+}, \\
\Delta_0 - \rmi K \Delta_2 &=& \frac{1}{\komega_0} \bra{\WW^+} \HP \ket{\VO} \bra{\WO}\HP \ket{\VV_+},  \\
\kappa_0 - \rmi K \kappa_2 &=& \frac{2}{\komega_0} \bra{\WO} \HP \ket{\VV_+} \braket{\WW^+}{ \SSS^{kk}_{\underline{+-}} }, \\
 P_1^2+P_3^2 &=& | \bra{\WW^+} \HP \ket{\VV_-} |^2. \label{Vall2}
\end{eqnarray} \esub

Here, we have used
\begin{align}
\HPI &= \mathbbm{1} - \sum_{m\in \{ -1,0,1 \} } \ket{\VV_{(m)}} \bra{\WW^{(m)}}, && \label{Vall} \\
P^{(m)}_{\hs {(n)}} &= \bra{\WW^{(m)}} \HP \ket{\VV_{(n)}}, &m,n &\in \{-1,0,1\}  \nn\\
\ket{ \uu^k_{(m)} } &=  (\HL- \rmi \, m\, \omega_0)^{-1} \HPI \HP \ket{\VV_{(m)}}, &  m &\in \{-1,1\}  \nn \\
\ket{\uu^{dw} }&= \HL^{-1} \HPI \HP \ket{\dd_\theta \uu_0}, && \nn \\
 \ket{\uu^{ww} }&= - \HL^{-1} \HPI \HP \ket{\dd^2_\theta \uu_0}, && \nn \\
\ket{\SSS^{kk}_{(np)}  }&= \ket{\dd_\theta \uu^k_{(n)}}  \bra{\WO} \HP \ket{\VV_{(p)}} 
+ (\HP\delta^{(m)}_{(n)} - P^{(m)}_{\hs (n)}) \ket{\dd_{(m)} \uu^k _{(p)}}  \nn  +  \rho_{(n)} \HP \ket{\VV_{(p)}} - \frac{1}{2} \ket{\bF''  \uu^k_{(n)} \uu^k_{(p)}}, & n,p &\in \{-1,1\} \nn \\
\ket{\uu^{kk}_{(mn)}} &= (\HL - \rmi (m+n) \omega_0 )^{-1} \HPI \ket{\SSS^{kk}_{(mn)}} 
    - \rmi m \bra{\WO} \HP \ket{\VV_{(n)}} \  (\HL - \rmi (m+n) \omega_0 )^{-1} \ket{\uu^k_{(m)}}. &  m,n &\in \{-1,1\} \nn 
\end{align} 
\end{widetext}

\end{document}